\documentclass[aps,pre,onecolumn,citesort,longbibliography]{revtex4-2}
\usepackage{amsfonts}
\usepackage{epstopdf, epsfig}
\usepackage{ulem}
\usepackage{amsthm}
\usepackage{amsmath}
\usepackage{mathtools}
\usepackage{amscd}
\usepackage{subfigure}
\usepackage{graphicx}
\usepackage{color}
\usepackage{bm}
\usepackage{natbib}
\usepackage{psfrag}
\usepackage{float}
\usepackage[colorlinks=true,linkcolor=blue]{hyperref}
\usepackage{geometry}
\usepackage[font=small,labelfont=bf,format=plain,justification=raggedright]{caption}
\usepackage{tabularx}
\usepackage{makecell}
\usepackage{booktabs}
\usepackage{paralist}
\usepackage[english]{babel}
\usepackage{mathrsfs}
\usepackage{cleveref}
\usepackage{comment}

\begin{document}

\title{Stabilization of the Rayleigh-B\'enard system\\ by injection of thermal inertial particles and bubbles}
\author{Saad Raza}
\email{saad.raza@univ-lille.fr}  
\author{Silvia C. Hirata}
\author{Enrico Calzavarini}
\affiliation{Universit$\acute{e}$ de Lille, Unit$\acute{e}$ de M$\acute{e}$canique de Lille - J. Boussinesq, UML ULR 7512, F 59000 Lille, France}

\date{May 2023}

\begin{abstract}
The effects of a dispersed particulate phase on the onset of Rayleigh-B\'enard (RB) convection in a fluid layer is studied theoretically by means of a two-fluid Eulerian modelization. The particles are non-Brownian, spherical, with inertia and heat capacity, and are assumed to interact with the surrounding fluid mechanically and thermally. 
We study both the cases of particles denser and lighter than the fluid that are injected uniformly at the system's horizontal boundaries with their settling terminal velocity and prescribed temperatures. The performed linear stability analysis shows that the onset of thermal convection is stationary, i.e., the system undergoes a pitchfork bifurcation as in the classical single-phase RB problem. Remarkably, the mechanical coupling due to the particle motion always stabilizes the system, increasing the critical Rayleigh number ($Ra_c$) of the convective onset. Furthermore, the particle to fluid heat capacity ratio provides an additional stabilizing mechanism, that we explore in full by addressing both the asymptotic limits of negligible and overwhelming particle thermal inertia.
The overall resulting stabilization effect on $Ra_c$ is significant: \textcolor{black}{for a particulate volume fraction of $0.1\%$} it reaches up to a factor 30 for the case of 
the lightest particle density (i.e.\ bubbles) and 60 for the heaviest one.  The present work extends the analysis performed by Prakhar \& Prosperetti (Physical Review Fluids 6, 083901, 2021) where the thermo-mechanical stabilization effect has been first demonstrated for highly dense particles. Here, by including the effect of the added-mass force in the model system, we succeed in exploring the full range of particle densities. Finally, we critically discuss the role of the particle injection boundary conditions which are adopted in this study and how their modification may lead to different dynamics, that deserve to be explored in the future.
\end{abstract}

\maketitle

\section{Introduction}

Virtually all fluids present in the natural environment contain dispersed matter, i.e., matter of a different composition from the one of the surrounding environment. This occurs in the form of solid particles, liquid drops or gaseous bubbles.  Examples include phenomena as diverse as sand and pollens in the near ground air, rain drops and ice crystals in clouds,  air bubbles entrained at the sea-air interface, volatile elements springing up in solidifying magma \textcolor{black}{and planktonic microorganisms dispersed in the ocean}\cite{degruyter2019volatiles, koyaguchi1990sedimentation, elkins2012magma, solomatov2007magma, chang2008natural, schwaiger2012ash3d, squires1995preferential, breuer2015iron,ArdeshiriPRE2016}.  
Such particle laden fluids are often set into motion by thermal differences present in the environment. However, while most of the time the role played by the dispersed phase is negligible for the overall dynamics of the fluid - think e.g. to the passive role of the grains of dust brought in suspension by a storm - there are situations where the mechanical agitation and/or the thermal coupling produced by the dispersed phase are relevant for the resulting flow dynamics \textcolor{black}{(one of such examples is bioconvection \cite{BeesMartinA2020AiB})}. 
Fluid-particle coupling phenomena are also of interest for industrial applications such as for optimization of mixing in bubble column reactors or for the design of particle based solar collectors \cite{dennis2013properties,pouransari2017effects,frankel2016settling,rahmani2018effects}. The widespread relevance of this topic has spurred extensive researches aimed at understanding the complex interplays governing the dynamics of particle-laden fluid flows \cite{brandt2022particle, kuruneru2017analysis, mathai2020bubbly}.\\  

To gain insight into particle-fluid systems, several approaches are possible, each corresponding to different physical conditions of the problem at hand  but also to distinct levels of abstraction in their description. At the most refined level we find the ``particle resolved" approach that treat the complete fluid-dynamical and thermal fluid-solid body interaction problem. This is the only sound modelization when the characteristic scales of the particulate phase are large as compared to the ones of the fluid. However, the major drawback of this approach is the complexity of its mathematical treatment even with state-of-the-art numerical methods, due to the high numbers of degrees of freedom involved \cite{Balachandar_2024}. When the particle sizes are of the same order or smaller than the typical spatial scales of the flow and of the heat transfer process, the particle description can be approximated by material points (dubbed ``point particles"). Their coupling to the fluid can be described by localized forces or source terms that satisfy global conservation laws of mass, momentum and energy. It falls in this case the so called Eulerian-Lagrangian modelization of particulate laden flows, where the fluid variables are \textcolor{black}{treated as continuous fields} evolving in the Eulerian frame, while the particles are described as individual entities in the Lagrangian frame. A further level of abstraction is represented by the two-fluid or Eulerian-Eulerian methods where both phases are described in terms of conservation equations for continuous and differentiable fields. The particulate state variables are typically their mass concentration, velocity and local temperature  but additional degrees of freedom can be introduced  (e.g. the local particle orientation, in case of non spherical particles) \cite{vie2016particle,xu2016three}. The latter approach has even a more restrictive domain of applications as it requires the particles not only to be tiny in size but also to be sufficiently numerous in order to be able to define continuous state variables in space and time. 
One appealing aspect of fully Eulerian models is the relative simplicity of their governing equations, which are expressed in the form of partial differential equations  that closely resemble the local conservation laws of fluids and transported scalar fields. This also implies that analytical approaches, as for instance the ones of hydrodynamic stability \cite{chandrasekhar2013hydrodynamic} \textcolor{black}{or fluctuating hydrodynamics \cite{de2006hydrodynamic}} can be straightforwardly adapted to this type of modelling.\\

In the  present work we aim at understanding how settling and rising thermal inertial particles affect the hydrodynamic stability of an immobile thermally stratified fluid layer. This will be achieved by adopting a fully Eulerian modelization and by performing a fluid dynamics linear stability analysis of the system's governing equations. 
The system we study in this article builds on the classical model of natural convection, known as Rayleigh-B\'enard (RB) model system, which is a layer of fluid between two horizontal planes kept at different constant temperatures, the above one being colder so that the fluid layer is slightly denser on top with respect to the bottom \cite{getling1998rayleigh}. If the system does not contain a pure fluid but rather a suspension of material particles, it goes under the name of particulate Rayleigh-B\'enard (pRB) system \cite{park2018rayleigh}. 
The research questions of interest in the pRB context are multiple. On one hand it is interesting to understand how the fluid flow affects the particle dynamics. In particular how it impacts on the  particle spatial distribution and clusters formation, or how it affects the settling speeds, the resulting sedimentation patterns forming at the walls and the possibility of particle resuspension and entrainement by the flow.
On the other hand it is pertinent to tackle how the feedback of the particulate phase can in turn influence the spatial structure and the temporal evolution of the flow, affecting its thermal stability, modulating the heat transfer across the system or disrupting the coherent flow structures that characterize the single phase RB flow. 
Many of these questions have been addressed in former studies and a comprehensive overview of them goes beyond the scope of this brief introduction. It is worth mentioning here the pioneering experimental studies by Solomatov et al. \cite{SolomatovEPSL1993} and Lavorel et al. \cite{lavorel2009sedimentation} focused on the settling dynamics of solid particles and to their resuspension in vigorously convective fluids, the more recent experiments on vapour droplets dynamics in a supersaturated RB cell (cloud chamber) \cite{Chandrakar_Cantrell_Krueger_Shaw_Wunsch_2020}, the studies on the dynamics of large non-isotropic particles in RB \cite{Jiang_Calzavarini_Sun_2020,JIANG2021TAML}. More frequent are the  numerical studies that have attempted a characterization of the one-way coupled dynamics of particles in RB  \cite{CalzavariniPF_RB2D, patovcka2020settling,patovcka2022residence,DenzelPRF2023,Xu_Xu_Xi_2024}.
When the so called two-way coupling is considered, i.e. the particle feedback on the fluid flow,  the dynamics and the  parameter space of the problem becomes much wider. 
Oresta et al. \cite{LakkarajuPNAS2013, oresta2013effects, oresta2014multiphase} showed that the presence of vapor bubbles or suspended particles in a cylindrical convective cell significantly influence flow and heat transfer. 

Studies with a similar Eulerian-Lagrangian point-particle numerical approach where conducted by Park et al. \cite{park2018rayleigh} and in \cite{gereltbyamba2019flow, demou2022turbulent, Yuhang&YangPF2023, SunWanSunPF2024}. This line of research goes even beyond the RB setting and extend to general convective turbulent flows, see e.g. \cite{zamansky2016turbulent}.
Numerical studies adopting particle resolved approach,
where both mechanical and thermal couplings between dispersed phase and the fluid are included, are on the other hand quite recent, with limitations in the number of particles   \cite{gu2018influence, takeuchi2019flow, ChenPropseretti2024}. This constraint however will become likely less severe in the forthcoming future as more computational power will be available.

In this work, we build upon the recent key findings of the work by Prakhar \& Prosperetti \cite{prakhar2021linear}, which demonstrated theoretically by means of a two-fluid modelization that the introduction of particles, whose density is much larger than the fluid one, has a sensible stabilizing effect on the onset of convection in the Rayleigh-B\'enard system. This stabilizing influence becomes increasingly pronounced with rising the particle concentration and the mass density and is primarily attributed to the mechanical interactions between the particles and the fluid. Furthermore, these authors find that the thermal inertia of the particles acts as an additional stabilization factor, and this regardless of the temperatures of the injected particles. 
One might argue that the physical origin of this mostly mechanical stabilization effect comes from the fact that collectively the falling particles acts as a widespread negative buoyancy force, and that the effect might be reversed for particles that are lighter than the fluid. 
\textcolor{black}{Indeed it is well known that a rising bubble front can destabilize a quiescent fluid layer \cite{ClimentPRL1999}, and that bubbles are very effective to enhance mixing \cite{ MathaiARFM2020}.}
On the other hand, the stabilization might be due to the enhanced dissipation produced by the dispersed phase, that would reduce the effective Rayleigh number (as the latter can be seen as a ratio between buoyant and dissipative forces), similarly to what occurs for the RB instability in superdiffusive media\cite{Barletta-anomalous}. According to the latter argument the particulate phase would lead to the fluid layer stabilization independently of its mass density and settling direction.
This open question represents the primary motivation of the present study. 

Our work extends the model adopted in \cite{prakhar2021linear} by considering particles of arbitrary mass density with respect to the fluid in order to encompass the cases of stone-like to bubble-like particles. This amount to take into account the role of the added mass hydrodynamic force.  Remarkably, we find that the mechanical stabilization effect persists even for particles which are lighter than the flow, although the effect tends to vanish when the particle mass density becomes negligible with respect to the fluid one (such as for the case of bubbles). In this study we also further explore the influence of the combined  thermo-mechanical coupling. The linear stability threshold for the  onset of convection  depends upon factors such as particle diameter and the specific heat capacity ratio of  the particle and fluid phases. A thermal/kinetic energy budget analysis is used to validate the stability results and to reveal the thermal and mechanical coupling contributions to system stabilization. 
Finally we critically discuss the role of the key assumption taken in this study for the particle injection modelization and how their variation could lead to different dynamics, that deserve to be explored in the future. 
\begin{figure*}[!htb]
\centering
\includegraphics[width=\textwidth]{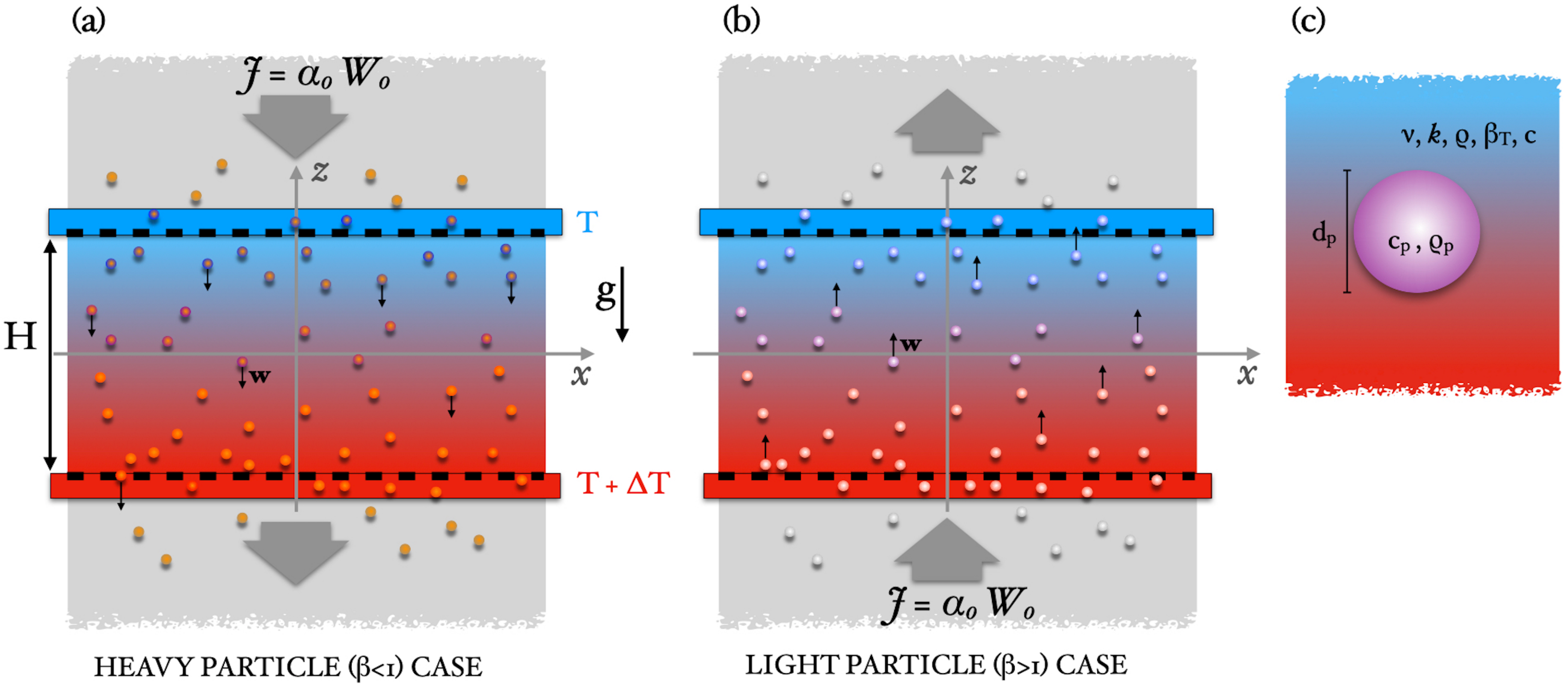}
\label{fig:visual}
\caption{\label{fig:visual}\textcolor{black}{Sketch of the particulate Rayleigh-B\'enard model system, for the case of heavy (a) and light particles (b). The fluid domain has a height $H$ and is infinitely extended in the horizontal direction. From the stability point of view only one lateral dimension is important , hence the system can be though and represented as two-dimensional. The horizontal boundaries are isothermal, with the bottom being warmer of $\Delta T>0$, and no-slip for the fluid velocity. The particles are injected either from top (for heavier than the fluid particles) or from the bottom walls (for light particles) at their terminal velocity with a prescribed volume flow rate. The overall set of parameters specifying the fluid and particle properties are in indicated in panel (c). They are for the fluid: $\nu$ kinematic viscosity, $\kappa$ thermal diffusivity, $\rho$ mass density, $\beta_T$ thermal expansion coefficient,$c$ the specific heat capacity at constant pressure. For the particle: $d_p$ the diameter,$c_p$ the specific heat capacity. $\rho_p$ the mass density.}}
\end{figure*}

\section{Particulate Rayleigh-B\'enard model system}
As we have already mentioned, in this study we adopt an Eulerian model system for the description of the dynamics of a non-Brownian suspension of particles in the Rayleigh-B\'enard (RB) setting. The particle volume concentration is assumed small everywhere so that the fluid can be considered incompressible and described by the conventional Boussinesq system of equations for the fluid velocity field  $\textbf{u}(\bm{x},t)$ and its temperature $T(\bm{x},t)$. 
However, due to the total conservation of momentum and thermal energy, the particulate phase can exert on the fluid both mechanical and thermal feedbacks. 
The particulate phase is characterized by the individual particle material properties, the mass density $\rho_p$,  particle diameter $d_p$, specific thermal capacity (at constant pressure)  $c_{p}$, and by the fields of volume concentration $\alpha(\bm{x},t)$, velocity $\textbf{w}(\bm{x},t)$ and temperature $T_p(\bm{x},t)$. 

The conservation equations of mass, momentum and heat for the fluid and particle phases read as follow: 
\begin{eqnarray}
      0 &=&  \nabla \cdot \textbf{u}, \label{eq:mass-fluid}\\
\frac{d \alpha}{dt}  &=& - \alpha (\nabla \cdot \textbf{w}),\label{eq:mass-particles}\\
 \frac{D\mathbf{u}}{Dt}  &=& \frac{-\nabla p}{\rho} +\nu\nabla^2 \mathbf{u}   + [1-\beta_T(T-T_r)]\mathbf{g} + \alpha \Bigg[ \bigg(\frac{D\mathbf{u}}{Dt} - \mathbf{g}\bigg) + \frac{\rho_p}{\rho} \bigg(\mathbf{g} - \frac{d\mathbf{w}}{dt}  \bigg) \Bigg], \label{eq:mom-fluid}\\
\frac{d \mathbf{w}}{dt} &=&  \beta\frac{D\mathbf{u}}{Dt}+\frac{\mathbf{u}-\mathbf{w}}{\tau_p}+(1-\beta)\mathbf{g},\label{eq:mom-particle}\\
   \frac{DT}{Dt}   &=& \kappa  \nabla^2  T  - \alpha E \frac{T - T_p}{\tau_T},\label{eq:temp-fluid}\\
     \frac{d T_p}{dt}&=& \dfrac{T-T_p}{\tau_{T}}.\label{eq:temp-particle}
\end{eqnarray}

Some observations are in order. First, $\frac{D}{Dt}() = \partial_t () + \textbf{u} \cdot \nabla()$ denotes the fluid material derivative, while $\frac{d}{dt}() = \partial_t() + \textbf{w} \cdot \nabla()$ is the particulate phase material derivative.
Second, we take into account the effect of three main hydrodynamics forces on the particle: the Stokes drag force, the fluid acceleration force with the added mass correction and the buoyancy. The drag is parameterized by the viscous response time  $\tau_p = d_p^2/(12 \nu \beta)$ where $\nu$ is the fluid viscosity. The added mass force intensity by the modified density ratio $\beta = 3\rho/(\rho + 2\rho_p)$ with $\rho$ the fluid mass density.
\textcolor{black}{Apart from the history force (i.e. unsteady Stokes drag) and Fax\'en correction, which are here ignored, eq. (\ref{eq:mom-particle}) is the well known Maxey-Riley-Gatignol equation \cite{maxey-riley-1983,gatignol-1983} for the dynamics of a material spherical  particle in an unsteady and inhomogeneous flow. The lift force, whose role can be relevant for bubbles ($\beta=3$), see e.g. \cite{PhysRevE.102.053102} , is here neglected.}
Third, the temperature inside each particle is assumed constant (lumped approximation) and its relaxation to the equilibrium is given by the timescale $ \tau_{T} =   d_p^2 E /(12  \kappa )$  with $\kappa$ the fluid thermal diffusivity and  $E=\rho_p c_{p}/(\rho c)$ with $c$ the fluid specific heat at constant pressure. \textcolor{black}{We note that the parameter $E$ describes the intensity of the thermal coupling between the particulate phase and the fluid (such coupling vanishes in the $E=0$ limit). The thermophoretic force on the particle is neglected, as normally done for non-Brownian particles.}
The remaining constants \textcolor{black}{represent} the fluid volumetric thermal expansion coefficient $\beta_T$ at the reference temperature $T_r$, the gravity vector $\textbf{g}$ and the pressure field $p(\textbf{x},t)$.

The system spatial domain is three-dimensional, confined by two-infinite parallel horizontal walls at coordinates $z = \pm H/2$, with $\hat{\textbf{z}}$ pointing upwards. \textcolor{black}{The boundaries are no-slip for the fluid velocity ($\textbf{u}=0$), and isothermal with a thermal gap of $\Delta T$ between them, the bottom wall being the warmest. In this way, when the thermal expansion coefficient of the fluid ($\beta_T$) is positive, an unstable density stratification is created. In order to help the reader to visualize the model system a schematic representation is provided in Figure \ref{fig:visual}.}

The particles are injected from one of the boundaries\textcolor{black}{, denoted as \textit{inlet},} at a constant volume concentration $\alpha_0$ and at their terminal velocity, $\mathbf{w}_0= (1-\beta)\tau_p\mathbf{g}$. 
Such a velocity can be easily computed by setting uniformly to zero the fluid velocity in Eq (\ref{eq:mom-particle}). 
 Particles with $\beta<1$, hereafter \textcolor{black}{denoted} as  \textit{heavy particles}, are injected from the top boundary, while $\beta>1$ ones, i.e. \textit{light particles}, are injected from the bottom. The particle inlet temperature is also prescribed at a specific value $T_p^*$. 
 \textcolor{black}{In other words the inlet particle condition is prescribed by a concentration particle flux whose intensity is $\mathcal{J}=\alpha_0 |\mathbf{w}_0|$, and by an analogous thermal flux $\mathcal{J}_T=\alpha_0 |\mathbf{w}_0| T_p^*$.}
 On the other hand, the particle accumulation on the opposite boundary, \textcolor{black}{the \textit{outlet}}, is neglected, as if they are removed from the domain as soon as they reach the opposite wall. \textcolor{black}{ We note that the corresponding outlet mass and thermal fluxes do not need to match those at the inlet. Importantly, the equations for the particulate phase are first-order in space (without a dissipation term in the form of a Laplacian) as a consequence just the inlet boundary condition is necessary for their solution.}   
 
\subsection{Dimensionless system}
In view of the stability study that \textcolor{black}{will be performed later,} it is convenient to adimensionalize the model system in terms of its height ($\mathbf{X}= \mathbf{x}/H$), the corresponding conductive time scale,  ($\mathcal{T} = t \kappa/H^2$) and the fluid density $\rho$. This leads to the introduction of the dimensionless fields:
$$\mathbf{U}=\mathbf{u}\frac{H}{\kappa}, \quad P = \frac{p H^4}{\rho \kappa^2 },
\quad\Theta = \frac{T-T_r}{\Delta T},
 \quad \mathbf{W}=\mathbf{w}\frac{H}{\kappa}, \quad\Theta_p = \frac{T_{p}-T_r}{\Delta T},$$
 respectively for the fluid velocity, pressure and temperature and for the particulate velocity and temperature.
\textcolor{black}{Without any loss of generality we take the reference temperature $T_r$ to be equal to the fixed temperature of the top (cold) wall.}
Keeping the same notation for the dimensionless material derivatives and the differential operator ($\nabla$), and making explicit the particles momentum feedback, equations (\ref{eq:mass-fluid})-(\ref{eq:temp-particle}) read:
\begin{eqnarray}
0 &=& \nabla \cdot \textbf{U},
\label{massfluidunperturb}\\
\frac{d \alpha}{d\mathcal{T}}  &=& - \alpha (\nabla \cdot \textbf{W}),\\
\frac{D\mathbf{U}}{D \mathcal{T}} &=& -\nabla P + Pr \nabla^2 \mathbf{U} + Pr Ra \Theta \hat{\mathbf{z}} 
+ \frac{\alpha}{2} \Bigg[ (\beta-1) \left(\frac{D\mathbf{U}}{D \mathcal{T}}+ \Lambda \hat{\mathbf{Z}}\right) -
    12 Pr (3-\beta )\frac{\mathbf{U} - \mathbf{W}}{{\Phi^2}}\Bigg],
    \label{momentumfluidunperturb}\\
\frac{d \mathbf{W}}{d\mathcal{T}}
 &=&  \beta \left( \frac{D\mathbf{U}}{D \mathcal{T}} + 12 Pr  \frac{\mathbf{U} - \mathbf{W}}{\Phi^2} \right) - {(1 - \beta)} \Lambda \hat{\mathbf{Z}},
\label{momentumparticleunperturb}\\
\frac{D\Theta}{D \mathcal{T}} &=& {\nabla^2 \Theta} - \alpha  12 \frac{\Theta - \Theta_p}{\Phi^2},
\label{energyfluidunperturb}\\
\frac{d \Theta_p}{d\mathcal{T}} &=& \frac{12}{E}\frac{\Theta - \Theta_p}{\Phi^2},
    \label{energyparticleunperturb}
\end{eqnarray}
where we have introduced the following dimensionless characteristic parameters, 
\begin{equation}
Ra=\frac{\beta_T \Delta T g H^3}{\nu \kappa}, \quad Pr = \frac{\nu }{\kappa}, \quad \Lambda = \frac{g H^3}{\kappa^2} = Ga Pr^2, \quad \Phi = \frac{d_p}{H}.\label{eq:control-param}
\end{equation}
Here $Ra$ represents the Rayleigh number which indicates the strength of thermally induced buoyancy as compared to the system mechanical and thermal dissipation, $Pr$ is the Prandtl number characteristic of the fluid phase, and $Ga$  the Galileo number, which parameterizes the ratio between the gravitational and viscous dissipative force.  Together with the modified fluid-to-particle density ratio $\beta$ and the particle global volume fraction $\alpha_0$ they constitute the full set of control parameters of the system. In conclusion the system state is identified by 7 parameters, three relative to the fluid ($Ra,Pr,\Lambda$), \textcolor{black}{two to the particulate phase ($\Phi,\beta$) and two related to the intensity of their mechanical and thermal couplings ($\alpha_0,E$)}.

In terms of the dimensionless variables the boundary conditions read, 
\textcolor{black}{
\begin{equation}
\mathbf{U} = 0;\quad  \Theta = 1 \quad \textrm{at} \quad  Z=- \frac{1}{2}, \quad \textrm{and}\quad  
\mathbf{U} = 0;\quad  \Theta = 0 \quad \textrm{at} \quad  Z= \frac{1}{2}
\label{eq:bc:fluid} 
\end{equation}
}
\begin{equation}
\alpha = \alpha_0; \qquad  \mathbf{W} = \mathbf{W}_0 =\frac{1-\beta}{\beta}\frac{\Lambda  \Phi^2}{12 Pr} \hat{\mathbf{Z}};\quad  \Theta_p =\Theta_p^*  \quad \textrm{at} \quad  Z=Z^*,\label{eq:bc:particle} 
\end{equation}
with $Z^*$ denoting the location of the inlet boundary condition. 
We observe that in the present model system the coupling between the fluid and the particles depends linearly on the local volume fraction $\alpha$. The parameter $E$, particle to fluid mass specific heat capacity ratio, controls the thermal coupling between the two phases. If $E\to 0$ then $\Theta_p \to \Theta$ meaning that the particle phase immediately adapts to the temperature of the fluid. In the opposite limit $E\to \infty$ the particles do not change their temperature and act as volume heat sources in the fluid. 
The parameter $\beta$ that varies in the range [0,3] discriminates between the cases of particles heavier than the fluid, $\beta<1$,
or lighter than the fluid, $\beta>1$, the limiting cases corresponding respectively to the infinitely heavy (ballistic limit) and infinitely light (which is a good approximation e.g. for air bubbles in water).
Finally we note that the parameter $\Lambda$ (or the Galileo number $Ga$) although it does not depend on the particle properties, it only becomes relevant when the coupling between the particle and the fluid is considered.

\section{Linear stability analysis}
\subsection{Conductive state}

We start to analyze a condition where the fluid is at rest, the particles are uniformly distributed in the domain with a volume concentration $\alpha_0$, and they continuously enter the system domain at the terminal velocity, $\bm{W}_0$, defined in (\ref{eq:bc:particle}).
Ref. \cite{oresta2013effects} observed that injecting particles with a different velocity (e.g., close to zero velocity) leads to the formation of a highly concentrated, nonuniform particle layer at the inlet boundary. Since this would complicate considerably the stability analysis \textcolor{black}{(see e.g. \cite{PhysRevE.102.053102} for the case of an isothermal layer)}, we adopt the same strict hypothesis: particles are injected at terminal velocity. This condition is realistic as long as the spatial distance across which the \textcolor{black}{particles} accelerates is much smaller than cell height. The order of magnitude of such a distance is \textcolor{black}{$|\textbf{w}_0| \tau_p$ (or in dimensionless terms $|\textbf{W}_0| \frac{\Phi^2}{12 Pr \beta}$ ).}

In such a stationary conductive state, usually denoted as base state, the pressure gradient is 
only $Z$-dependent function given by 
\begin{equation}
    -\bm{\nabla} P  = Pr Ra \Theta_0 \hat{\mathbf{Z}}+ \frac{\alpha_0(\beta-1)\Lambda \hat{\mathbf{Z}}}{2}+  12 Pr (3-\beta )\frac{{\mathbf{W}_0}}{{\Phi^2}},
    \label{fluid_conductive}
\end{equation}
 in which the $\Theta_0 = \Theta_0(Z)$ is the base fluid temperature field, which can be computed according to the following steps.
First, the particle temperature equations in the conductive state using equation (\ref{energyparticleunperturb}) reads,
\begin{equation}
\Theta_0 = W_0 \frac{E \Phi^2}{12} \partial_Z \Theta_{p0} + \Theta_{p0},
\label{eq15}
\end{equation}
here $\Theta_{p0} = \Theta_{p0}(Z)$ is the undistributed particle temperature. Second, upon elimination of the term $(\Theta - \Theta_p)$ from equations (\ref{energyfluidunperturb}) and (\ref{energyparticleunperturb}) yields,

\begin{equation}
     \partial_Z^2 \Theta_0 - W_0 E \alpha_0 \ \partial_Z \Theta_{p0} = 0,
     \label{eq16}
\end{equation}

Third and finally, by eliminating $\Theta_0$ from equations (\ref{eq15}) and (\ref{eq16}) we get,

\begin{equation}
{\partial_Z( W_0 \frac{E \Phi^2}{12}\partial_Z^2 \Theta_{p0} + \partial_Z \Theta_{p0} - W_0 E \alpha_0\ \Theta_{p0}) = 0}. 
\label{eq17}
\end{equation}
This third order linear differential equation can be solved analytically and, by means of the boundary conditions for the inlet particle temperature and for the fluid temperature on top and bottom walls, one \textcolor{black}{gets} the explicit expression:

\begin{equation}
{\Theta_{p0} = \Theta_{p}^* + C_1\big[1- e^{k_1(Z-Z^*)}\big]+ C_2 \big[1- e^{k_2(Z-Z^*)}\big]},
\label{Particle_conductive}
\end{equation}

here $C_1$ and $C_2$ are integration constants and
$$k_{1,2} = \frac{6}{W_0 E \Phi^2 }\bigg({-1\pm{\sqrt{1+\frac{W_0^2 E^2 \Phi^2 \alpha_0}{3}}}}\bigg)$$

Using equations (\ref{fluid_conductive}) and (\ref{Particle_conductive}), the base state temperature field in the fluid phase is therefore determined as

\begin{equation}
     {\Theta_0 = \Theta_{p}^* + C_1\big[1- (1+lk_1)e^{k_1(Z-Z^*)}\big]+ C_2 \big[1- (1+lk_2)e^{k_2(Z-Z^*)}\big]}.
     \label{fluid_baseT}
\end{equation}
where $l  = W_0 E \Phi^2/12$.
The integration constants are determined by the fluid temperature boundary conditions at the top and bottom of the cell, respectively, $Z = \pm\frac{1}{2}$. Their complete expression is:

\begin{equation}
C_1 =\frac{{e^{\left(\frac{1}{2} + Z^*\right) k_1} \left( -e^{\left(\frac{1}{2} + Z^*\right) k_2} - (1 + k_2 l) \left( e^{k_2} (-1 + \Theta_{p}^*) - \Theta_{p}^* \right) \right)}}{{e^{\left(\frac{1}{2} + Z^*\right) k_1} (-1 + e^{k_2}) (1 + k_2 l) + (1 + k_1 l) \left( -e^{\left(\frac{1}{2} + Z^*\right) k_2} (-1 + e^{k_1}) + (e^{k_1} - e^{k_2}) (1 + k_2 l) \right)}}
\end{equation}

\begin{equation}
C_2 = \frac{{e^{\left(\frac{1}{2} + Z^*\right) k_2} \left( -e^{\left(\frac{1}{2} + Z^*\right) k_1} - (1 + k_1 l) \left( e^{k_1} (-1 + \Theta_{p}^*) - \Theta_{p}^* \right) \right)}}{{e^{\left(\frac{1}{2} + Z^*\right) k_2} (-1 + e^{k_1}) (1 + k_1 l) + (1 + k_2 l) \left( -e^{\left(\frac{1}{2} + Z^*\right) k_1} (-1 + e^{k_2}) - (e^{k_1} - e^{k_2}) (1 + k_1 l) \right)}}.
\end{equation}

\begin{figure}[!htb]
\centering
\includegraphics[scale=0.5]{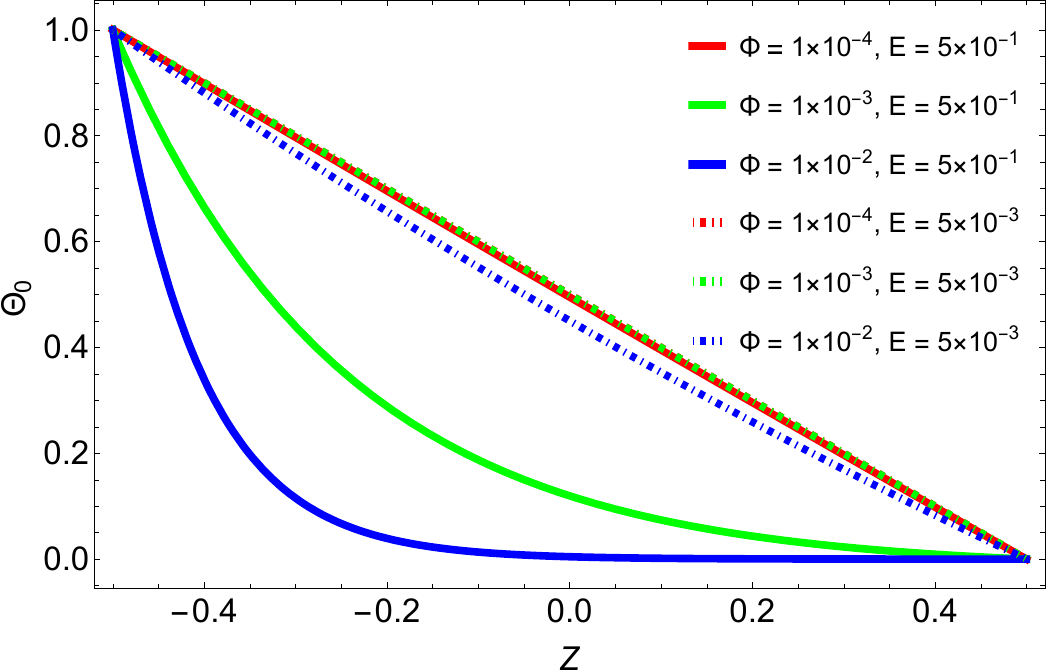}
\caption{The fluid temperature vertical distribution in the cell  diameter for heavy particles ($\Theta_{p}^* = 0$) \textcolor{black}{ with respect to different values of the particle diameter $\Phi$.} The particles to fluid heat capacity ratio $E = 5\times10^{-1}$ (solid lines) and  $E = 5\times10^{-3}$ (dashed lines), the particle to fluid density ratio $\beta = 0.5$. The particle volume fraction is $\alpha_0 = 10^{-3}$, and the Galileo number $\Lambda = 48 \times 10^{10}$.}
\label{BasePhi_Theta0}
\end{figure}

\begin{figure}[htb!]
\includegraphics[scale=0.5]{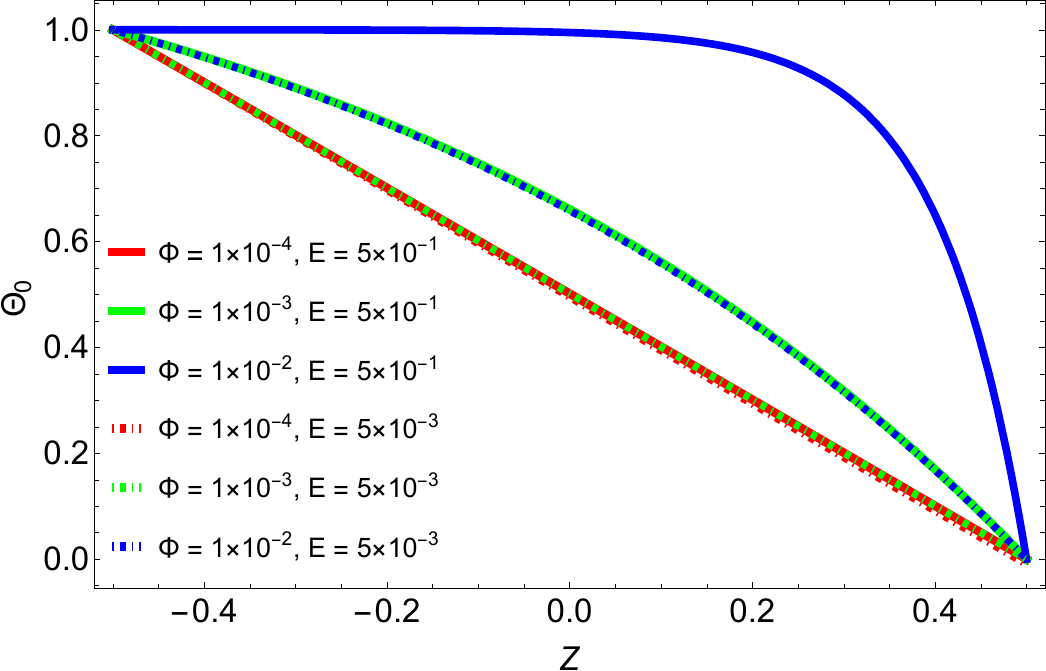}
\caption{Same as previous figure for light particles with $\beta = 1.5$ and $\Theta_{p}^* = 1$.
}
\label{Base}
\end{figure}

\begin{figure}[htb!]
    \centering
    \includegraphics[scale = 0.35]{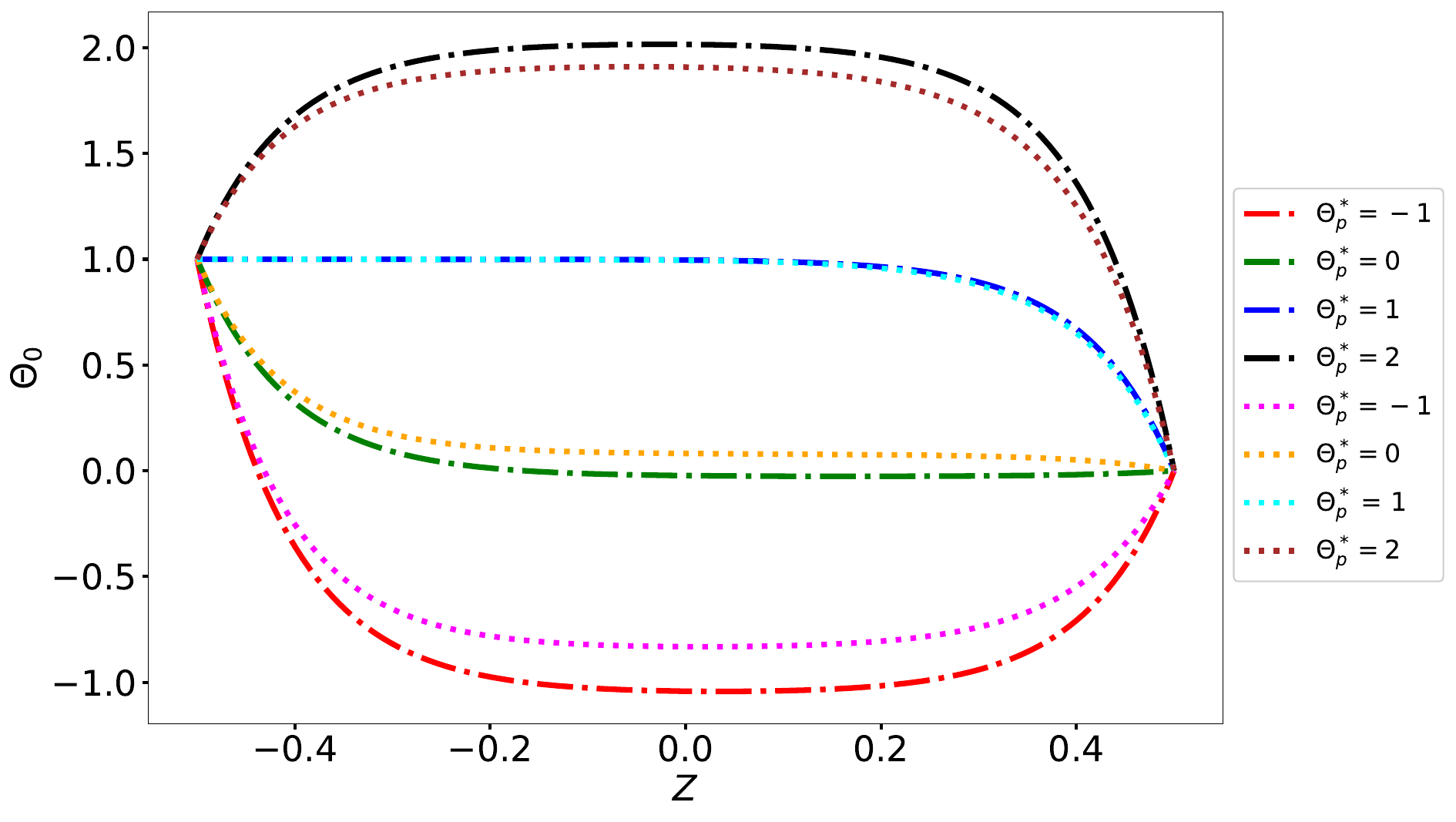}
    \caption{The fluid temperature vertical distribution in the cell for $\Theta_{p}^* = -1, 0, 1, 2$  for heavy $\beta = 0.5$ (dash dot lines) and light $\beta = 1.5$ (\textcolor{black}{dotted lines}) particles .The dimensionless particle diameter $\Phi = 10^{-2}$, the particles to fluid heat capacity ratio $E = 5\times10^{-1}$, the particle to fluid density ratio $\beta =0.5$.}
    \label{Thetainfluence_heavy}
\end{figure}

 The base fluid temperature vertical profile (eq. \ref{fluid_baseT}) is plotted in Figure \ref{BasePhi_Theta0} for the case of heavy particles with $\beta=0.5$ and different values of the heat capacity ratio $E$ and particle diameter $\Phi$. In this case, particles are injected from above with the same temprature as the one of the top horizontal boundary temperature ($\Theta_p^*=0$). One may observe that for small values of $E$ the profile remains close to linear, which is what is expected in the single-phase Rayleigh-B\'enard system, but  larger gradients appear for large values of $E$ and $\Phi$. In the case where both $E$ and $\Phi$ assume the largest values,  it may be observed that the fluid temperature remains almost constant on the upper part of the system. As a consequence, a region of strong unstable thermal stratification forms at the bottom.  Figure \ref{Base} shows the base fluid temperature vertical distribution for the case of light particles with $\beta=1.5$, injected from the bottom with the hot wall temperature ($\Theta_p^*=1$). The temperature field appears to be nearly the upside-down mirrored copy of the ones just observed for the heavy particle case. 
Additionally, the influence of inlet particles temperature for both heavy (solid lines) and light (dashed lines) particles is presented in Figure \ref{Thetainfluence_heavy}.  In this case temperature gradients can appear also at the particle inlet wall due to the local heating/cooling produced by the particulate phase on the fluid.

\subsection{Linearization}
We now proceed to linearize  equations (\ref{massfluidunperturb}-\ref{energyparticleunperturb}) assuming a small departure from the previously calculated base state. According to the standard linear stability approach \cite{chandrasekhar2013hydrodynamic}, the dependent vector and scalar fields in the governing equations are constructed as a superposition of the base state and infinitesimal perturbations 

\begin{eqnarray}
\quad  P &=&  P_0+ \epsilon P',\quad  \mathbf{U} = \bm{U}_0 + \epsilon \bm{U}', \quad \quad \Theta = \Theta_0 + \epsilon \Theta',  \\
\alpha &=& {\alpha}_0 + \epsilon {\alpha}',\quad  \mathbf{W}= W_0\mathbf{\hat{Z}}+ \epsilon \bm{W}' \quad \Theta_p = \Theta_{p0} + \epsilon \Theta_p', 
\end{eqnarray}
where $\epsilon<<1$, the primed quantities represent the perturbations, and  base state fields is denoted by the index zero. Substituting the above relations in the governing equations leads to a new system of equation at different orders in $\epsilon$. Upon deducting the base-state relations and disregarding higher-order terms in the perturbed quantities, the fluid-phase governing equations at the leading order become,

\begin{equation}
      0 =  \nabla \cdot \bm{U}',
    \label{eq27}
\end{equation}

\begin{equation}
\begin{aligned}
\partial_\mathcal{T} \mathbf{U'} &= -\nabla P' + Pr \nabla^2 \bm{U'} + Pr Ra \Theta' \hat{\mathbf{Z}} + \frac{\alpha_0(\beta -1)}{2}  \partial_\mathcal{T} \bm{U}' \\
&\quad- \alpha_0 \frac{6 Pr (3 -\beta) }{\Phi^2}(\bm{U'- W'}) + \left( \frac{(\beta -1)}{2} \Lambda  + 6 Pr (3 -\beta) \frac{ W_0}{\Phi^2}\right) \alpha' \mathbf{\hat{Z}},
\end{aligned}
\label{eq28}
\end{equation}

\begin{equation}
   \partial_\mathcal{T} {\Theta'} + U'_Z \partial_Z \Theta_0 =   \nabla^2 \Theta'   - 12 \alpha_0 \frac{(\Theta' - \Theta_{p}')}{\Phi^2} -  12 \alpha' \frac{(\Theta_0 - \Theta_{p0})}{\Phi^2},
\label{eq29}
\end{equation}

Adopting the same procedure as above the corresponding particle-phase governing equations become, 

\begin{equation}
     \partial_\mathcal{T} \alpha' + W_0 \partial_Z \alpha'  = - \alpha_0(\nabla \cdot \bm{W}'),
     \label{eq30}
\end{equation}

\begin{equation}
\partial_\mathcal{T} \bm{W}' + W_0 {\partial_Z \bm{W}'} = \beta \partial_\mathcal{T} \bm{U}' + 12 Pr \beta \frac{(\bm{U}'-\bm{W}')}{\Phi^2},
 \label{eq31}
\end{equation}

\begin{equation}
   \partial_\mathcal{T}  \Theta'_p + W_0 \partial_Z \Theta'_p + {W_Z}'  \nabla \Theta_{p0} = \frac{12}{E}\frac{\Theta' - \Theta'_p}{\Phi^2}.
     \label{eq32}
\end{equation}

Subject to the following boundary conditions

$${U_Z'} = 0  \quad \quad \Theta' = 0   \quad \quad \textbf{at} \quad Z = \pm\frac{1}{2}$$
 
$$\partial_Z{U'_Z} = 0  \quad \quad \textbf{at} \quad Z = \pm\frac{1}{2}$$
and for the particles
 $$\quad {W_Z'} = 0 \quad \quad 
\Theta_p' = 0\quad \quad  \textbf{at} \quad Z = Z^*$$
where $Z^*=1/2$ for heavy particles ($\rho_p > \rho$) and $Z^*=-1/2$ for light particles ($\rho_p <\rho$).

\subsection{Problem reduction}
The boundary condition for the particle velocity fluctuation at the inlet wall, $\bm{W}'=0$,   implies that $\partial_X \bm{W}'=0$, and by means of (\ref{eq31})  also that $\partial_Z \bm{W}'=0$ at the same wall ($Z=Z^*$). This has a consequence that $\nabla \cdot \bm{W}'=0$ at the inlet.
Now, as already observed in \cite{prakhar2021linear}, taking the divergence of Eq.~(\ref{eq31}) we get
\begin{equation}
 \partial_\mathcal{T} {(\nabla \cdot \bm{W}')} - W_0\partial_z {(\nabla \cdot \bm{W}')} =  - 12 Pr \beta  \frac{(\nabla \cdot \bm{W}') }{\Phi^2},
 \label{eq33a}
\end{equation}
This equation has solution 
\begin{equation}
\nabla\cdot \bm{W}'(Z,\mathcal{T}) = \nabla\cdot \bm{W}'|_{Z=Z^*}\ e^{\frac{ 12 Pr \beta}{\Phi^2 W_0} (1/2-Z)}. 
\end{equation}
It follows that $\nabla \cdot \bm{W}'=0$ everywhere in the system. The latter relation has an important consequence because through equation (\ref{eq30}), \textcolor{black}{which becomes an advection equation,} one obtains $\alpha'=\alpha'|_{Z=Z^*} = 0$.  Hence the chosen inlet condition for the particles imposes that the particle concentration is always constant throughout the system. In other words the phenomenon of particle clustering \textcolor{black}{cannot} occur in the present conditions. This \textcolor{black}{is} an important point that will be further discussed in the concluding section of this work.  

As next step we eliminate the pressure field from the fluid momentum equation (\ref{eq28}). This is achieved by taking the double curl of the fluid momentum equation. 

\begin{equation}
\begin{aligned}
\partial_\mathcal{T} \nabla^2 \bm{U'} &=   Pr \nabla^4 \bm{U'} - Pr Ra (\nabla( \partial_Z \Theta') -\nabla^2\Theta' \bm{\hat{Z}}) + \frac{\alpha_0(\beta -1)}{2}  \partial_\mathcal{T} \nabla^2\bm{U}' \\
& + \alpha_0 \frac{6 Pr (3 -\beta) }{\Phi^2}(-\nabla^2\bm{U}' + \nabla^2 \bm{W}' - \nabla(\nabla \cdot \bm{W}'))\\
& -  \left( \frac{(\beta -1)}{2} \Lambda  + 6 Pr (3 -\beta) \frac{ W_0}{\Phi^2}\right) (\nabla (\partial_Z \alpha') -\nabla^2\alpha' \bm{\hat{Z}}),
\end{aligned}
\label{eq43}
\end{equation}

by substituting $\nabla \cdot W' = 0$ and $\alpha' = 0$ the problem is now reduced to four equations for the unknowns $U_z'$, $W_z'$, $\Theta'$, $\Theta_p'$.

\begin{eqnarray}
\partial_{\mathcal{T}} \nabla^2 \bm{U}' &=& Pr \nabla^4 \bm{U}' - Pr \, Ra \left(\nabla (\partial_Z \Theta') - \nabla^2 \Theta' \bm{\hat{Z}}\right) \nonumber \\
 &+& \frac{\alpha_0 (\beta - 1)}{2} \partial_{\mathcal{T}} \nabla^2 \bm{U}' + \alpha_0 \frac{6 Pr (3 - \beta)}{\Phi^2} \left(-\nabla^2 \bm{U}' + \nabla^2 \bm{W}'\right), \\
\partial_{\mathcal{T}} \bm{W}'  &=& - W_0 \partial_Z \bm{W}'+\beta \partial_{\mathcal{T}} \bm{U}' + 12 Pr \beta \frac{(\bm{U}' - \bm{W}')}{\Phi^2}, \\
\partial_{\mathcal{T}} \Theta'  &=& - U'_Z \partial_Z \Theta_0+ \nabla^2 \Theta' - 12 \alpha_0 \frac{\Theta' - \Theta_{p}'}{\Phi^2}, \\
\partial_{\mathcal{T}} \Theta'_p  &=& - W_0 \partial_Z \Theta'_p - {W_Z}' \nabla \Theta_{p0}+\frac{12}{E} \frac{\Theta' - \Theta'_p}{\Phi^2}.
\end{eqnarray}

\section{Modal Analysis}

In the modal approach, the perturbation is assumed to have a monochromatic wave like behaviour in the homogeneous direction, which allows the decomposition of the perturbed quantities in Fourier modes of the form

\textcolor{black}{
\begin{equation}
    \xi'(X,Y,Z,\mathcal{T}) = \xi_n(Z)\; e^{(i k X+\lambda \mathcal{T})}+c.c.
    \label{eq44}
\end{equation}
where $ \xi'=\{\mathbf{U}',\Theta',P',\alpha',\Theta_p',\mathbf{W}'\}$, $c.c.$ is the complex conjugate, and $\xi_n$ is the normal mode amplitude varying in the non-homogeneous direction $Z$. Accordingly to the temporal stability approach, $k$ is the real wave number, and $\lambda=\lambda_r+ i\; \lambda_i$, where $\lambda_r$ is the temporal growth rate of the perturbation and $\lambda_i$ is its oscillation frequency.} 
By substituting the above prescribed transformations we get the final linearized non-dimensional system,

\begin{equation}
\begin{split}
(D^2 - k^2)\left[\left(1-\frac{\alpha_0(\beta-1)}{2}\right)\lambda U_n - (D^2 - k^2) Pr U_n\right] + Pr Ra k^2\Theta_n \\
+ \frac{ 6 \alpha_0 Pr (3-\beta )}{\Phi^2}(D^2 - k^2)(U_n - W_n) = 0,
\end{split}
\label{momentumpert_fluid}
\end{equation}

\begin{equation}
\lambda {W_n} - \frac{(1-\beta)}{\beta}\frac{\Lambda  \Phi^2}{12 Pr} D {W_n} - 12 Pr \beta \frac{({U_n- W_n})}{\Phi^2}-\beta \lambda {U_n} = 0,
\label{momentumpert_particle}
\end{equation}

\begin{equation}
    \lambda\Theta_n + {U_n}D \Theta_0 - (D^2 - k^2)\Theta_n + 12 \alpha_0 \frac{(\Theta_n- \Theta_{pn}) }{\Phi^2}= 0,
     \label{energypert_fluid}
\end{equation}

\begin{equation}
\lambda \Theta_{pn}-\frac{(1-\beta)}{\beta}\frac{\Lambda  \Phi^2}{12 Pr} D\Theta_{pn} + {W_n}D \Theta_{p0}-\frac{12}{E}\frac{(\Theta_n- \Theta_{pn})}{\Phi^2} = 0,
\label{energypert_particle}
\end{equation}

subject to the following boundary conditions for the fluid

$$ U_n = DU_n = 0, \quad \quad \Theta_n = 0  \quad \quad  \textbf{at} \quad Z= \pm\frac{1}{2},$$

and for the particles 

 $$\text{heavy ($\beta<1$):} \quad {W_n} = 0 \quad \quad 
\Theta_{pn} = 0\quad \quad  \textbf{at} \quad Z = \frac{1}{2},$$
 $$\text{light ($\beta>1$)} \quad {W_n} = 0 \quad \quad 
\Theta_{pn} = 0\quad \quad  \textbf{at} \quad Z = -\frac{1}{2}.$$

\textcolor{black}{where $D$ represents the derivative with respect to $Z$.} The linearized boundary value problem (BVP) (\ref{momentumpert_fluid})-(\ref{energypert_particle}) is solved by means of a shooting method.
\textcolor{black}{The values of $Ra$ and $\lambda$ are numerically obtained for prescribed values of the wave number $k$. Then, by minimizing $Ra$ with respect to $k$, the critical condition $(Ra_c, \lambda_c,k_c)$ is determined.}
The same approach and algorithm has been employed in previous studies \cite{ali2022soret,hirata2015convective} and is discussed in detail in \cite{Alvesetal}. We also employed the Galerkin method to validate the results obtained from the shooting method. Notably, both methods produced matching results, reinforcing the accuracy of our findings. For the sake of conciseness, we omit here the details of the numerical procedure.

\section{Results and discussion}
\textcolor{black}{This section describes the main results on the onset of convection in the pRB model system, obtained by means of the linear stability analysis. Before venturing into this, it is worth briefly discussing the system's behavior in some limiting cases, which offer an easier insight. Notably, we first examine the case of small particles ($\Phi \to 0$) where a perturbative solution of the system is possible. Second, we consider what happens in the limiting cases of the thermal coupling, i.e. when particles are thermally ineffective for the fluid, $E=0$, (pure mechanical coupling) and the opposite case when their particle thermal inertia is overwhelming, $E=+\infty$. The linear stability analysis will then focus on trends as compared to different parameters: the particle to fluid mass density ratio ($\beta$), the particle size ($\Phi$), the particulate volume flux ($\alpha_0 W_0$) and finally the particulate temperature ($\Theta_p^*$) (that determines the particulate inlet heat flux).}

\subsection{Perturbative solution in the $\Phi^2\to 0$ limit}
If equations (\ref{momentumparticleunperturb}) and (\ref{energyparticleunperturb}) are multiplied by $\Phi^2$ then we can take the limit for $\Phi\rightarrow 0$ and the solution $\mathbf{W} = \mathbf{U}$ is readily obtained. This implies that for small values of $\Phi^2$ the value of $W$ should be close to the one of the fluid velocity $\mathbf{U}$. In the limit of small but non-vanishing $\Phi^2$ a perturbative solution of the above equations can indeed be obtained. We consider that the solution of $\mathbf{W}$ and $\Theta_p$ will be of the form  

$$\mathbf{W} \simeq \mathbf{U} + \Phi^2 \mathbf{W}_1\  , \  \Theta_p \simeq \Theta + \Phi^2 \Theta_{p1} \textrm{, and} \ \alpha = \alpha_0 + \Phi^2 \alpha_1.$$

Substituting these expressions into the equations for $\mathbf{W}$, $\Theta_p$ and $\alpha$ at the leading order in $\Phi^2$ we obtain the following relations for the particle's variables:
\begin{eqnarray}
\mathbf{W} &=& \mathbf{U} +  \Phi^2 \frac{\beta-1}{12 Pr \beta}  \left( \frac{D\mathbf{U}}{D \mathcal{T}} + \Lambda \hat{\mathbf{Z}} \right),\\
\Theta_p &=& \Theta + \Phi^2 \frac{E}{12} \left( \frac{D\Theta}{D \mathcal{T}} \right),\\
\frac{D\alpha_0}{D\mathcal{T}} &=& 0, \qquad 
\frac{D\alpha_1}{D\mathcal{T}} =  \frac{\alpha_0}{12 Pr} \ \frac{1-\beta}{ \beta}  \nabla \mathbf{U} : \nabla \mathbf{U}.
\end{eqnarray}
 Taking this into account the fluid equations reduce to:
\begin{eqnarray}
    \frac{D\mathbf{U}}{D \mathcal{T}} &=& -\nabla P + Pr \nabla^2 \mathbf{U} + Pr Ra \Theta \hat{\mathbf{z}} 
    + (\alpha_0 + \alpha_1 \Phi^2) \Bigg[\frac{3 (\beta-1)}{2 \beta} \left(\frac{D\mathbf{U}}{D \mathcal{T}}+ \Lambda \hat{\mathbf{Z}}\right)\Bigg],\\
\frac{D\Theta}{D \mathcal{T}} &=& \nabla^2 \Theta + (\alpha_0 + \alpha_1 \Phi^2) E \frac{D\Theta}{D \mathcal{T}}.
\end{eqnarray}
Introducing the boundary condition ($\alpha = \alpha_0$) the above equations simplify to:
\begin{eqnarray}
   \frac{D\mathbf{U}}{D \mathcal{T}} &=& -\nabla P' + \frac{Pr}{ \left(1 - \alpha_0 \frac{3(\beta-1)}{2 \beta} \right) } \nabla^2 \mathbf{U} + \frac{Pr Ra}{\left(1 - \alpha_0 \frac{3(\beta-1)}{2 \beta} \right)} \Theta \hat{\mathbf{z}},\\
\frac{D\Theta}{D \mathcal{T}} &=& \frac{1}{1-\alpha_0 E}{\nabla^2 \Theta},
\end{eqnarray}
where $P'$ is a redefined pressure. Now redefining the time as $\mathcal{\tilde{T}} = \mathcal{T}/(1-\alpha_0 E)$. We obtain 
\begin{eqnarray}
   \frac{D\mathbf{\tilde{U}}}{D \mathcal{\tilde{T}}} &=& -\nabla \tilde{P} + \frac{Pr(1-\alpha_0 E)}{ \left(1 - \alpha_0 \frac{3(\beta-1)}{2 \beta} \right) } \nabla^2 \mathbf{\tilde{U}} + \frac{Pr Ra (1-\alpha_0 E)^2}{\left(1 - \alpha_0 \frac{3(\beta-1)}{2 \beta} \right)} \Theta \hat{\mathbf{z}},\\
\frac{D\Theta}{D \mathcal{\tilde{T}}} &=& \nabla^2 \Theta,
\label{eq52}
\end{eqnarray}
which has the form of the usual Boussinesq system. It is linearly unstable for $Ra (1-\alpha_0 E) \gtrsim 1708$ at any value of $Pr$ and $\beta$. This correction is always tiny in the range of parameters considered in this study. In fact we will consider at most $\alpha_0 E = 10^{-4}$.


\subsection{Thermal coupling limiting cases}

In this section we will present the thermal coupling limiting cases (i) $E\rightarrow 0$ and (ii) $E \rightarrow \infty $\\

Case (i) $E \rightarrow 0 $\\

When the thermal specific heat capacity ratio ($E$) number is very small, fluid and particulate temperatures are strongly coupled so that $\Theta_p \approx \Theta$. Fluid and particle momentum  equations are given by (\ref{momentumpert_fluid}) and (\ref{momentumpert_particle}). The fluid energy equation will become,

\begin{equation}
    \lambda\Theta_n + {U_n}D \Theta_0 - (D^2 - k^2)\Theta_n = 0,
     \label{}
\end{equation}

and particle energy equation can be discarded.\\

Case (ii) $E \rightarrow \infty $\\

In case of extreme heat capacity ratio ($E \rightarrow \infty$) the temperature of the particulate phase does not change and also in this case particle energy equation can be discarded. We have to reconstruct the base state which follows from equations (\ref{energyfluidunperturb}) and (\ref{energyparticleunperturb}) as,

\begin{equation}
    \partial_z^2 \Theta_0 - \frac{12 \alpha_0}{\Phi^2} (\Theta_0 - \Theta_p^*) = 0 
    \label{base_fluid}
\end{equation}

\begin{figure}[!htb]
\begin{center}
\subfigure[$E = \infty$]{\includegraphics[width=0.45\textwidth]{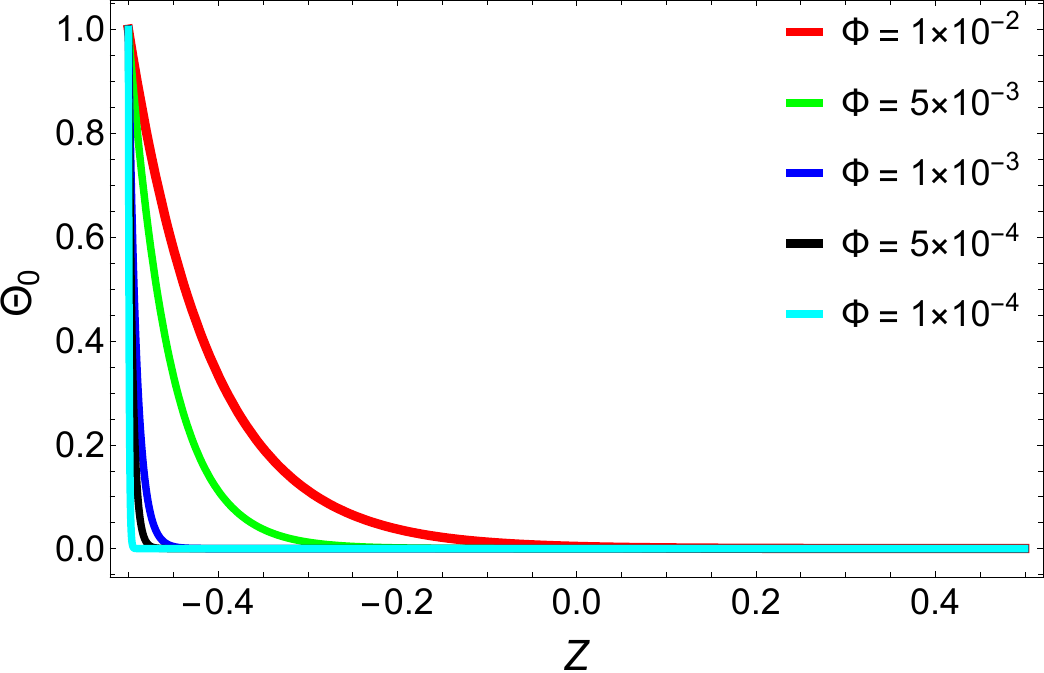}}$\quad$
\subfigure[$E = \infty$]{\includegraphics[width=0.45\textwidth]{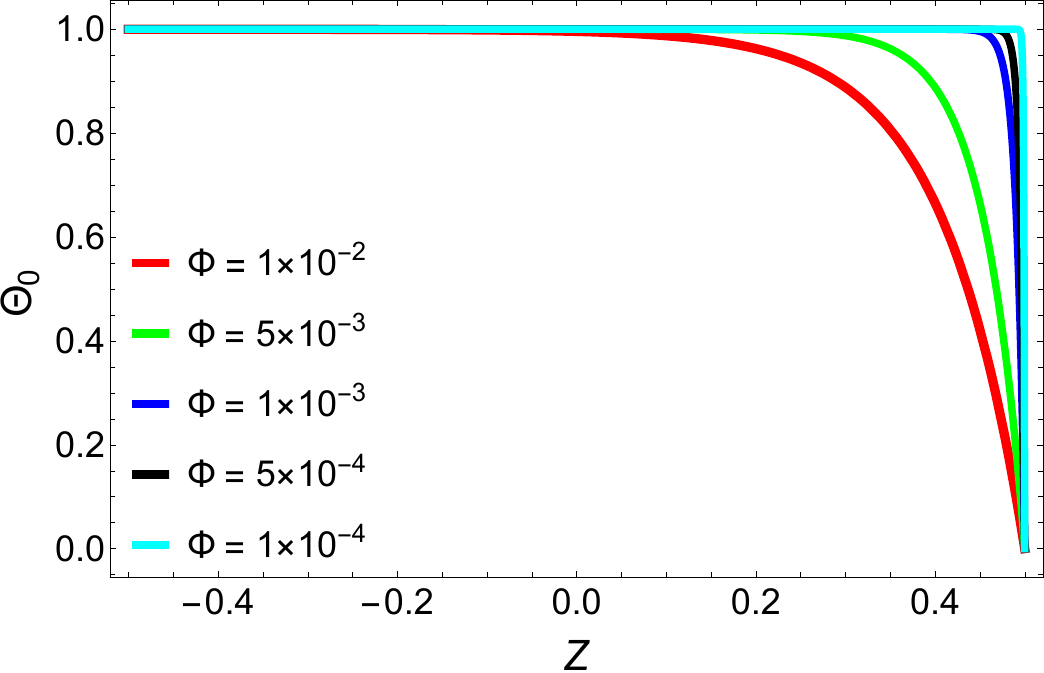}}
\caption{The fluid temperature vertical distribution in the cell for different particle diameter $\Phi$ (a) heavy particles ($\beta = 0.5, \Theta_{p}^* = 0$) and (b) the light particles ($\textcolor{black}{\beta = 1.5}, \Theta_{p}^* = 1$).}
\label{Base_infinity}
\end{center}
\end{figure}

The expression for the fluid base state, as illustrated in Figure \ref{Base_infinity}, is derived using equation (\ref{base_fluid}). After substitution of the boundary conditions we get,

\begin{equation}
\small
        \Theta_0 = \frac{e^{-\frac{2 \sqrt{3\alpha_0 z}}{\Phi}} \left(-e^{\frac{\sqrt{3 \alpha_0}}{\Phi}} + e^{\frac{2 \sqrt{3 \alpha_0 z}}{\Phi}}\right) \left(\left(e^{\frac{2 \sqrt{3\alpha_0}}{\Phi}} + e^{\frac{(1 + 2z) \sqrt{3 \alpha_0}}{\Phi}}\right) (-1 + \Theta_{p}^*) - \Theta_{p}^* - e^{\frac{(3 + 2z) \sqrt{3 \alpha_0}}{\Phi}} \Theta_{p}^*\right)}{-1 + e^{\frac{4  \sqrt{3 \alpha_0}}{\Phi}}}
\end{equation}

\subsection{Linear properties of the instability}

In this section we present the effects of both heavy and light particles on the stability of the RB system. In addition to the Rayleigh and Prandtl numbers $Ra$ and $Pr$, the presence of a particulate phase also introduces the following dimensionless parameters: the global particle volume fraction $\alpha_0$, the added mass-adjusted fluid-to-particle density ratio $\beta$, the dimensionless particle diameter $\Phi$, the heat capacity ratio $E$, the Galileo number $Ga=\Lambda/Pr^2$, and the dimensionless injection temperature of the particles $\Theta_p^*$.  

Parameter values for some representative systems of heavy and light particles are given in Table \ref{parameters}. The dimensionless parameters were obtained by considering a layer of height $H=0.1m$. In order to simplify the analysis, we chose to take water as a representative working fluid and fix in all the calculations $\alpha_0=10^{-3}$, $\Lambda=48\times10^{10}$ and $Pr=5$. \textcolor{black}{Even though} we cannot prove that the principle of exchange of stabilities holds for the present problem, we found that the least stable modes have \textcolor{black}{pure imaginary eigenvalues ($\lambda_i=0$)}. In other words, the system undergoes a pitchfork bifurcation giving rise to stationary convection for all cases studied.

\begin{table}[ht]
    \centering
    \begin{tabular}{|c|c|c|c|}
    \hline\hline 
    \multicolumn{4}{|c|}{\textbf{glass/water}}   \\
    \hline\hline 
    $\rho_p$ &$2.5 g/cm^3$ &$\beta$ &$0.5$ \\
    \hline
    $c_p$ &$0.84J/g$ &$E$ &$0.5$\\
    \hline
    $\rho$ &$1 g/cm^3$ &$\Lambda$ &$48\times 10^{10}$ \\
    \hline
    $c$ &$4.1813J/g$ &$Pr$ &$5$\\
       \hline
    $\kappa$ &$1.43\times10^{-7}m^2/s$ & &\\
    \hline
    $\nu$ &$10^{-6}m^2/s$ & &\\

    \hline\hline
    \multicolumn{4}{|c|}{\textbf{polypropylene/water}} \\
    \hline\hline 
    $\rho_p$ &$0.86 g/cm^3$ &$\beta$ &$1.1$ \\
    \hline
    $c_p$ & 1.92 $J/g$ &$E$ &$4 \times 10^{-4}$\\
    \hline
    $\rho$ &$1 g/cm^3$ &$\Lambda$ &$48\times 10^{10}$ \\
    \hline
    $c$ &$4.1813J/g$ &$Pr$ &$5$\\
       \hline
    $\kappa$ &$1.43\times10^{-7}m^2/s$ & &\\
    \hline
    $\nu$ &$10^{-6}m^2/s$ & &\\ 
    \hline\hline
     \multicolumn{4}{|c|}{\textbf{ice crystals/water}} \\
    \hline\hline 
    $\rho_p$ &$0.92 g/cm^3$ &$\beta$ &$1.6$ \\
    \hline
    $c_p$ &$2.09 J/g$ &$E$ &$0.45$\\
    \hline
    $\rho$ &$1 g/cm^3$ &$\Lambda$ &$48\times 10^{10}$ \\
    \hline
    $c$ &$4.1813J/g$ &$Pr$ &$5$\\
       \hline
    $\kappa$ &$1.43\times10^{-7}m^2/s$ & &\\
    \hline
    $\nu$ &$10^{-6}m^2/s$ & &\\
    \hline\hline
     \multicolumn{4}{|c|}{\textbf{air bubbles/water}}\\
    \hline\hline 
    $\rho_p$ &$0.001225 g/cm^3$ &$\beta$ &$3$ \\
    \hline
    $c_p$ &$1.005 J/g$ &$E$ &$3 \times 10^{-4}$\\
    \hline
    $\rho$ &$1 g/cm^3$ &$\Lambda$ &$48\times 10^{10}$ \\
    \hline
    $c$ &$4.1813J/g$ &$Pr$ &$5$\\
       \hline
    $\kappa$ &$1.43\times10^{-7}m^2/s$ & &\\
    \hline
    $\nu$ &$10^{-6}m^2/s$ & &\\
    \hline
     \end{tabular}
 \caption{Dimensional and dimensionless parameter values for some representative systems. The parameters are defined as follows:
$\rho_p$: Density of the particle material,
$c_p$: Specific heat capacity of the particle material,
$\rho$: Density of the fluid,
$c$: Specific heat capacity of the fluid,
$\kappa$: Thermal diffusivity of the fluid,
$\nu$: Kinematic viscosity of the fluid,
$\beta$:  modified fluid-to-particle density
ratio,
$E$: Particle-to-fluid thermal heat capacity ratio,
$\Lambda$: Galileo number,
$Pr$: Prandtl number.}
    \label{parameters}
\end{table}

\begin{figure}[htb!]
\centering
\includegraphics[width=0.7\textwidth]{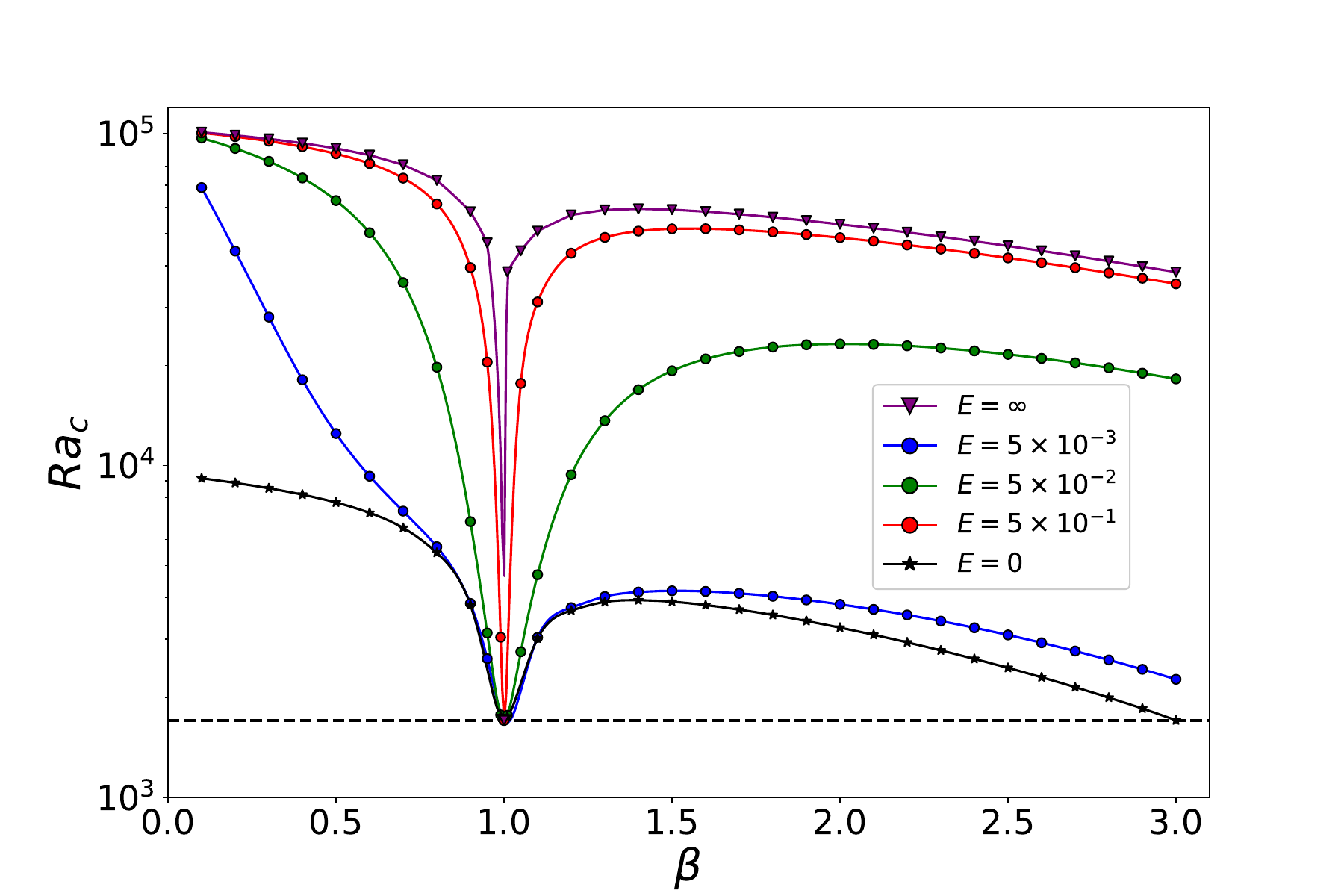}
\includegraphics[width=0.6\textwidth]{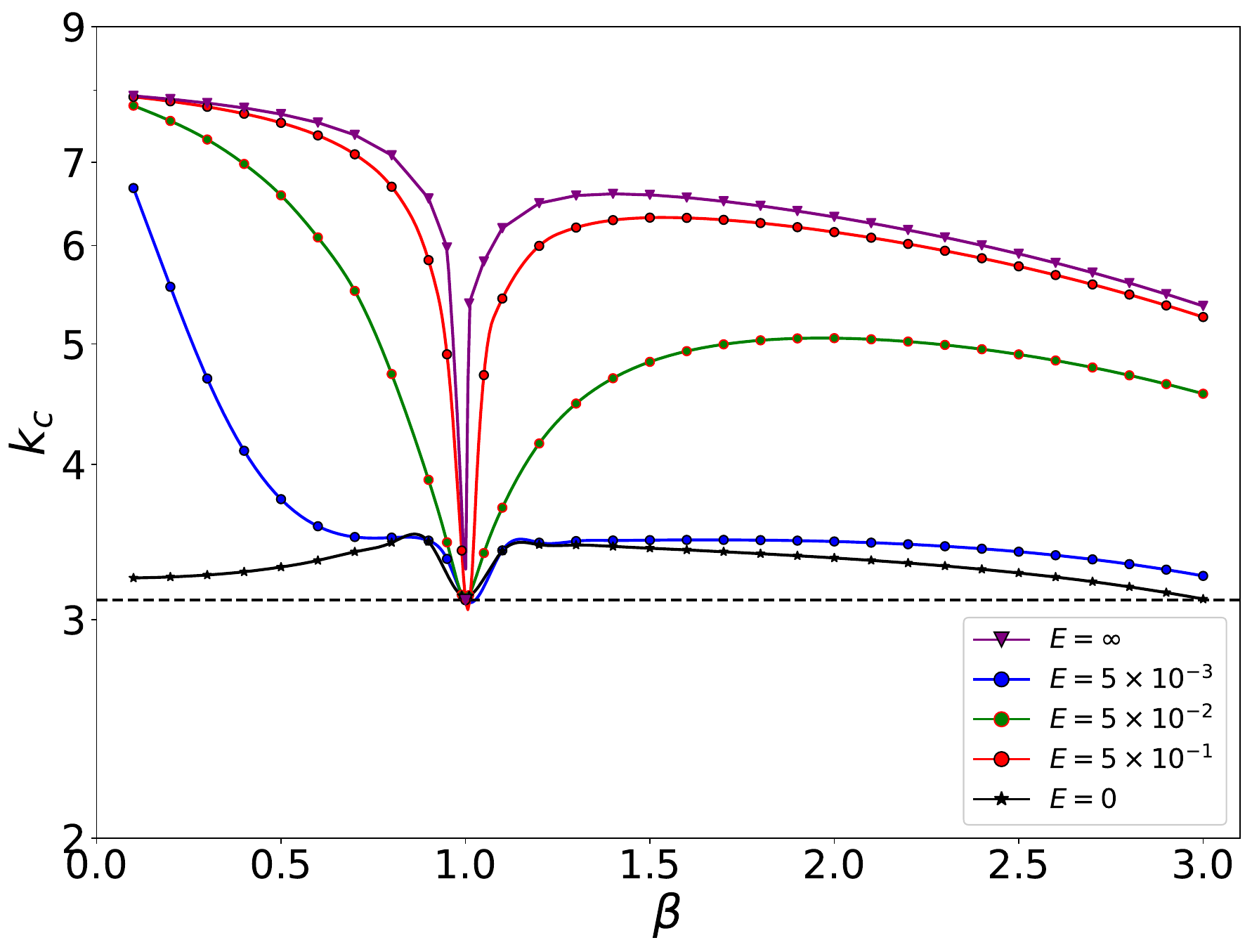}
\caption{(a) Critical Rayleigh number and (b) corresponding wave number as function of the modified density ratio $\beta$. Results obtained for fixed $\Phi = 0.01$ and $\Theta_{p}^*=0$ ($\beta<1$), $\Theta_{p}^*=1$ ($\beta>1$). The horizontal dashed lines correspond to the single-phase Rayleigh-B\'enard thresholds.}
\label{fig:beta-dependence}
\end{figure}

Let us first focus on the influence of  $\beta$ on the stability of the particulate RB system. The case $\beta=1$ corresponds to neutrally buoyant particles, and at this particular value our model presents a singularity \textcolor{black}{as the inlet particle flux can not be different from zero}. Heavy particles ($\beta<1$) are injected from the top with the cold wall temperature ({$\Theta_p^*=0$}), and light particles ($\beta>1$) are injected from the bottom with the hot wall temperature ({$\Theta_p^*=1$}), unless specified otherwise. Figure \ref{fig:beta-dependence} shows the critical thresholds as functions of $\beta$, for different values of the heat capacity ratio $E$. The limiting case $E =0$ \textcolor{black}{corresponds} to instant thermal coupling, i.e. the particle and fluid temperature fields are the same. On the other hand, when $E \rightarrow \infty$ the particles temperature remains constant $(\Theta_p = \Theta_p^*)$ and they act as internal heat source. One can remark that the introduction of particles, either heavy or light, \textcolor{black}{stabilizes} the system with respect to the single-phase RB threshold $Ra_c \simeq 1708$ and $k_c=3.11$. The system becomes increasingly stable as the density of heavy particles increases (i.e., as $\beta$ decreases from 1), and for large values of $E$ the critical thresholds tend to the asymptotic values $Ra_c \sim 10^5$ and $k_c\sim 8$. On the contrary, as light particles become lighter (i.e., as $\beta$ increases from 1), for large $E$ the system experiences a sharp stabilization followed by mild decrease in $Ra_c$, until the limiting value $\beta=3$. As it will be evident in the energy budget analysis of section \ref{energybudget}, at this value of $\beta$, the stabilizing role played by the particles is entirely due to the thermal coupling. Indeed, the Stokes drag term in equation (\ref{momentumpert_fluid}) vanishes for $\beta=3$. Note that in the neutral stability state, the added mass term of this equation (which multiplies $\alpha_0 (\beta-1)/2$) does not affect the linear stability results since the bifurcation is stationary ($\lambda =0$).
\textcolor{black}{Finally, by comparing the trends of $Ra_c$ and $k_c$ versus $\beta$ (panels (a) and (b) of figure \ref{fig:beta-dependence}) an approximate proportionality relation is clearly noticeable. This is likely related to the form of the fluid temperature base state which is characterized by strong gradients near the top/bottom walls (respectively for heavy/light particles see Fig \ref{BasePhi_Theta0} and Fig. \ref{Base}). This form of the  temperature profile effectively reduces the height of the thermally unstable layer in the system. As a consequence the convective rolls at the onset appears only in these layers which are characterized by smaller horizontal wave vectors as compared to the RB case (as the rolls have approximately a unit aspect ratio) and by a Rayleigh number that roughly increases by a factor $(H/\lambda_c)^3$ (i.e. the ratio between the usual Rayleigh number based on the cell height and the effective Rayleigh number based on the of the roll height). That is why when $k_c$ increases, $Ra_c$ also increases.}

\begin{figure}[htb!]
\centering
\includegraphics[scale=0.63]{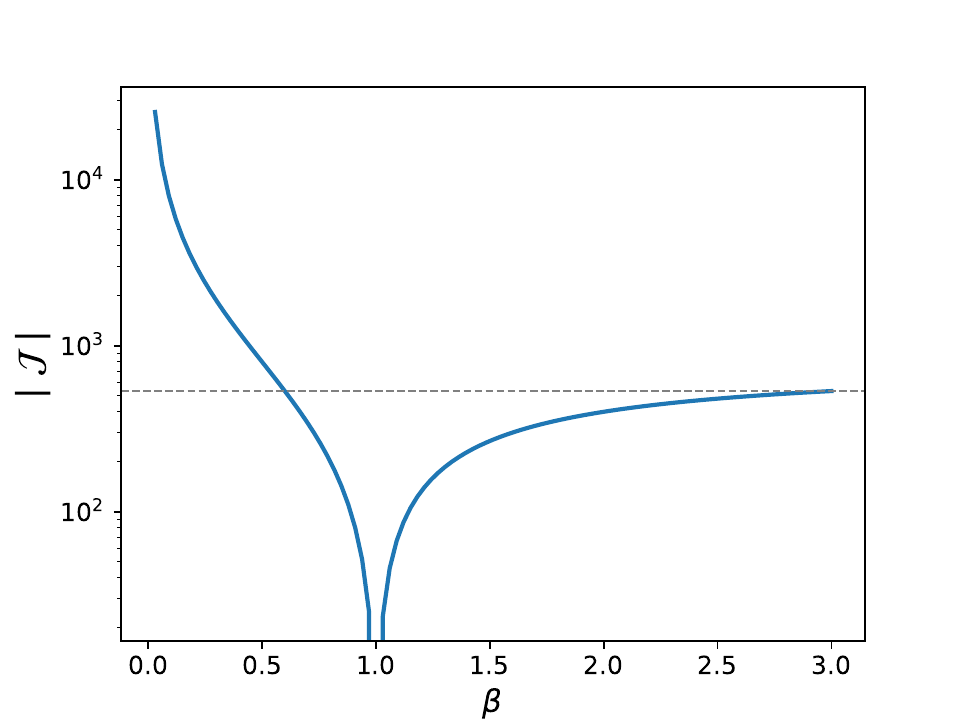}
\includegraphics[scale=0.34]{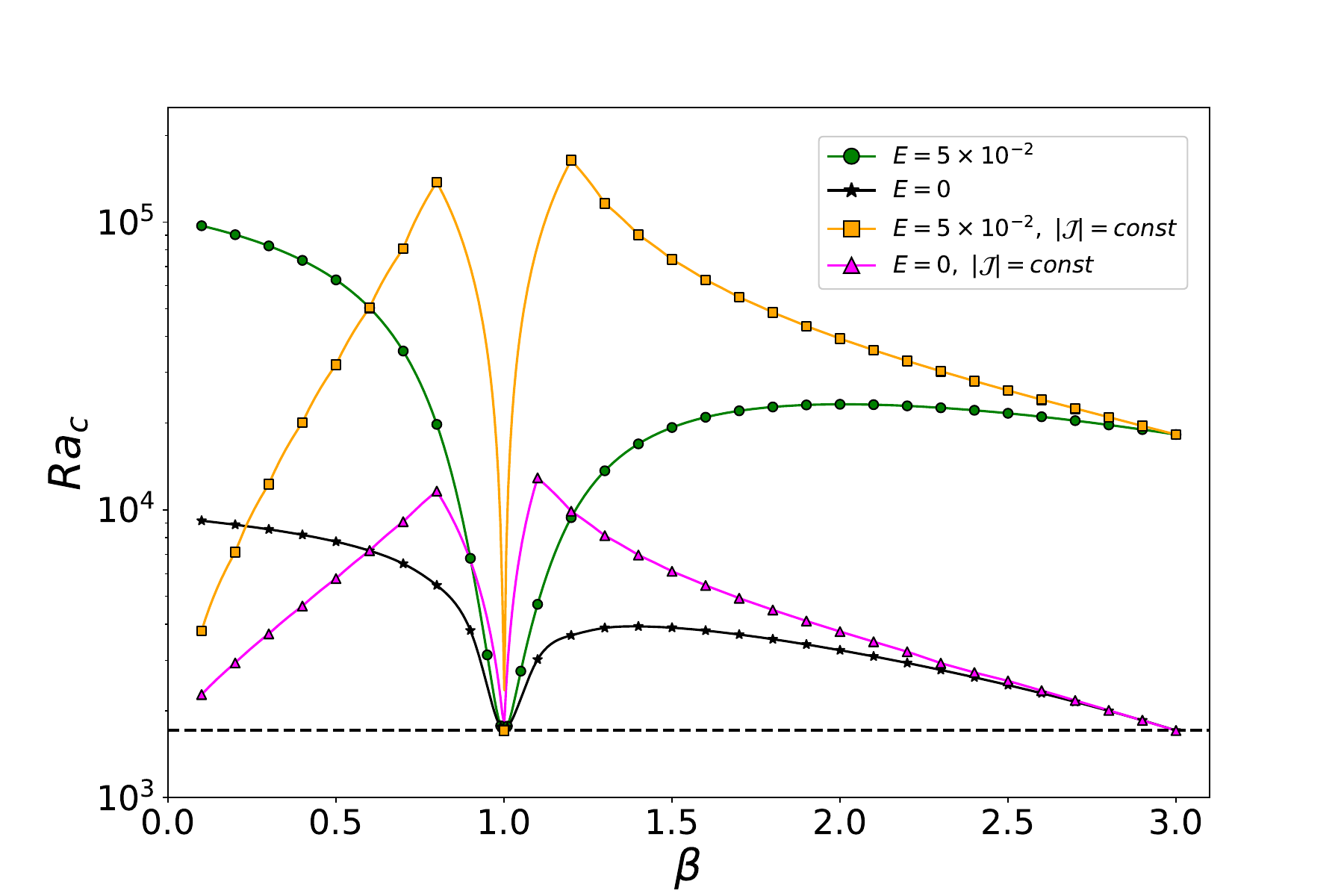}
\caption{\textcolor{black}{((a) Intensity of the particulate volumetric inlet flux $\mathcal{J}=\alpha_0 W_0$ as a function of $\beta$. The horizontal dashed line corresponds to the reference flux intensity taken for the calculations at fixed flux presented in panel (b).
(b) Critical Rayleigh number versus  $\beta$ for the cases of variable ($\mathcal{J}=\alpha_0 W_0$) and fixed inlet particle flux. In the second case, the fixed flux is taken equal to the case of $\beta=3$ and $\beta=0.6$.
All the other conditions are the same as in figure \ref{fig:beta-dependence}. The horizontal dashed line correspond to the single-phase Rayleigh-B\'enard threshold.}}
\label{fig:Jbeta}
\end{figure}

\textcolor{black}{One may wonder if the inlet particulate flux affects the stability of the overall system. This question is appropriate because the special choice of injecting particles at their terminal velocity $\bm{W_0}$ and at a prescribed fixed concentration $\alpha_0$ implies that the particle volumetric flux $\mathcal{J}=\alpha_0 \bm{W_0}$ is a function of $\beta$. In figure \ref{fig:Jbeta}(a), blue line, we show the dependence of the intensity of the particulate flux $\left| \mathcal{J} \right|$ as a function of $\beta$. Its qualitative behavior reflects some (but not all) of the features of the $Ra_c(\beta)$ curve of figure \ref{fig:beta-dependence}. The flux intensity is moderate for the case of bubbles, null for neutral particles, and progressively increasing and even diverging in the limit of very heavy particles. To better understand the impact of the particulate inlet flux on the instability as a function of the $\beta$-type of particle we carry on an additional stability calculation where the inlet flux is kept constant. The particles are still inserted in the system at their terminal velocity, $\bm{W_0}$, but their volume concentration $\alpha_0$ is adjusted so that $\left| \mathcal{J} \right|$ is the same for all $\beta$. In particular, we fix $\left| \mathcal{J} \right|$ to the value adopted for bubbles ($\beta=3$) or equivalently to the case of heavy particles with $\beta=0.6$ (corresponding to the dashed horizontal line in \ref{fig:Jbeta}(a).  The results for two representative cases ($E=0,0.05$) are shown in figure \ref{fig:Jbeta}(b). The curves, at fixed flux, have quite different trends. The extreme values of $\beta$ point to a mild stabilization of the system, while a stronger stabilization (larger $Ra_c$ values are attained for cases approaching neutral particles. The fact that these curves are not flat, i.e. independent of $\beta$, confirms that the stabilization effect can not be completely ascribed to the intensity of the particulate inlet flux, and that particulate hydrodynamics forces and feedback do play a role in the stability of this model system.}

\begin{figure}[htb!]
\begin{center}
\begin{minipage}[b!]{0.49\textwidth}
\includegraphics[scale=0.29]{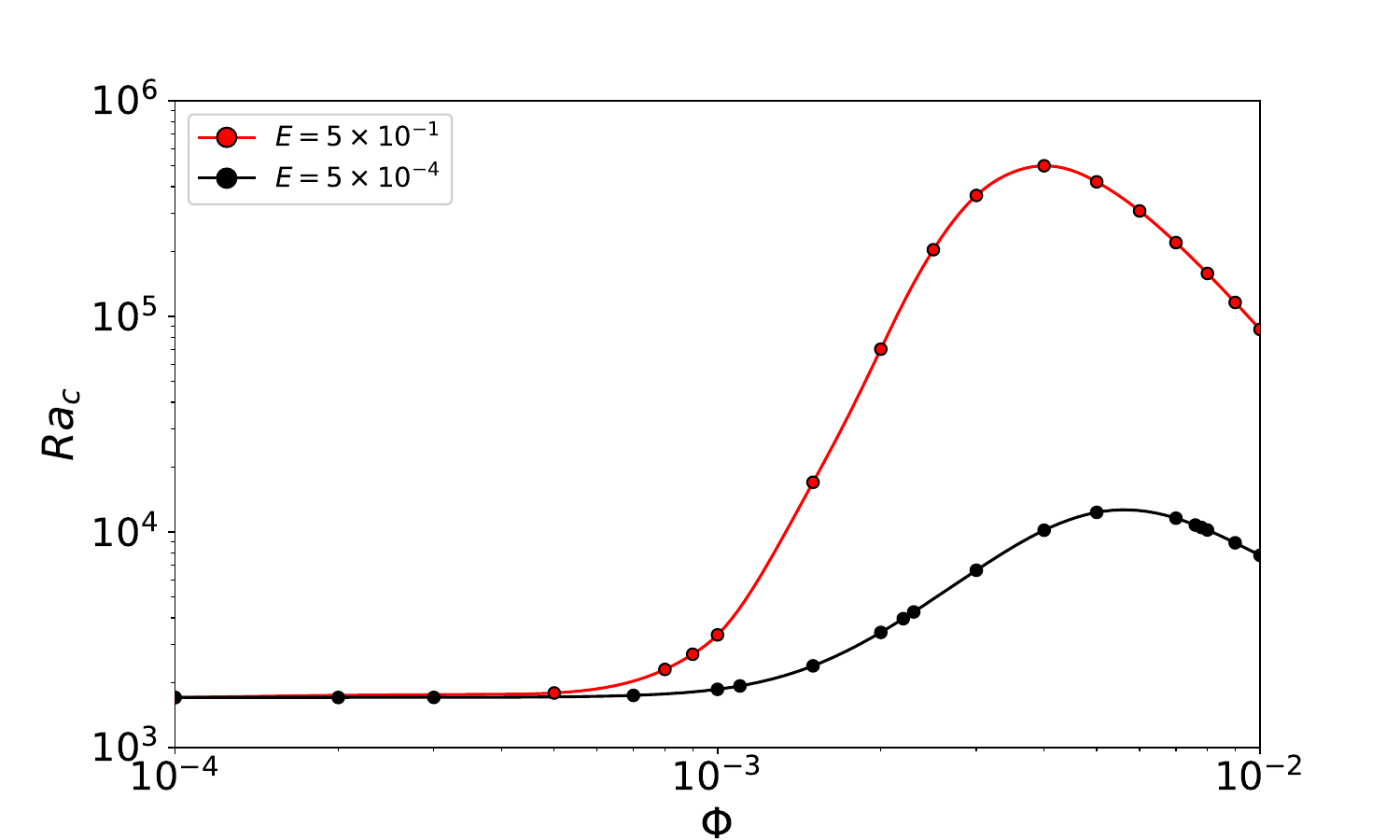}
\caption*{(a) $\beta=0.5$}
\end{minipage}
\hfill
\begin{minipage}[b!]{0.49\textwidth}
\includegraphics[scale=0.29]{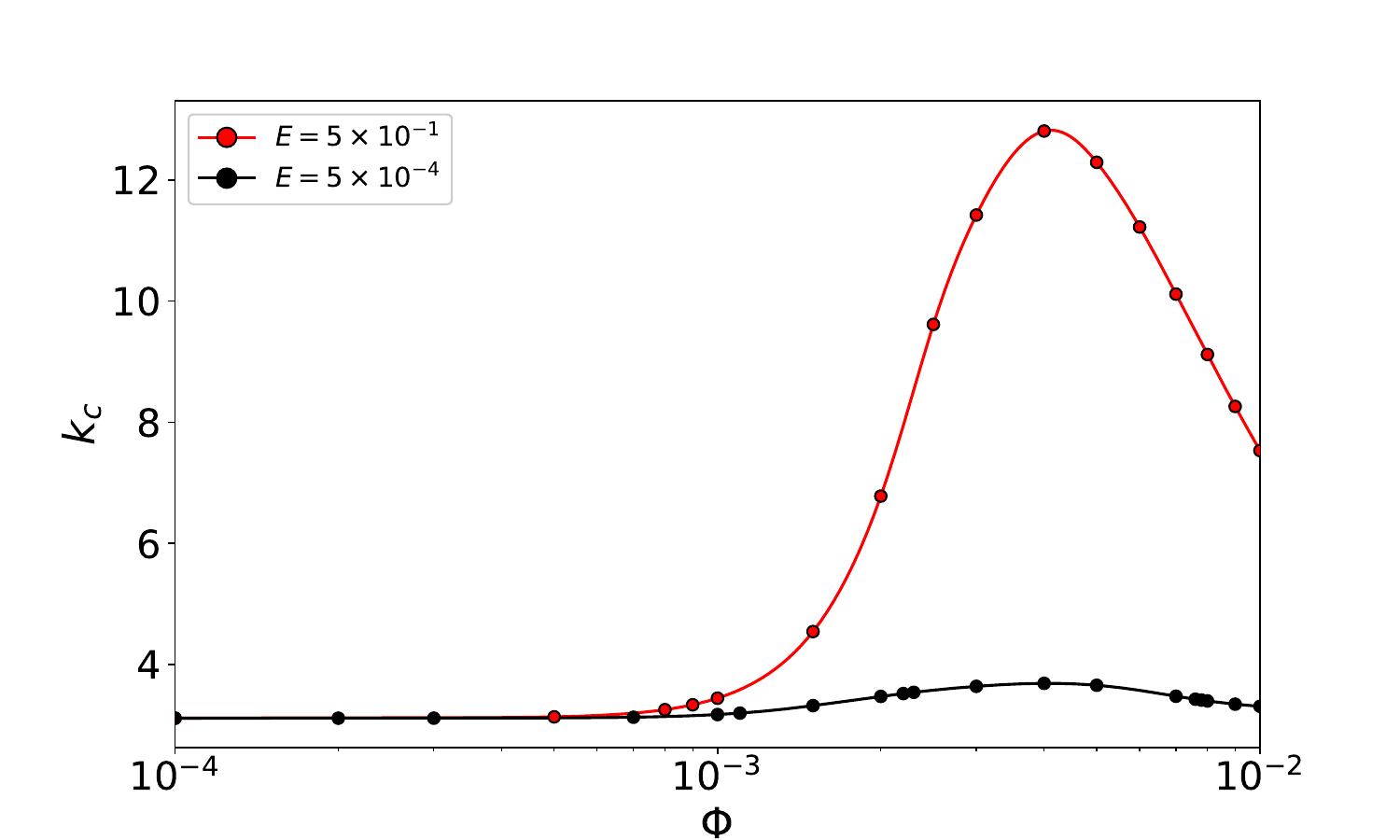}
\caption*{(b) $\beta=0.5$}
\end{minipage}
\begin{minipage}[b!]{0.49\textwidth}
\includegraphics[scale=0.29]{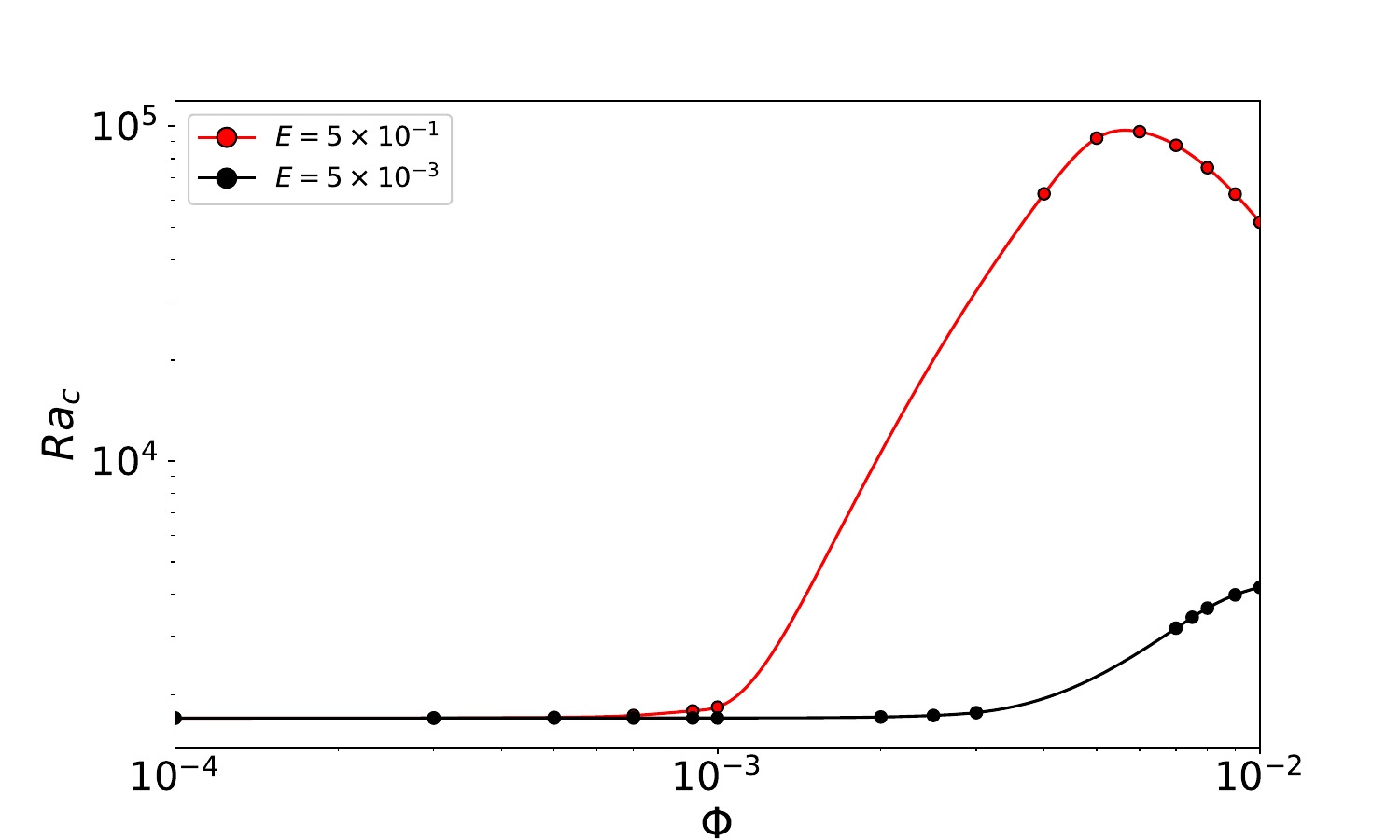}
\caption*{(c) $\beta=1.5$}
\end{minipage}
\hfill
\begin{minipage}[b!]{0.49\textwidth}
\includegraphics[scale=0.29]{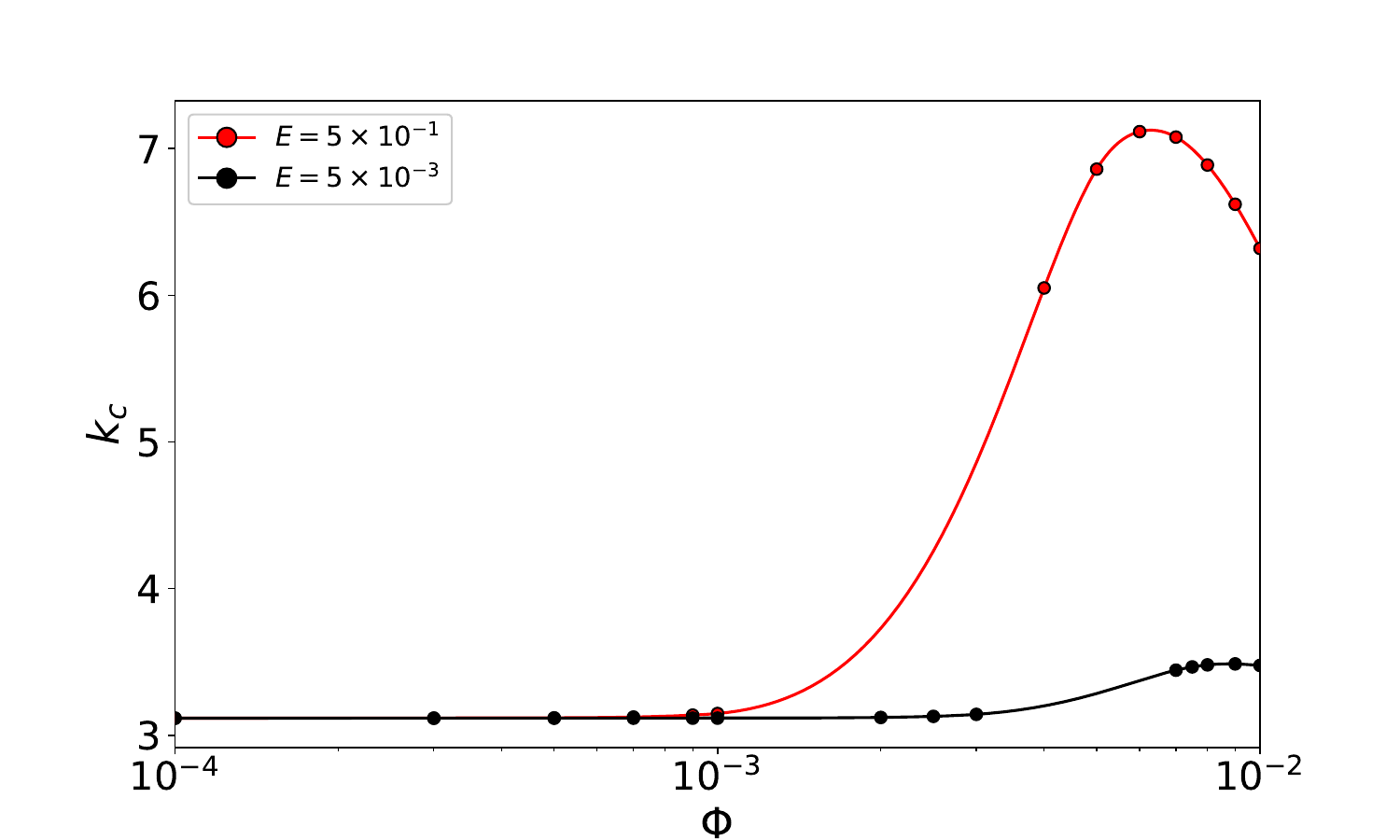}
\caption*{(d) $\beta=1.5$}
\end{minipage}
\end{center}
\caption{Evolution of the critical thresholds with particle diameter $\Phi$, obtained for : (a), (b) heavy particles with $\beta=0.5$ and $\Theta_{p}^*=0$; (c), (d) light particles with $\beta=1.5$ and $\Theta_{p}^*=1$.}
\label{phi}
\end{figure}

\begin{figure}[htb!]
\begin{center}
\begin{minipage}[t]{0.4\textwidth}
\includegraphics[width=\textwidth]{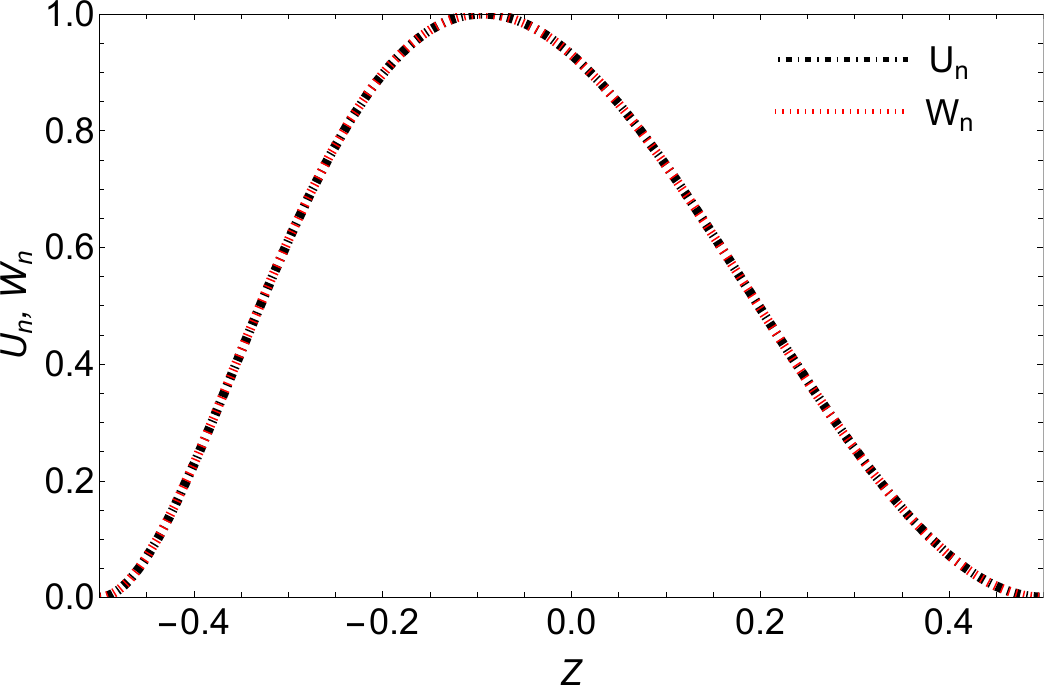}
\caption*{(a) $\Phi = 1 \times 10 ^{-3}$}
\end{minipage}
\hfill
\begin{minipage}[t]{0.4\textwidth}
\includegraphics[width=\textwidth]{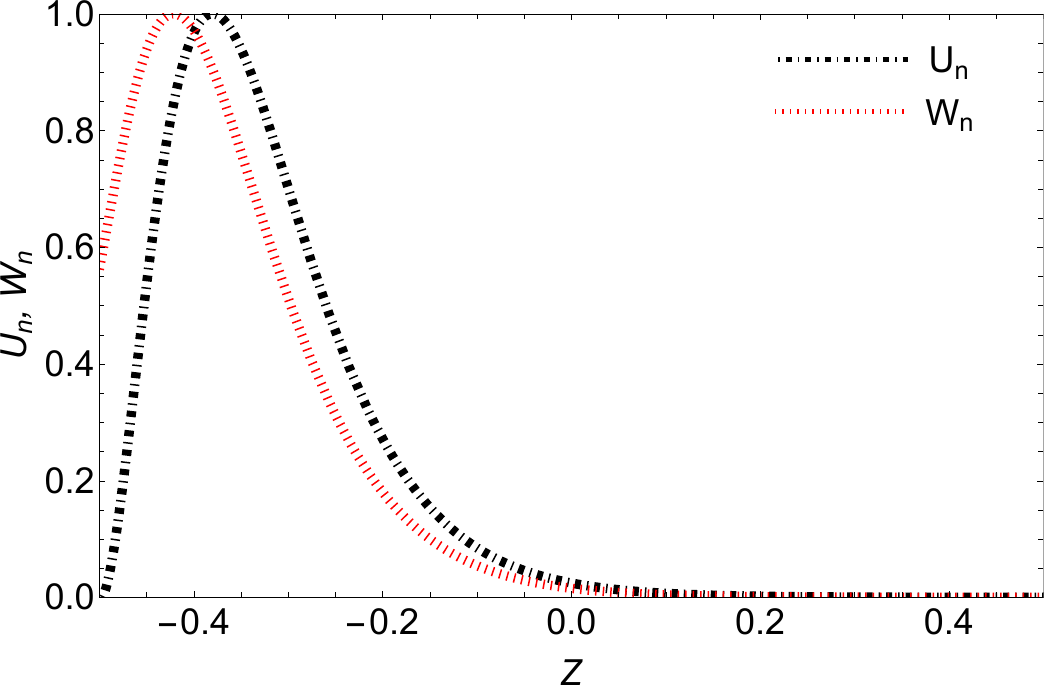}
\caption*{(b) $\Phi = 4 \times 10 ^{-3}$}
\end{minipage}


\begin{minipage}[t]{0.4\textwidth}
\centering
\includegraphics[width=\textwidth]{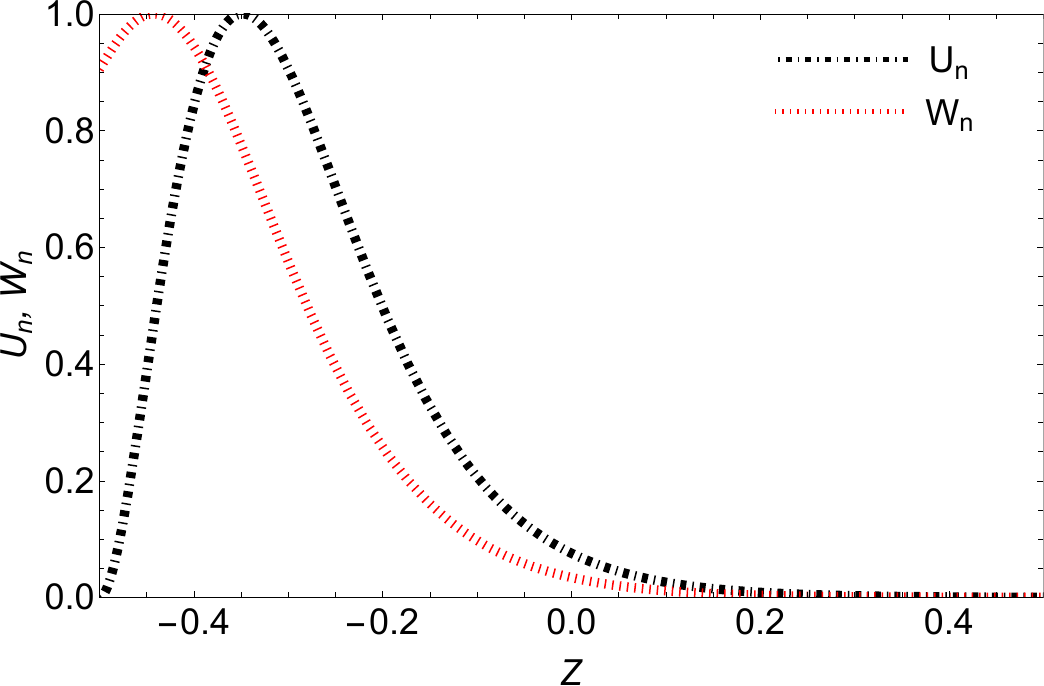}
\caption*{(c) $\Phi = 6 \times 10 ^{-3}$}
\end{minipage}
\end{center}
\caption{\textcolor{black}{Fluid and particles vertical velocity profiles at neutral conditions computed for parameters $\beta = 0.5$ and $E = 0.5$, illustrating changes in system behavior at three values of $\Phi$: (a) just before the critical "jump" in $Ra_c$ ($\Phi = 10^{-3}$), (b) at the inflection point where $Ra_c$ reaches its peak ($\Phi = 4 \times 10^{-3}$), (c) and after the gradual decrease in $Ra_c$ ($\Phi = 6 \times 10^{-3}$).}}

\label{eigen_heavy_phi}
\end{figure}

Figure \ref{phi} illustrates the effects of the dimensionless particle diameter $\Phi=d_p/H$ on the stability thresholds of heavy and light particles. The trends are similar for the two cases and for all values of the heat capacity ratio $E$: the critical Rayleigh number (wave number) remains close to the single-phase value for small $\Phi$, then it quickly rises, reaches a maximum and then starts to decrease. This sharp increase on $Ra_c$ can be explained as follows. From the particle momentum equation (\ref{momentumpert_particle}) in neutral conditions, and for $\lambda=0$, one may infer that the magnitude of the velocity difference $|U_n'-W_n'|$ grows with $\Phi$. This difference appears on the last term of the fluid momentum equation (\ref{momentumpert_fluid}), which accounts for the drag exerted by the particles on the fluid (Stokes drag). Therefore, as $\Phi$ increases, this term gains importance, and as a consequence the flow is stabilized. This explanation can be confirmed by inspection of the eigenvectors of Figure \ref{eigen_heavy_phi}, computed for $\beta=0.5$, $E=0.5$ and three values of $\Phi$: just before the ``jump" observed on $Ra_c$ ($\Phi=10^{-3}$), at the inflection point where $Ra_c$ reaches a maximum ($\Phi=4\times10^{-3}$), and after a smooth decrease on $Ra_c$ ($\Phi=6\times10^{-3}$). Before the jump, the particles diameter is relatively small, and the fluid and particle vertical velocity profiles are nearly the same. However, a significant difference can be observed on the eigenvectors obtained for larger values of $\Phi$, meaning that the velocity difference $|U_z'-W_z'|$ is large enough to play a stabilizing role through the last term of equation (\ref{momentumpert_fluid}). After the jump, the critical Rayleigh number decreases smoothly with increasing $\Phi$. A possible explanation for this destabilization lies in the non-trivial role played by $\Phi$ on the fluid/particle thermal coupling, as $\Phi$ appears not only on the coefficients of equations (\ref{energypert_fluid}) and (\ref{energypert_particle}), but also on the base state expressions $\Theta_0$ and $\Theta_{p0}$. This point will be further discussed in section \ref{energybudget}. We remark that the destabilization of the flow with increasing particle diameter was also observed by Prakhar and Prosperetti (see Figs. 2 and 3 of \cite{prakhar2021linear}) with a similar model and for $\Phi \ge 0.01$.

\begin{figure}[htb]
\begin{center}
\subfigure[$\Phi = 1 \times 10 ^{-3}$]{\label{fig:iso1}\includegraphics[width=0.45\textwidth]{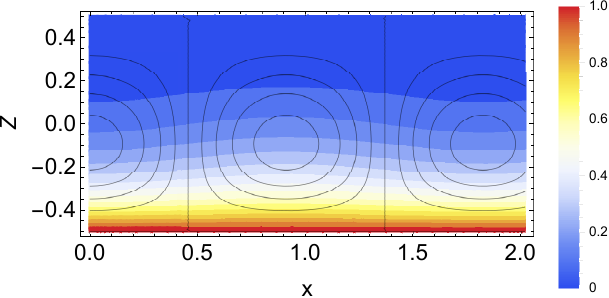}}
\subfigure[$\Phi = 4 \times 10 ^{-3}$]{\label{fig:iso1}\includegraphics[width=0.45\textwidth]{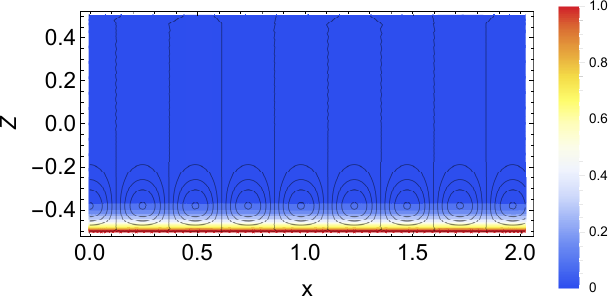}}
\subfigure[$\Phi = 6 \times 10 ^{-3}$]{\label{fig:iso1}\includegraphics[width=0.45\textwidth]{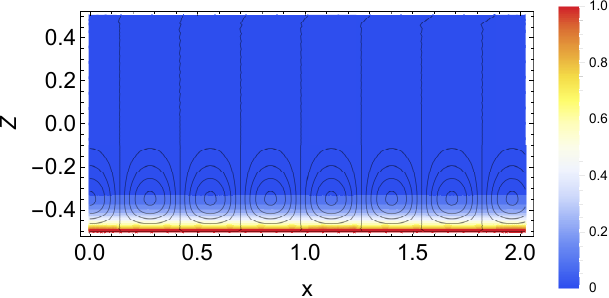}}
\caption{\textcolor{black}{Iso-contours of the fluid velocity field and heatmap of the temperature for the set of parameters of Figure \ref{eigen_heavy_phi}, showing the effect of particle diameter $\Phi$ on flow characteristics. The panels illustrate cases with (a) $\Phi = 1 \times 10^{-2}$, (b) $\Phi = 4 \times 10^{-3}$, and (c) $\Phi = 6 \times 10^{-2}$, all at $E = 5 \times 10^{-1}$ and $\beta = 0.5$.}} 
\label{contour_heavy}
\end{center}
\end{figure}

The increase in $Ra_c$ with the dimensionless particle diameter is accompanied by an increase on the critical wave number, as shown in Figure \ref{phi}(b). \textcolor{black}{ This can be explained as follows. As $\Phi$ increases, the linear fluid temperature base profile turns into a nonlinear profile with important thermal gradients on the bottom of the layer (see Figure \ref{BasePhi_Theta0}). As a consequence, convective rolls emerge in the bottom with a shorter wavelength (i.e. higher wave number). Then, convective motion which begins in the bottom layer drives the movement in the upper (less unstable) part of the layer, as illustrated in Figure \ref{contour_heavy}.}


\begin{figure}[htb]
\begin{center}
\begin{minipage}[b!]{0.49\textwidth}
\includegraphics[scale=0.29]{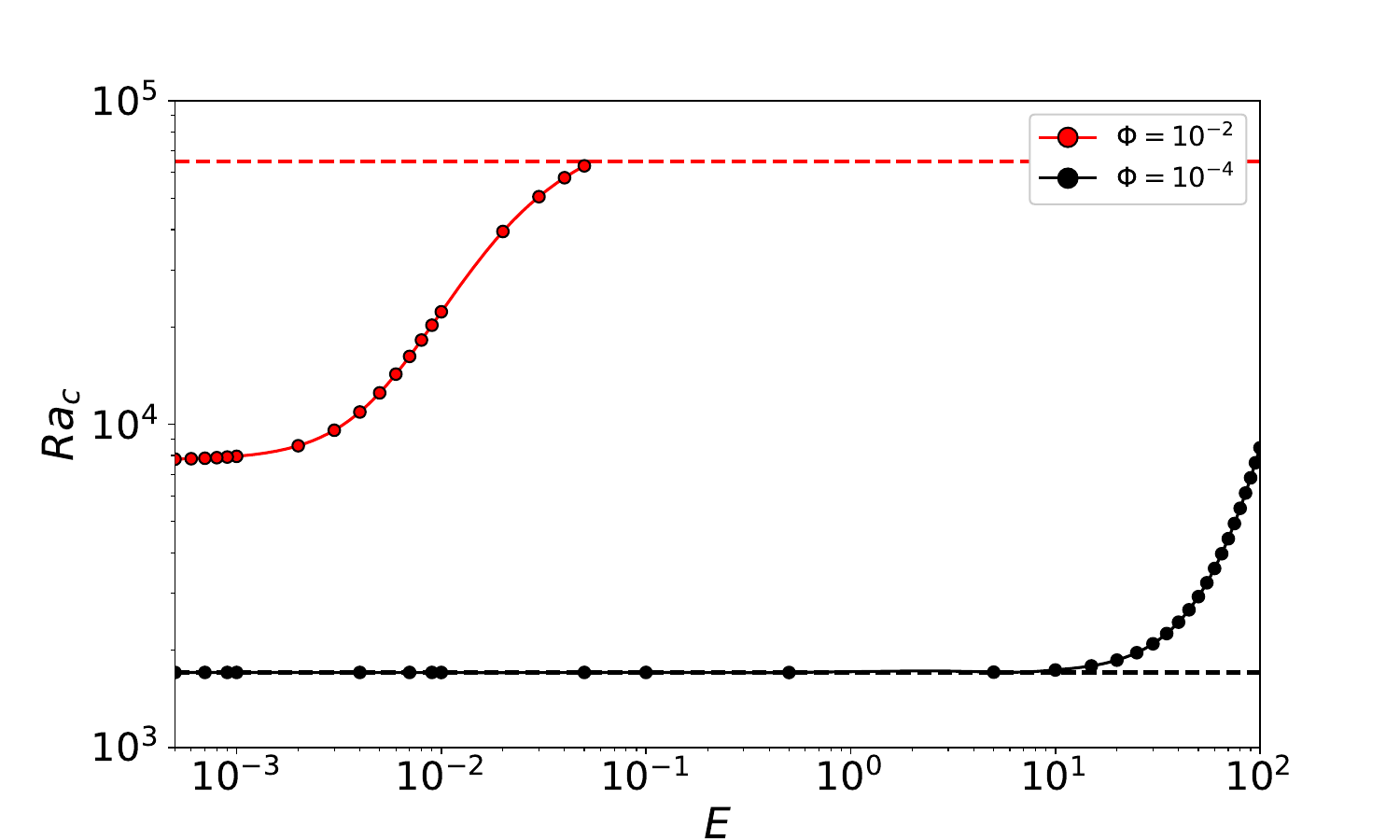}
\caption*{(a)}
\end{minipage}
\hfill
\begin{minipage}[b!]{0.49\textwidth}
\includegraphics[scale=0.29]{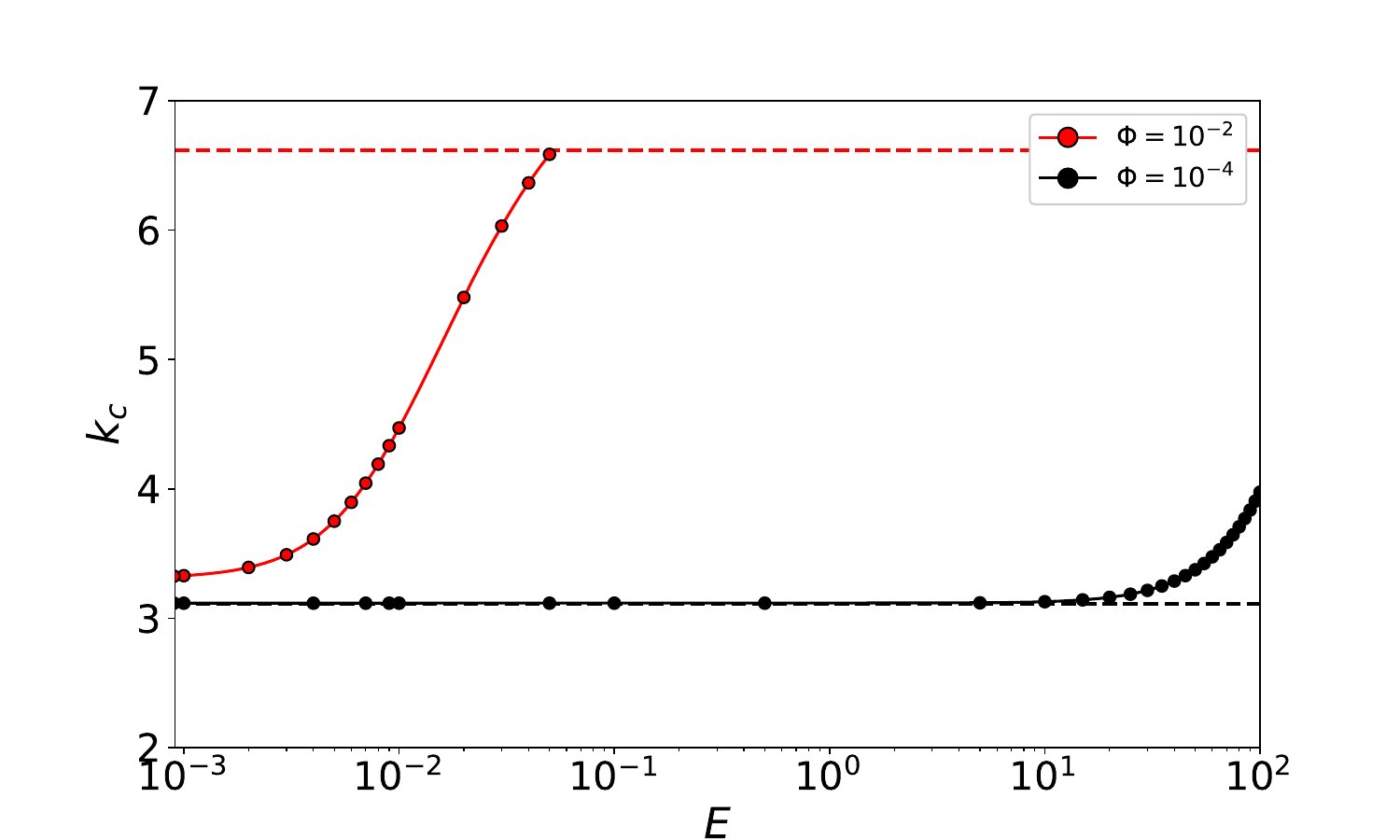}
\caption*{(b)}
\end{minipage}
\hfill
\begin{minipage}[b!]{0.49\textwidth}
\includegraphics[scale=0.29]{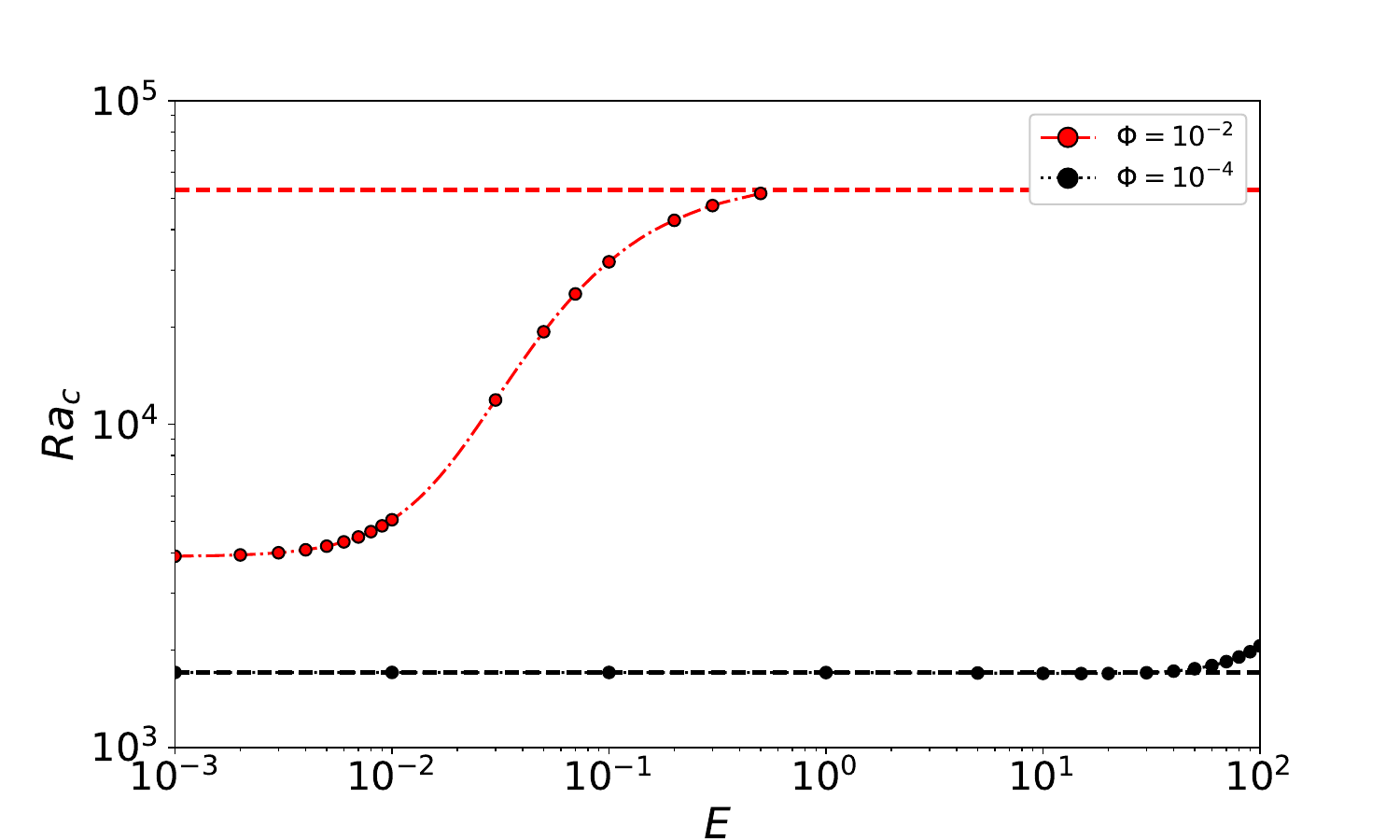}
\caption*{(c)}
\end{minipage}
\hfill
\begin{minipage}[b!]{0.49\textwidth}
\includegraphics[scale=0.29]{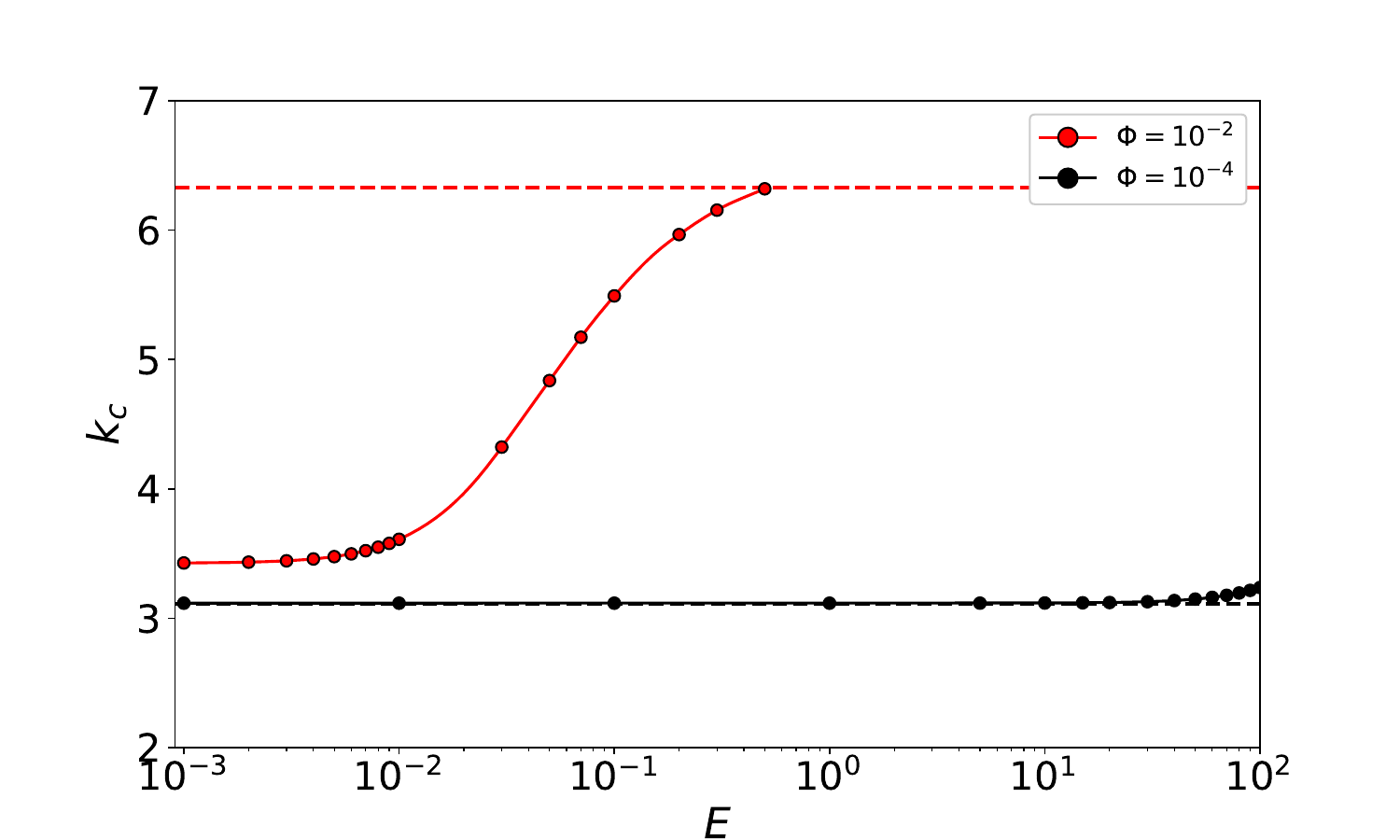}
\caption*{(d)}
\end{minipage}
\end{center}
\caption{Evolution of the critical thresholds with the heat capacity ratio $E$, obtained for: (a), (b) heavy particles with $\beta=0.5$ and $\Theta_{p}^*=0$; (c), (d) light particles with $\beta=1.5$ and $\Theta_{p}^*=1$. The red horizontal dashed line represents the limiting case  $E \rightarrow \infty$ , where the stability threshold remains constant, illustrating the asymptotic behavior of  $Ra_c$. The horizontal black dashed line represents single-phase Rayleigh-B\'enard thresholds.}
\label{E}
\end{figure}

The stabilizing effect of the heat capacity ratio $E$ illustrated in Figure \ref{E} can be easily understood. The fluid/particle temperature difference increases with $E$. Indeed, the case $E\to0$ corresponds to instant local thermal equilibrium between the fluid and the particles, i.e. $\Theta'=\Theta_p'$, and to a linear base fluid temperature profile. When cold heavy particles ($\Theta^*_{p}=0$) are being injected from above with a high thermal inertia (i.e. high $E$), particles are still cold when they get to the bottom, hence contributing to homogenize the fluid temperature within the layer. With smaller thermal gradients, the fluid gets stabilized. The inverse reasoning can be done for hot light particles being injected from the below ($\Theta^*_{p}=1$). From Figures \ref{E}(a),(c) one may also note that the stabilization provoked by $E$ arrives earlier for larger particles. Following the behavior of $Ra_c$, a sharp increase is also observed on the critical wave number in Figures \ref{E}(b),(d), which can be explained from the base fluid temperature profile as before. 

\begin{figure}[htb!]
    \begin{center}
    \includegraphics[width=0.49\textwidth]{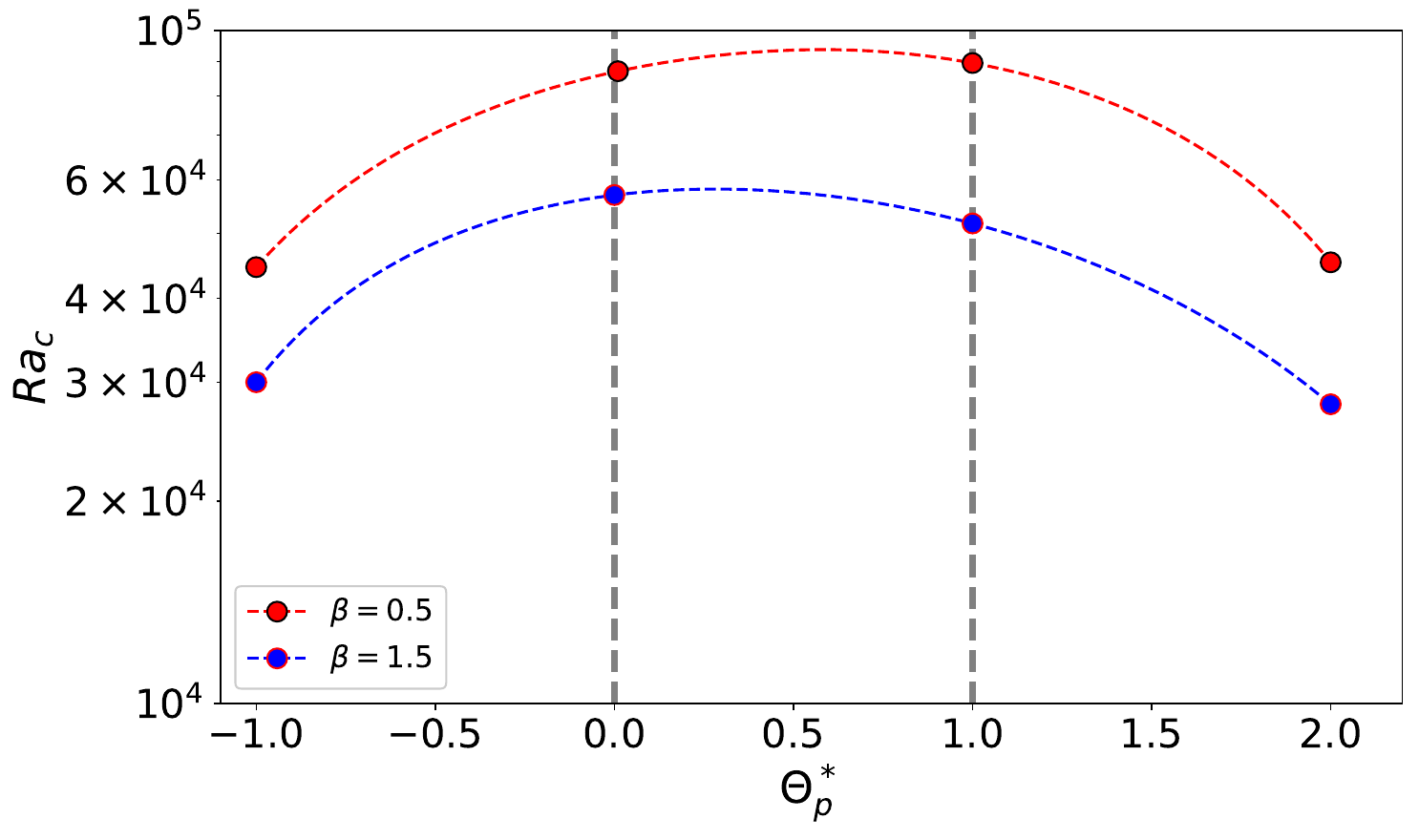}
    \caption{Evolution of the critical Rayleigh number with particle injection temperature for heavy and light particles for fixed $E = 5\times10^{-1}$ and $\Phi=10^{-2}$.}
    \label{Thetastar}
    \end{center}
\end{figure}

Figure \ref{Thetastar} shows the effect of particle injection temperature for heavy and light particles, with $E=0.5$ and $\Phi=0.01$. The observed trend is the same for both cases : the critical Rayleigh number increases, reaches a maximum and then starts to decrease. This behavior can be understood by inspecting the base temperature profiles of Figure \ref{Thetainfluence_heavy}. For $\Theta^*_{p}=-1$ one may observe a large region in the lower part of the system where the 
undisturbed vertical temperature gradient is destabilizing. By increasing the particle temperature to $\Theta^*_{p}=0$, the extent of this region decreases and as a consequence, the system becomes more stable. This is the cause of the increase on the critical Rayleigh number observed in Figure \ref{Thetastar}. For $\Theta^*_{p}=1$, the unstable part of the undisturbed temperature gradient is now located on the top of the cell (cf. Figure \ref{Thetainfluence_heavy}). The extent of this unstable region grows by further increasing particle temperature, which causes destabilization and hence, the decrease in $Ra_c$ observed for $\Theta^*_{p}=2$.

\subsection{Energy budget analysis}
\label{energybudget}

An \textit{a posteriori} analysis of the energy transferred between the base state and the critical mode is here employed to identify the physical mechanisms leading to flow instability and to validate the overall energy conservation of our neutral mode. To that end, we employ a methodology similar to the one presented in \cite{ali2022soret}, and numerically evaluate all terms in the spatially averaged linearized energy and momentum conservation equations. 

From equation (\ref{energypert_fluid}), the following relationship for the spatially averaged disturbance thermal energy $e_\Theta$ is obtained 
\begin{equation*}
    \lambda e_\Theta = e_\Theta^{th} + e_\Theta^{diff} + e_\Theta^{fp}
\end{equation*}
with
\begin{eqnarray}
   e_\Theta &=&  \int_{-1/2}^{1/2} \mathrm{Re}\left[|\Theta_n(z)|^2\right] dz,\\
   e_\Theta^{th} &=&  \int_{-1/2}^{1/2}  \mathrm{Re}\left[(U_n D\Theta_0) \bar{\Theta}_n\right]dz,\\
   e_\Theta^{diff} &=&  \int_{-1/2}^{1/2}  \mathrm{Re}\left[(D^2 -k^2) (\Theta_n \times\bar{\Theta}_n )\right]dz,\\
   e_\Theta^{fp} &=&  -\frac{\alpha_0 12}{\Phi^2}\int_{-1/2}^{1/2}  \mathrm{Re}\left[(\Theta_n -\Theta_{pn}) (\bar{\Theta}_n )\right]dz,
\end{eqnarray}
where overbars denote the complex conjugate, $e_\Theta^{th}$ is the energy due to thermal advection, $e_\Theta^{diff}$ corresponds to the thermal dissipation energy. \textcolor{black}{The energy exchange between particles and fluid as a result of drag forces is represented by $e_\Theta^{fp}$. It quantifies how particle motion is impacted by the fluid resistance, which results in transfers of energy between particles and the surrounding fluid.} The sign of the integrands determines whether the local energy transfer acts as a destabilizing (positive) or a stabilizing (negative) contribution. If the rate of change of the total energy $e_\Theta$ is positive, the basic flow is unstable, and vice-versa. Hence, the energy budget can also be used to verify the linear stability results since the rate of change of the total energy must vanish for the neutral modes. In our computations, such a condition is verified at the fifth digit. 
By normalizing the different contributions by the absolute value of the dissipation energy, we obtain at neutral conditions $(\lambda = 0)$ :
\begin{equation}
   E_\Theta^{th}  + E_\Theta^{fp} = 1,
   \label{thermalbudget}
\end{equation}
where $E_\Theta^{th} = {e_\Theta^{th}}/{|e_\Theta^{diff}|}$ and $E _\Theta^{fp} = {e_\Theta^{fp}}/{|e_\Theta^{diff}|}$.

\begin{figure}[htb!]
\centering
\begin{minipage}[b!]{0.49\textwidth}
\centering
\includegraphics[scale=0.29]{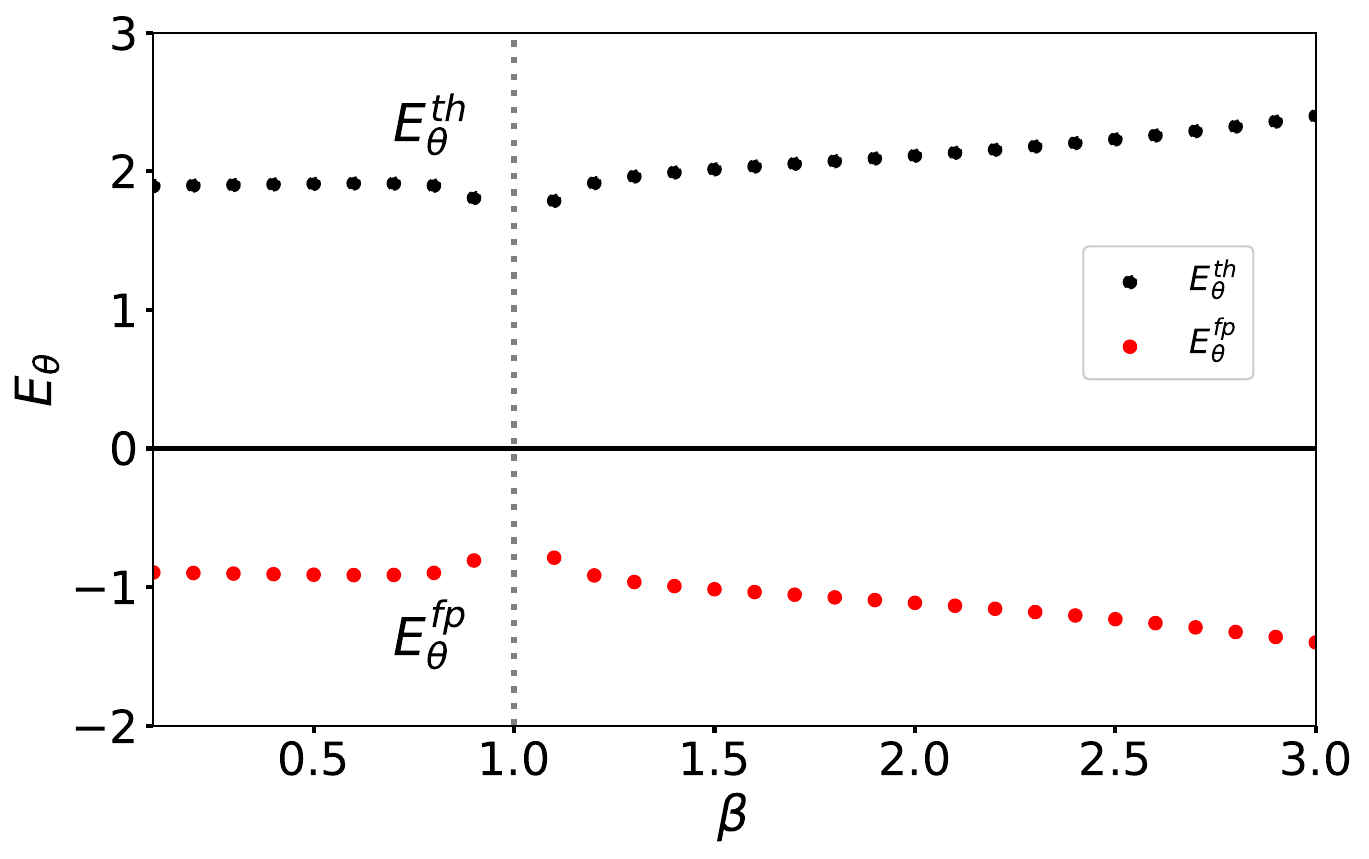}
\caption*{(a)}
\end{minipage}
\hfill
\begin{minipage}[b!]{0.49\textwidth}
\centering
\includegraphics[scale=0.29]{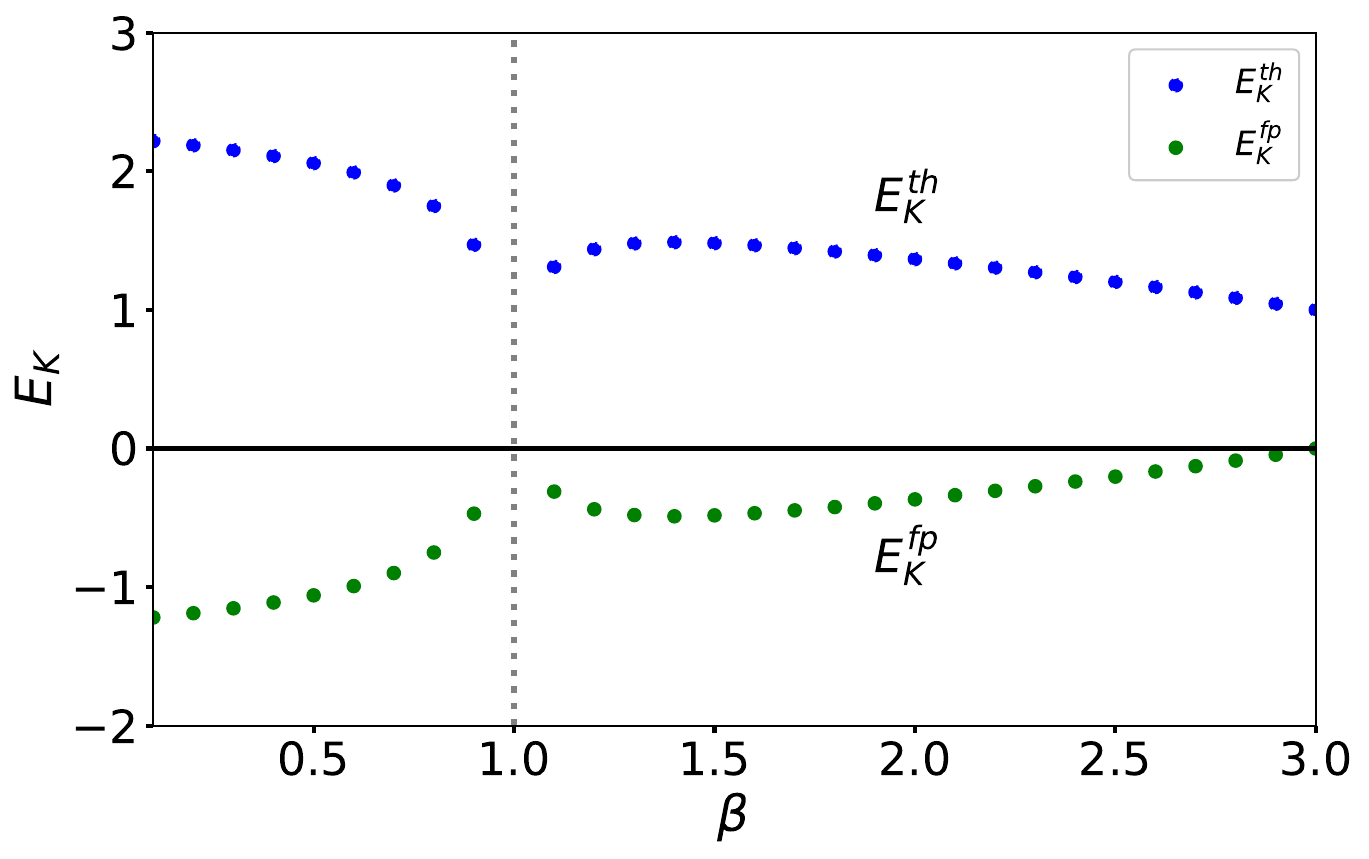}
\caption*{(b)}
\end{minipage}
\caption{(a) Thermal and (b) Kinetic energy budgets for $E = 0.5$ and $\Phi = 0.01$. 
}
\label{Energybeta}
\end{figure}

By following the same procedure, equation (\ref{momentumpert_fluid}) leads to the ensuing relationship for the rate of change of the fluctuating kinetic energy $e_K$ :
\begin{equation}
    \lambda e_K= e^{th} + e^{diff} + e^{fp},
\end{equation}
with
\begin{eqnarray}
   e_K^{th} &=&  Pr Ra \int_{-1/2}^{1/2} \mathrm{Re}\left[(\Theta_n k^2) \bar{U}_n\right]dz,\\
   e_K^{diff} &=& -Pr \int_{-1/2}^{1/2} \mathrm{Re}\left[(D^2 -k^2)^2 (U_n \times \bar{U}_n )\right]dz,\\
   e_K^{fp} &=&  \frac{ 6 \alpha_0 Pr (3-\beta )}{\Phi^2}\int_{-1/2}^{1/2} \mathrm{Re}\left[(D^2 -k^2) (U_n - W_n)\times \bar{U}_n )\right]dz,
\end{eqnarray}
After normalization, we obtain at neutral conditions $(\lambda = 0)$ :
\begin{equation}
   E_K^{th}  + E_K^{fp} = 1,
   \label{kineticbudget}
\end{equation}
where $E_K^{th} = {e_K^{th}}/{|e_K^{diff}|}$ and $E _K^{fp} = {e_K^{fp}}/{|e_K^{diff}|}$. 
Figure \ref{Energybeta} reports the total thermal and kinetic energy budgets for the neutral modes as a function of the density ratio $\beta$. Results show that both the thermal and mechanical fluid/particle coupling contribute to the stabilization of the base state. Heavy particles present a more pronounced mechanical effect than light particles, while the thermal stabilization effect is more important for light particles. As the particles become lighter, the mechanical stabilization effect decreases, and as mentioned before, for $\beta=3$ the contribution of Stokes drag is zero ($E_K^{fp}=0$).

\begin{figure}[htb!]
\begin{center}
\begin{minipage}[b!]{0.49\textwidth}
\includegraphics[scale=0.29]{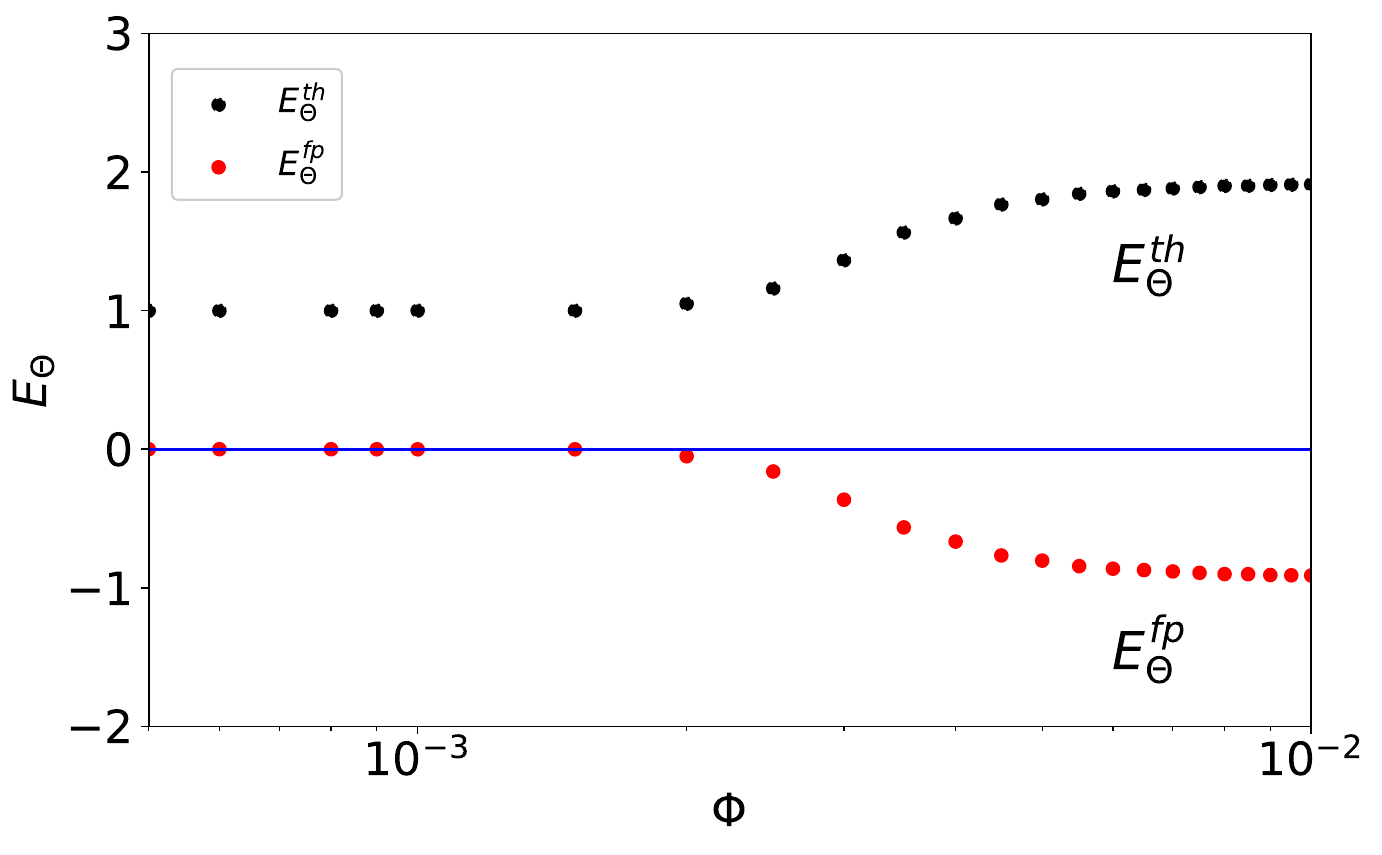}
\caption*{(a)}
\end{minipage}
\hfill
\begin{minipage}[b!]{0.49\textwidth}
\includegraphics[scale=0.29]{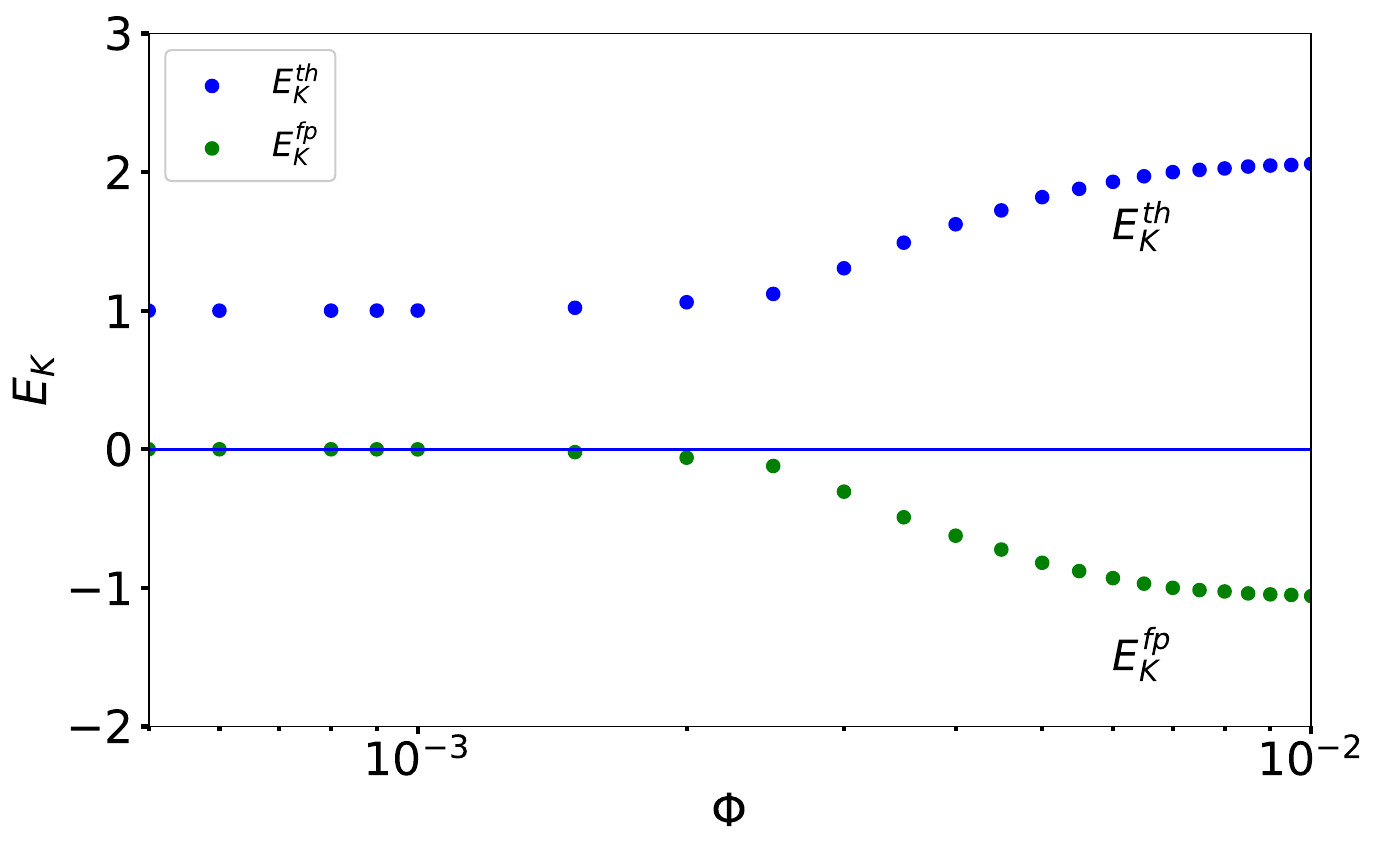}
\caption*{(b)}
\end{minipage}
\end{center}
\caption{(a) Thermal and (b) Kinetic energy budgets for $E = 0.5$ and $\beta=0.5$.
}
\label{Energyphi}
\end{figure}

The total energy budgets for varying $\Phi$ are presented in Figure \ref{Energyphi}. One may note that the curves present roughly three regions with different slopes : $\Phi \lesssim 10^{-3}$, $10^{-3}\lesssim\Phi\lesssim4\times10^{-3}$, and $\Phi\gtrsim4\times10^{-3}$, corresponding to the three different slopes of Figure \ref{phi}(a) for $E=0.5$.

 \section{Conclusions}
We theoretically studied the effects of a \textcolor{black}{diluted} dispersed particulate phase on the onset of Rayleigh-B\'enard (RB) convection in a fluid layer by means of a two-fluid Eulerian modelization. The particles are macroscopic, spherical, with inertia and heat capacity, and are assumed to interact with the surrounding fluid mechanically and thermally. 
We examine both the cases of particles denser and lighter than the fluid that are injected uniformly at the system's top and bottom walls respectively, with their settling terminal Stokes velocity and prescribed temperatures. The presented linear stability analysis shows that the onset of thermal convection is stationary, i.e., the system undergoes a pitchfork bifurcation as in the classical single-phase RB problem. Remarkably, the particle presence always stabilizes the system, increasing the critical Rayleigh number ($Ra_c$) of the convective onset.  
The limiting cases $E \to 0$ (when particles instantly adapt to fluid temperature) and $E \to \infty$ (when particles remain at their inlet temperature) were discussed in detail. 
The overall resulting stabilization effect on $Ra_c$ is significant, reaching \textcolor{black}{for a particulate volume fraction of $0.1\%$} up to a factor 30 for the case of the lightest density particles and 60 for the heaviest ones.  
Particle diameter and inlet temperature have a non-monotonic effect due to nonlinear particle/fluid interactions that we have analyzed in detail. A thermal and kinetic energy budget analysis was also carried out, clarifying the role of the different thermal and mechanical contributions to the heavy and light particle systems. 

In spite of the fact that the present model system accounts for the compressibility of the particulate velocity field and so can accommodate for the phenomenon of particle clustering, this aspect has not been explored in this study. Indeed, by simultaneously imposing the particle Stokes velocity and the particulate volume concentration at the inlet, we constrained the divergence of the particle velocity field to zero throughout the system. Although this assumption has the advantage of reducing the dimensionality of the problem (from eight to four scalar equations), this assumption is very restrictive and may have important consequences on the stability. Relaxing such a condition is possible for instance by imposing an inlet volume (or mass) flux of particles, however only at the price of more complex calculations. 
\textcolor{black}{This may reconcile the apparent counterintuitive nature of our current results, where any kind of particle heavier/lighter than the fluid is capable to increase the system stability. Although we are not aware of experiments in the exact setting described by our pRB model system, as we already mentioned, it is well known that bubbles rising in a isothermal still fluid produce mechanical destabilizations that lead to enhance mixing and to convective like movements \cite{ClimentPRL1999, MathaiARFM2020,PhysRevE.102.053102}.}
For this reason, it will be of primary interest to check how different boundary conditions, in particular resembling to the ones that could be realized in a laboratory experiment, may have consequences on the overall hydrodynamic stability of the system. \textcolor{black}{Furthermore, as pointed out in \cite{MazzitelliPRE2009, Nakamura_Yoshikawa_Tasaka_Murai_2021} the inclusion of the lift force for bubbles may also play key role in the fluid layer destabilization.}
These aspects will be the subject of future works.   

\textit{Acknowledgments} We acknowledge useful discussions with Najib M. Ouarzazi. 


\begin{thebibliography}{57}%
\makeatletter
\providecommand \@ifxundefined [1]{%
 \@ifx{#1\undefined}
}%
\providecommand \@ifnum [1]{%
 \ifnum #1\expandafter \@firstoftwo
 \else \expandafter \@secondoftwo
 \fi
}%
\providecommand \@ifx [1]{%
 \ifx #1\expandafter \@firstoftwo
 \else \expandafter \@secondoftwo
 \fi
}%
\providecommand \natexlab [1]{#1}%
\providecommand \enquote  [1]{``#1''}%
\providecommand \bibnamefont  [1]{#1}%
\providecommand \bibfnamefont [1]{#1}%
\providecommand \citenamefont [1]{#1}%
\providecommand \href@noop [0]{\@secondoftwo}%
\providecommand \href [0]{\begingroup \@sanitize@url \@href}%
\providecommand \@href[1]{\@@startlink{#1}\@@href}%
\providecommand \@@href[1]{\endgroup#1\@@endlink}%
\providecommand \@sanitize@url [0]{\catcode `\\12\catcode `\$12\catcode
  `\&12\catcode `\#12\catcode `\^12\catcode `\_12\catcode `\%12\relax}%
\providecommand \@@startlink[1]{}%
\providecommand \@@endlink[0]{}%
\providecommand \url  [0]{\begingroup\@sanitize@url \@url }%
\providecommand \@url [1]{\endgroup\@href {#1}{\urlprefix }}%
\providecommand \urlprefix  [0]{URL }%
\providecommand \Eprint [0]{\href }%
\providecommand \doibase [0]{https://doi.org/}%
\providecommand \selectlanguage [0]{\@gobble}%
\providecommand \bibinfo  [0]{\@secondoftwo}%
\providecommand \bibfield  [0]{\@secondoftwo}%
\providecommand \translation [1]{[#1]}%
\providecommand \BibitemOpen [0]{}%
\providecommand \bibitemStop [0]{}%
\providecommand \bibitemNoStop [0]{.\EOS\space}%
\providecommand \EOS [0]{\spacefactor3000\relax}%
\providecommand \BibitemShut  [1]{\csname bibitem#1\endcsname}%
\let\auto@bib@innerbib\@empty
\bibitem [{\citenamefont {Degruyter}\ \emph {et~al.}(2019)\citenamefont
  {Degruyter}, \citenamefont {Parmigiani}, \citenamefont {Huber},\ and\
  \citenamefont {Bachmann}}]{degruyter2019volatiles}%
  \BibitemOpen
  \bibfield  {author} {\bibinfo {author} {\bibfnamefont {W.}~\bibnamefont
  {Degruyter}}, \bibinfo {author} {\bibfnamefont {A.}~\bibnamefont
  {Parmigiani}}, \bibinfo {author} {\bibfnamefont {C.}~\bibnamefont {Huber}},\
  and\ \bibinfo {author} {\bibfnamefont {O.}~\bibnamefont {Bachmann}},\
  }\bibfield  {title} {\bibinfo {title} {How do volatiles escape their shallow
  magmatic hearth?},\ }\href
  {https://doi.org/http://doi.org/10.1098/rsta.2018.0017} {\bibfield  {journal}
  {\bibinfo  {journal} {Philos. Trans. Royal Soc. A}\ }\textbf {\bibinfo
  {volume} {377}},\ \bibinfo {pages} {20180017} (\bibinfo {year}
  {2019})}\BibitemShut {NoStop}%
\bibitem [{\citenamefont {Koyaguchi}\ \emph {et~al.}(1990)\citenamefont
  {Koyaguchi}, \citenamefont {Hallworth}, \citenamefont {Huppert},\ and\
  \citenamefont {Sparks}}]{koyaguchi1990sedimentation}%
  \BibitemOpen
  \bibfield  {author} {\bibinfo {author} {\bibfnamefont {T.}~\bibnamefont
  {Koyaguchi}}, \bibinfo {author} {\bibfnamefont {M.~A.}\ \bibnamefont
  {Hallworth}}, \bibinfo {author} {\bibfnamefont {H.~E.}\ \bibnamefont
  {Huppert}},\ and\ \bibinfo {author} {\bibfnamefont {R.~S.~J.}\ \bibnamefont
  {Sparks}},\ }\bibfield  {title} {\bibinfo {title} {Sedimentation of particles
  from a convecting fluid},\ }\href
  {https://doi.org/https://doi.org/10.1038/343447a0} {\bibfield  {journal}
  {\bibinfo  {journal} {Nature}\ }\textbf {\bibinfo {volume} {343}},\ \bibinfo
  {pages} {447} (\bibinfo {year} {1990})}\BibitemShut {NoStop}%
\bibitem [{\citenamefont {Elkins-Tanton}(2012)}]{elkins2012magma}%
  \BibitemOpen
  \bibfield  {author} {\bibinfo {author} {\bibfnamefont {L.~T.}\ \bibnamefont
  {Elkins-Tanton}},\ }\bibfield  {title} {\bibinfo {title} {Magma oceans in the
  inner solar system},\ }\href
  {https://doi.org/https://doi.org/10.1146/annurev-earth-042711-105503}
  {\bibfield  {journal} {\bibinfo  {journal} {Annu. Rev. Earth Planet. Sci.}\
  }\textbf {\bibinfo {volume} {40}},\ \bibinfo {pages} {113} (\bibinfo {year}
  {2012})}\BibitemShut {NoStop}%
\bibitem [{\citenamefont {Solomatov}(2007)}]{solomatov2007magma}%
  \BibitemOpen
  \bibfield  {author} {\bibinfo {author} {\bibfnamefont {V.}~\bibnamefont
  {Solomatov}},\ }\bibfield  {title} {\bibinfo {title} {9.04 - magma oceans and
  primordial mantle differentiation},\ }in\ \href
  {https://doi.org/https://doi.org/10.1016/B978-044452748-6.00141-3} {\emph
  {\bibinfo {booktitle} {Treatise on Geophysics}}},\ Vol.~\bibinfo {volume}
  {40}\ (\bibinfo {year} {2007})\ pp.\ \bibinfo {pages} {91--119}\BibitemShut
  {NoStop}%
\bibitem [{\citenamefont {Chang}\ \emph {et~al.}(2008)\citenamefont {Chang},
  \citenamefont {Mills},\ and\ \citenamefont {Hernandez}}]{chang2008natural}%
  \BibitemOpen
  \bibfield  {author} {\bibinfo {author} {\bibfnamefont {B.~H.}\ \bibnamefont
  {Chang}}, \bibinfo {author} {\bibfnamefont {A.~F.}\ \bibnamefont {Mills}},\
  and\ \bibinfo {author} {\bibfnamefont {E.}~\bibnamefont {Hernandez}},\
  }\bibfield  {title} {\bibinfo {title} {Natural convection of microparticle
  suspensions in thin enclosures},\ }\href
  {https://doi.org/https://doi.org/10.1016/j.ijheatmasstransfer.2007.11.030}
  {\bibfield  {journal} {\bibinfo  {journal} {Int. J. Heat Mass Transf.}\
  }\textbf {\bibinfo {volume} {51}},\ \bibinfo {pages} {1332} (\bibinfo {year}
  {2008})}\BibitemShut {NoStop}%
\bibitem [{\citenamefont {Schwaiger}\ \emph {et~al.}(2012)\citenamefont
  {Schwaiger}, \citenamefont {Denlinger},\ and\ \citenamefont
  {Mastin}}]{schwaiger2012ash3d}%
  \BibitemOpen
  \bibfield  {author} {\bibinfo {author} {\bibfnamefont {H.~F.}\ \bibnamefont
  {Schwaiger}}, \bibinfo {author} {\bibfnamefont {R.~P.}\ \bibnamefont
  {Denlinger}},\ and\ \bibinfo {author} {\bibfnamefont {L.~G.}\ \bibnamefont
  {Mastin}},\ }\bibfield  {title} {\bibinfo {title} {Ash3d: A finite-volume,
  conservative numerical model for ash transport and tephra deposition},\
  }\href {https://doi.org/https://doi.org/10.1029/2011JB008968} {\bibfield
  {journal} {\bibinfo  {journal} {J. Geophys. Res.: Solid Earth}\ }\textbf
  {\bibinfo {volume} {117}},\ \bibinfo {pages} {B04204} (\bibinfo {year}
  {2012})}\BibitemShut {NoStop}%
\bibitem [{\citenamefont {Squires}\ and\ \citenamefont
  {Yamazaki}(1995)}]{squires1995preferential}%
  \BibitemOpen
  \bibfield  {author} {\bibinfo {author} {\bibfnamefont {K.~D.}\ \bibnamefont
  {Squires}}\ and\ \bibinfo {author} {\bibfnamefont {H.}~\bibnamefont
  {Yamazaki}},\ }\bibfield  {title} {\bibinfo {title} {Preferential
  concentration of marine particles in isotropic turbulence},\ }\href
  {https://doi.org/https://doi.org/10.1016/0967-0637(95)00079-8} {\bibfield
  {journal} {\bibinfo  {journal} {Deep-Sea Res. Part I}\ }\textbf {\bibinfo
  {volume} {42}},\ \bibinfo {pages} {1989} (\bibinfo {year}
  {1995})}\BibitemShut {NoStop}%
\bibitem [{\citenamefont {Breuer}\ \emph {et~al.}(2015)\citenamefont {Breuer},
  \citenamefont {Rueckriemen},\ and\ \citenamefont {Spohn}}]{breuer2015iron}%
  \BibitemOpen
  \bibfield  {author} {\bibinfo {author} {\bibfnamefont {D.}~\bibnamefont
  {Breuer}}, \bibinfo {author} {\bibfnamefont {T.}~\bibnamefont
  {Rueckriemen}},\ and\ \bibinfo {author} {\bibfnamefont {T.}~\bibnamefont
  {Spohn}},\ }\bibfield  {title} {\bibinfo {title} {Iron snow, crystal floats,
  and inner-core growth: modes of core solidification and implications for
  dynamos in terrestrial planets and moons},\ }\href
  {https://doi.org/https://doi.org/10.1186/s40645-015-0069-y} {\bibfield
  {journal} {\bibinfo  {journal} {Prog. Earth Planet. Sci.}\ }\textbf {\bibinfo
  {volume} {2}},\ \bibinfo {pages} {1} (\bibinfo {year} {2015})}\BibitemShut
  {NoStop}%
\bibitem [{\citenamefont {Ardeshiri}\ \emph {et~al.}(2016)\citenamefont
  {Ardeshiri}, \citenamefont {Benkeddad}, \citenamefont {Schmitt},
  \citenamefont {Souissi}, \citenamefont {Toschi},\ and\ \citenamefont
  {Calzavarini}}]{ArdeshiriPRE2016}%
  \BibitemOpen
  \bibfield  {author} {\bibinfo {author} {\bibfnamefont {H.}~\bibnamefont
  {Ardeshiri}}, \bibinfo {author} {\bibfnamefont {I.}~\bibnamefont
  {Benkeddad}}, \bibinfo {author} {\bibfnamefont {F.~G.}\ \bibnamefont
  {Schmitt}}, \bibinfo {author} {\bibfnamefont {S.}~\bibnamefont {Souissi}},
  \bibinfo {author} {\bibfnamefont {F.}~\bibnamefont {Toschi}},\ and\ \bibinfo
  {author} {\bibfnamefont {E.}~\bibnamefont {Calzavarini}},\ }\bibfield
  {title} {\bibinfo {title} {Lagrangian model of copepod dynamics: Clustering
  by escape jumps in turbulence},\ }\href
  {https://doi.org/10.1103/PhysRevE.93.043117} {\bibfield  {journal} {\bibinfo
  {journal} {Phys. Rev. E}\ }\textbf {\bibinfo {volume} {93}},\ \bibinfo
  {pages} {043117} (\bibinfo {year} {2016})}\BibitemShut {NoStop}%
\bibitem [{\citenamefont {Bees}(2020)}]{BeesMartinA2020AiB}%
  \BibitemOpen
  \bibfield  {author} {\bibinfo {author} {\bibfnamefont {M.~A.}\ \bibnamefont
  {Bees}},\ }\bibfield  {title} {\bibinfo {title} {Advances in bioconvection},\
  }\href {https://doi.org/10.1146/annurev-fluid-010518-040558} {\bibfield
  {journal} {\bibinfo  {journal} {Annu. Rev. Fluid Mech.}\ }\textbf {\bibinfo
  {volume} {52}},\ \bibinfo {pages} {449} (\bibinfo {year} {2020})}\BibitemShut
  {NoStop}%
\bibitem [{\citenamefont {Dennis}(2013)}]{dennis2013properties}%
  \BibitemOpen
  \bibfield  {author} {\bibinfo {author} {\bibfnamefont {J.~S.}\ \bibnamefont
  {Dennis}},\ }\bibfield  {title} {\bibinfo {title} {Properties of stationary
  (bubbling) fluidised beds relevant to combustion and gasification systems},\
  }\href {https://doi.org/https://doi.org/10.1533/9780857098801.1.77}
  {\bibfield  {journal} {\bibinfo  {journal} {Fluidized Bed Technol. Near-Zero
  Emiss. Combust. Gasification}\ ,\ \bibinfo {pages} {77}} (\bibinfo {year}
  {2013})}\BibitemShut {NoStop}%
\bibitem [{\citenamefont {Pouransari}\ and\ \citenamefont
  {Mani}(2017)}]{pouransari2017effects}%
  \BibitemOpen
  \bibfield  {author} {\bibinfo {author} {\bibfnamefont {H.}~\bibnamefont
  {Pouransari}}\ and\ \bibinfo {author} {\bibfnamefont {A.}~\bibnamefont
  {Mani}},\ }\bibfield  {title} {\bibinfo {title} {Effects of preferential
  concentration on heat transfer in particle-based solar receivers},\ }\href
  {https://doi.org/https://doi.org/10.1115/1.4035163} {\bibfield  {journal}
  {\bibinfo  {journal} {J. Sol. Energy Eng.}\ }\textbf {\bibinfo {volume}
  {139}},\ \bibinfo {pages} {021008} (\bibinfo {year} {2017})}\BibitemShut
  {NoStop}%
\bibitem [{\citenamefont {Frankel}\ \emph {et~al.}(2016)\citenamefont
  {Frankel}, \citenamefont {Pouransari}, \citenamefont {Coletti},\ and\
  \citenamefont {Mani}}]{frankel2016settling}%
  \BibitemOpen
  \bibfield  {author} {\bibinfo {author} {\bibfnamefont {A.}~\bibnamefont
  {Frankel}}, \bibinfo {author} {\bibfnamefont {H.}~\bibnamefont {Pouransari}},
  \bibinfo {author} {\bibfnamefont {F.}~\bibnamefont {Coletti}},\ and\ \bibinfo
  {author} {\bibfnamefont {A.}~\bibnamefont {Mani}},\ }\bibfield  {title}
  {\bibinfo {title} {Settling of heated particles in homogeneous turbulence},\
  }\href {https://doi.org/doi:10.1017/jfm.2016.102} {\bibfield  {journal}
  {\bibinfo  {journal} {J. Fluid Mech.}\ }\textbf {\bibinfo {volume} {792}},\
  \bibinfo {pages} {869} (\bibinfo {year} {2016})}\BibitemShut {NoStop}%
\bibitem [{\citenamefont {Rahmani}\ \emph {et~al.}(2018)\citenamefont
  {Rahmani}, \citenamefont {Geraci}, \citenamefont {Iaccarino},\ and\
  \citenamefont {Mani}}]{rahmani2018effects}%
  \BibitemOpen
  \bibfield  {author} {\bibinfo {author} {\bibfnamefont {M.}~\bibnamefont
  {Rahmani}}, \bibinfo {author} {\bibfnamefont {G.}~\bibnamefont {Geraci}},
  \bibinfo {author} {\bibfnamefont {G.}~\bibnamefont {Iaccarino}},\ and\
  \bibinfo {author} {\bibfnamefont {A.}~\bibnamefont {Mani}},\ }\bibfield
  {title} {\bibinfo {title} {Effects of particle polydispersity on radiative
  heat transfer in particle-laden turbulent flows},\ }\href
  {https://doi.org/https://doi.org/10.1016/j.ijmultiphaseflow.2018.03.011}
  {\bibfield  {journal} {\bibinfo  {journal} {Int. J. Multiphase Flow}\
  }\textbf {\bibinfo {volume} {104}},\ \bibinfo {pages} {42} (\bibinfo {year}
  {2018})}\BibitemShut {NoStop}%
\bibitem [{\citenamefont {Brandt}\ and\ \citenamefont
  {Coletti}(2022)}]{brandt2022particle}%
  \BibitemOpen
  \bibfield  {author} {\bibinfo {author} {\bibfnamefont {L.}~\bibnamefont
  {Brandt}}\ and\ \bibinfo {author} {\bibfnamefont {F.}~\bibnamefont
  {Coletti}},\ }\bibfield  {title} {\bibinfo {title} {Particle-laden
  turbulence: progress and perspectives},\ }\href
  {https://doi.org/https://doi.org/10.1146/annurev-fluid-030121-021103}
  {\bibfield  {journal} {\bibinfo  {journal} {Annu. Rev. Fluid Mech.}\ }\textbf
  {\bibinfo {volume} {54}},\ \bibinfo {pages} {159} (\bibinfo {year}
  {2022})}\BibitemShut {NoStop}%
\bibitem [{\citenamefont {Kuruneru}\ \emph {et~al.}(2017)\citenamefont
  {Kuruneru}, \citenamefont {Sauret}, \citenamefont {Vafai}, \citenamefont
  {Saha},\ and\ \citenamefont {Gu}}]{kuruneru2017analysis}%
  \BibitemOpen
  \bibfield  {author} {\bibinfo {author} {\bibfnamefont {S.~T.~W.}\
  \bibnamefont {Kuruneru}}, \bibinfo {author} {\bibfnamefont {E.}~\bibnamefont
  {Sauret}}, \bibinfo {author} {\bibfnamefont {K.}~\bibnamefont {Vafai}},
  \bibinfo {author} {\bibfnamefont {S.~C.}\ \bibnamefont {Saha}},\ and\
  \bibinfo {author} {\bibfnamefont {Y.~T.}\ \bibnamefont {Gu}},\ }\bibfield
  {title} {\bibinfo {title} {Analysis of particle-laden fluid flows, tortuosity
  and particle-fluid behaviour in metal foam heat exchangers},\ }\href
  {https://doi.org/https://doi.org/10.1016/j.ces.2017.07.027} {\bibfield
  {journal} {\bibinfo  {journal} {Chem. Eng. Sci.}\ }\textbf {\bibinfo {volume}
  {172}},\ \bibinfo {pages} {677} (\bibinfo {year} {2017})}\BibitemShut
  {NoStop}%
\bibitem [{\citenamefont {Mathai}\ \emph
  {et~al.}(2020{\natexlab{a}})\citenamefont {Mathai}, \citenamefont {Lohse},\
  and\ \citenamefont {Sun}}]{mathai2020bubbly}%
  \BibitemOpen
  \bibfield  {author} {\bibinfo {author} {\bibfnamefont {V.}~\bibnamefont
  {Mathai}}, \bibinfo {author} {\bibfnamefont {D.}~\bibnamefont {Lohse}},\ and\
  \bibinfo {author} {\bibfnamefont {C.}~\bibnamefont {Sun}},\ }\bibfield
  {title} {\bibinfo {title} {Bubbly and buoyant particle--laden turbulent
  flows},\ }\href
  {https://doi.org/https://doi.org/10.1146/annurev-conmatphys-031119-050637}
  {\bibfield  {journal} {\bibinfo  {journal} {Annu. Rev. Condens. Matter
  Phys.}\ }\textbf {\bibinfo {volume} {11}},\ \bibinfo {pages} {529} (\bibinfo
  {year} {2020}{\natexlab{a}})}\BibitemShut {NoStop}%
\bibitem [{\citenamefont {Balachandar}(2024)}]{Balachandar_2024}%
  \BibitemOpen
  \bibfield  {author} {\bibinfo {author} {\bibfnamefont {S.}~\bibnamefont
  {Balachandar}},\ }\href
  {https://doi.org/https://doi.org/10.1017/9781009160452} {\emph {\bibinfo
  {title} {Fundamentals of Dispersed Multiphase Flows}}}\ (\bibinfo
  {publisher} {Cambridge University Press},\ \bibinfo {year}
  {2024})\BibitemShut {NoStop}%
\bibitem [{\citenamefont {Vié}\ \emph {et~al.}(2016)\citenamefont {Vié},
  \citenamefont {Pouransari}, \citenamefont {Zamansky},\ and\ \citenamefont
  {Mani}}]{vie2016particle}%
  \BibitemOpen
  \bibfield  {author} {\bibinfo {author} {\bibfnamefont {A.}~\bibnamefont
  {Vié}}, \bibinfo {author} {\bibfnamefont {H.}~\bibnamefont {Pouransari}},
  \bibinfo {author} {\bibfnamefont {R.}~\bibnamefont {Zamansky}},\ and\
  \bibinfo {author} {\bibfnamefont {A.}~\bibnamefont {Mani}},\ }\bibfield
  {title} {\bibinfo {title} {Particle-laden flows forced by the disperse phase:
  comparison between lagrangian and eulerian simulations},\ }\href
  {https://doi.org/https://doi.org/10.1016/j.ijmultiphaseflow.2015.10.010}
  {\bibfield  {journal} {\bibinfo  {journal} {Int. J. Multiphase Flow}\
  }\textbf {\bibinfo {volume} {79}},\ \bibinfo {pages} {144} (\bibinfo {year}
  {2016})}\BibitemShut {NoStop}%
\bibitem [{\citenamefont {Xu}\ \emph {et~al.}(2016)\citenamefont {Xu},
  \citenamefont {Zhao}, \citenamefont {Shi},\ and\ \citenamefont
  {Yan}}]{xu2016three}%
  \BibitemOpen
  \bibfield  {author} {\bibinfo {author} {\bibfnamefont {A.}~\bibnamefont
  {Xu}}, \bibinfo {author} {\bibfnamefont {T.~S.}\ \bibnamefont {Zhao}},
  \bibinfo {author} {\bibfnamefont {L.}~\bibnamefont {Shi}},\ and\ \bibinfo
  {author} {\bibfnamefont {X.~H.}\ \bibnamefont {Yan}},\ }\bibfield  {title}
  {\bibinfo {title} {Three-dimensional lattice {B}oltzmann simulation of
  suspensions containing both micro- and nanoparticles},\ }\href
  {https://doi.org/https://doi.org/10.1016/j.ijheatfluidflow.2016.08.001}
  {\bibfield  {journal} {\bibinfo  {journal} {Int. J. Heat Fluid Flow}\
  }\textbf {\bibinfo {volume} {62}},\ \bibinfo {pages} {560} (\bibinfo {year}
  {2016})}\BibitemShut {NoStop}%
\bibitem [{\citenamefont
  {Chandrasekhar}(1961)}]{chandrasekhar2013hydrodynamic}%
  \BibitemOpen
  \bibfield  {author} {\bibinfo {author} {\bibfnamefont {S.}~\bibnamefont
  {Chandrasekhar}},\ }\href@noop {} {\emph {\bibinfo {title} {Hydrodynamic and
  Hydromagnetic Stability}}}\ (\bibinfo  {publisher} {Oxford University
  Press},\ \bibinfo {year} {1961})\BibitemShut {NoStop}%
\bibitem [{\citenamefont {De~Zarate}\ and\ \citenamefont
  {Sengers}(2006)}]{de2006hydrodynamic}%
  \BibitemOpen
  \bibfield  {author} {\bibinfo {author} {\bibfnamefont {J.~M.~O.}\
  \bibnamefont {De~Zarate}}\ and\ \bibinfo {author} {\bibfnamefont {J.~V.}\
  \bibnamefont {Sengers}},\ }\href@noop {} {\emph {\bibinfo {title}
  {Hydrodynamic Fluctuations in Fluids and Fluid Mixtures}}}\ (\bibinfo
  {publisher} {Elsevier},\ \bibinfo {year} {2006})\BibitemShut {NoStop}%
\bibitem [{\citenamefont {Getling}(1998)}]{getling1998rayleigh}%
  \BibitemOpen
  \bibfield  {author} {\bibinfo {author} {\bibfnamefont {A.~V.}\ \bibnamefont
  {Getling}},\ }\href {https://doi.org/10.1142/3097} {\emph {\bibinfo {title}
  {{R}ayleigh-{B}\'enard Convection: Structures and Dynamics}}},\ Vol.~\bibinfo
  {volume} {11}\ (\bibinfo  {publisher} {World Scientific},\ \bibinfo {year}
  {1998})\BibitemShut {NoStop}%
\bibitem [{\citenamefont {Park}\ \emph {et~al.}(2018)\citenamefont {Park},
  \citenamefont {O’Keefe},\ and\ \citenamefont {Richter}}]{park2018rayleigh}%
  \BibitemOpen
  \bibfield  {author} {\bibinfo {author} {\bibfnamefont {H.~J.}\ \bibnamefont
  {Park}}, \bibinfo {author} {\bibfnamefont {K.}~\bibnamefont {O’Keefe}},\
  and\ \bibinfo {author} {\bibfnamefont {D.~H.}\ \bibnamefont {Richter}},\
  }\bibfield  {title} {\bibinfo {title} {{R}ayleigh-{B}\'enard turbulence
  modified by two-way coupled inertial, nonisothermal particles},\ }\href
  {https://doi.org/10.1103/PhysRevFluids.3.034307} {\bibfield  {journal}
  {\bibinfo  {journal} {Phys. Rev. Fluids}\ }\textbf {\bibinfo {volume} {3}},\
  \bibinfo {pages} {034307} (\bibinfo {year} {2018})}\BibitemShut {NoStop}%
\bibitem [{\citenamefont {Solomatov}\ \emph {et~al.}(1993)\citenamefont
  {Solomatov}, \citenamefont {Olson},\ and\ \citenamefont
  {Stevenson}}]{SolomatovEPSL1993}%
  \BibitemOpen
  \bibfield  {author} {\bibinfo {author} {\bibfnamefont {V.~S.}\ \bibnamefont
  {Solomatov}}, \bibinfo {author} {\bibfnamefont {P.}~\bibnamefont {Olson}},\
  and\ \bibinfo {author} {\bibfnamefont {D.~J.}\ \bibnamefont {Stevenson}},\
  }\bibfield  {title} {\bibinfo {title} {Entrainment from a bed of particles by
  thermal convection},\ }\href@noop {} {\bibfield  {journal} {\bibinfo
  {journal} {Earth Planet. Sci. Lett.}\ }\textbf {\bibinfo {volume} {120}},\
  \bibinfo {pages} {387} (\bibinfo {year} {1993})}\BibitemShut {NoStop}%
\bibitem [{\citenamefont {Lavorel}\ and\ \citenamefont
  {Le~Bars}(2009)}]{lavorel2009sedimentation}%
  \BibitemOpen
  \bibfield  {author} {\bibinfo {author} {\bibfnamefont {G.}~\bibnamefont
  {Lavorel}}\ and\ \bibinfo {author} {\bibfnamefont {M.}~\bibnamefont
  {Le~Bars}},\ }\bibfield  {title} {\bibinfo {title} {Sedimentation of
  particles in a vigorously convecting fluid},\ }\href
  {https://doi.org/10.1103/PhysRevE.80.046324} {\bibfield  {journal} {\bibinfo
  {journal} {Phys. Rev. E}\ }\textbf {\bibinfo {volume} {80}},\ \bibinfo
  {pages} {046324} (\bibinfo {year} {2009})}\BibitemShut {NoStop}%
\bibitem [{\citenamefont {Chandrakar}\ \emph {et~al.}(2020)\citenamefont
  {Chandrakar}, \citenamefont {Cantrell}, \citenamefont {Krueger},
  \citenamefont {Shaw},\ and\ \citenamefont
  {Wunsch}}]{Chandrakar_Cantrell_Krueger_Shaw_Wunsch_2020}%
  \BibitemOpen
  \bibfield  {author} {\bibinfo {author} {\bibfnamefont {K.~K.}\ \bibnamefont
  {Chandrakar}}, \bibinfo {author} {\bibfnamefont {W.}~\bibnamefont
  {Cantrell}}, \bibinfo {author} {\bibfnamefont {S.}~\bibnamefont {Krueger}},
  \bibinfo {author} {\bibfnamefont {R.~A.}\ \bibnamefont {Shaw}},\ and\
  \bibinfo {author} {\bibfnamefont {S.}~\bibnamefont {Wunsch}},\ }\bibfield
  {title} {\bibinfo {title} {Supersaturation fluctuations in moist turbulent
  {R}ayleigh–{B}\'enard convection: a two-scalar transport problem},\ }\href
  {https://doi.org/10.1017/jfm.2019.895} {\bibfield  {journal} {\bibinfo
  {journal} {J. Fluid Mech.}\ }\textbf {\bibinfo {volume} {884}},\ \bibinfo
  {pages} {A19} (\bibinfo {year} {2020})}\BibitemShut {NoStop}%
\bibitem [{\citenamefont {Jiang}\ \emph {et~al.}(2020)\citenamefont {Jiang},
  \citenamefont {Calzavarini},\ and\ \citenamefont
  {Sun}}]{Jiang_Calzavarini_Sun_2020}%
  \BibitemOpen
  \bibfield  {author} {\bibinfo {author} {\bibfnamefont {L.}~\bibnamefont
  {Jiang}}, \bibinfo {author} {\bibfnamefont {E.}~\bibnamefont {Calzavarini}},\
  and\ \bibinfo {author} {\bibfnamefont {C.}~\bibnamefont {Sun}},\ }\bibfield
  {title} {\bibinfo {title} {Rotation of anisotropic particles in
  {R}ayleigh–{B}\'enard turbulence},\ }\href
  {https://doi.org/10.1017/jfm.2020.539} {\bibfield  {journal} {\bibinfo
  {journal} {J. Fluid Mech.}\ }\textbf {\bibinfo {volume} {901}},\ \bibinfo
  {pages} {A8} (\bibinfo {year} {2020})}\BibitemShut {NoStop}%
\bibitem [{\citenamefont {Jiang}\ \emph {et~al.}(2021)\citenamefont {Jiang},
  \citenamefont {Wang}, \citenamefont {Liu}, \citenamefont {Sun},\ and\
  \citenamefont {Calzavarini}}]{JIANG2021TAML}%
  \BibitemOpen
  \bibfield  {author} {\bibinfo {author} {\bibfnamefont {L.}~\bibnamefont
  {Jiang}}, \bibinfo {author} {\bibfnamefont {C.}~\bibnamefont {Wang}},
  \bibinfo {author} {\bibfnamefont {S.}~\bibnamefont {Liu}}, \bibinfo {author}
  {\bibfnamefont {C.}~\bibnamefont {Sun}},\ and\ \bibinfo {author}
  {\bibfnamefont {E.}~\bibnamefont {Calzavarini}},\ }\bibfield  {title}
  {\bibinfo {title} {Rotational dynamics of bottom-heavy rods in turbulence
  from experiments and numerical simulations},\ }\href
  {https://doi.org/https://doi.org/10.1016/j.taml.2021.100227} {\bibfield
  {journal} {\bibinfo  {journal} {Theor. Appl. Mech. Lett.}\ }\textbf {\bibinfo
  {volume} {11}},\ \bibinfo {pages} {100227} (\bibinfo {year}
  {2021})}\BibitemShut {NoStop}%
\bibitem [{\citenamefont {Calzavarini}\ \emph {et~al.}(2020)\citenamefont
  {Calzavarini}, \citenamefont {Jiang},\ and\ \citenamefont
  {Sun}}]{CalzavariniPF_RB2D}%
  \BibitemOpen
  \bibfield  {author} {\bibinfo {author} {\bibfnamefont {E.}~\bibnamefont
  {Calzavarini}}, \bibinfo {author} {\bibfnamefont {L.}~\bibnamefont {Jiang}},\
  and\ \bibinfo {author} {\bibfnamefont {C.}~\bibnamefont {Sun}},\ }\bibfield
  {title} {\bibinfo {title} {Anisotropic particles in two-dimensional
  convective turbulence},\ }\href
  {https://doi.org/https://doi.org/10.1063/1.5141798} {\bibfield  {journal}
  {\bibinfo  {journal} {Phys. Fluids}\ }\textbf {\bibinfo {volume} {32}},\
  \bibinfo {pages} {023305} (\bibinfo {year} {2020})}\BibitemShut {NoStop}%
\bibitem [{\citenamefont {Pato\v{c}ka}\ \emph {et~al.}(2020)\citenamefont
  {Pato\v{c}ka}, \citenamefont {Calzavarini},\ and\ \citenamefont
  {Tosi}}]{patovcka2020settling}%
  \BibitemOpen
  \bibfield  {author} {\bibinfo {author} {\bibfnamefont {V.}~\bibnamefont
  {Pato\v{c}ka}}, \bibinfo {author} {\bibfnamefont {E.}~\bibnamefont
  {Calzavarini}},\ and\ \bibinfo {author} {\bibfnamefont {N.}~\bibnamefont
  {Tosi}},\ }\bibfield  {title} {\bibinfo {title} {Settling of inertial
  particles in turbulent {R}ayleigh-{B}\'enard convection},\ }\href
  {https://doi.org/https://doi.org/10.1103/PhysRevFluids.5.114304} {\bibfield
  {journal} {\bibinfo  {journal} {Phys. Rev. Fluids}\ }\textbf {\bibinfo
  {volume} {5}},\ \bibinfo {pages} {114304} (\bibinfo {year}
  {2020})}\BibitemShut {NoStop}%
\bibitem [{\citenamefont {Pato\v{c}ka}\ \emph {et~al.}(2022)\citenamefont
  {Pato\v{c}ka}, \citenamefont {Tosi},\ and\ \citenamefont
  {Calzavarini}}]{patovcka2022residence}%
  \BibitemOpen
  \bibfield  {author} {\bibinfo {author} {\bibfnamefont {V.}~\bibnamefont
  {Pato\v{c}ka}}, \bibinfo {author} {\bibfnamefont {N.}~\bibnamefont {Tosi}},\
  and\ \bibinfo {author} {\bibfnamefont {E.}~\bibnamefont {Calzavarini}},\
  }\bibfield  {title} {\bibinfo {title} {Residence time of inertial particles
  in {3D} thermal convection: implications for magma reservoirs},\ }\href
  {https://doi.org/https://doi.org/10.1016/j.epsl.2022.117622} {\bibfield
  {journal} {\bibinfo  {journal} {Earth Planet. Sci. Lett.}\ }\textbf {\bibinfo
  {volume} {591}},\ \bibinfo {pages} {117622} (\bibinfo {year}
  {2022})}\BibitemShut {NoStop}%
\bibitem [{\citenamefont {Denzel}\ \emph {et~al.}(2023)\citenamefont {Denzel},
  \citenamefont {Bragg},\ and\ \citenamefont {Richter}}]{DenzelPRF2023}%
  \BibitemOpen
  \bibfield  {author} {\bibinfo {author} {\bibfnamefont {C.~J.}\ \bibnamefont
  {Denzel}}, \bibinfo {author} {\bibfnamefont {A.~D.}\ \bibnamefont {Bragg}},\
  and\ \bibinfo {author} {\bibfnamefont {D.~H.}\ \bibnamefont {Richter}},\
  }\bibfield  {title} {\bibinfo {title} {Stochastic model for the residence
  time of solid particles in turbulent {R}ayleigh-{B}\'enard flow},\ }\href
  {https://doi.org/10.1103/PhysRevFluids.8.024307} {\bibfield  {journal}
  {\bibinfo  {journal} {Phys. Rev. Fluids}\ }\textbf {\bibinfo {volume} {8}},\
  \bibinfo {pages} {024307} (\bibinfo {year} {2023})}\BibitemShut {NoStop}%
\bibitem [{\citenamefont {Xu}\ \emph {et~al.}(2024)\citenamefont {Xu},
  \citenamefont {Xu},\ and\ \citenamefont {Xi}}]{Xu_Xu_Xi_2024}%
  \BibitemOpen
  \bibfield  {author} {\bibinfo {author} {\bibfnamefont {A.}~\bibnamefont
  {Xu}}, \bibinfo {author} {\bibfnamefont {B.-R.}\ \bibnamefont {Xu}},\ and\
  \bibinfo {author} {\bibfnamefont {H.-D.}\ \bibnamefont {Xi}},\ }\bibfield
  {title} {\bibinfo {title} {Particle transport and deposition in wall-sheared
  thermal turbulence},\ }\href {https://doi.org/10.1017/jfm.2024.936}
  {\bibfield  {journal} {\bibinfo  {journal} {J. Fluid Mech.}\ }\textbf
  {\bibinfo {volume} {999}},\ \bibinfo {pages} {A15} (\bibinfo {year}
  {2024})}\BibitemShut {NoStop}%
\bibitem [{\citenamefont {Lakkaraju}\ \emph {et~al.}(2013)\citenamefont
  {Lakkaraju}, \citenamefont {Stevens}, \citenamefont {Oresta}, \citenamefont
  {Verzicco}, \citenamefont {Lohse},\ and\ \citenamefont
  {Prosperetti}}]{LakkarajuPNAS2013}%
  \BibitemOpen
  \bibfield  {author} {\bibinfo {author} {\bibfnamefont {R.}~\bibnamefont
  {Lakkaraju}}, \bibinfo {author} {\bibfnamefont {R.~J. A.~M.}\ \bibnamefont
  {Stevens}}, \bibinfo {author} {\bibfnamefont {P.}~\bibnamefont {Oresta}},
  \bibinfo {author} {\bibfnamefont {R.}~\bibnamefont {Verzicco}}, \bibinfo
  {author} {\bibfnamefont {D.}~\bibnamefont {Lohse}},\ and\ \bibinfo {author}
  {\bibfnamefont {A.}~\bibnamefont {Prosperetti}},\ }\bibfield  {title}
  {\bibinfo {title} {Heat transport in bubbling turbulent convection},\ }\href
  {https://doi.org/10.1073/pnas.1217546110} {\bibfield  {journal} {\bibinfo
  {journal} {Proc. Natl. Acad. Sci. U.S.A.}\ }\textbf {\bibinfo {volume}
  {110}},\ \bibinfo {pages} {9237} (\bibinfo {year} {2013})}\BibitemShut
  {NoStop}%
\bibitem [{\citenamefont {Oresta}\ and\ \citenamefont
  {Prosperetti}(2013)}]{oresta2013effects}%
  \BibitemOpen
  \bibfield  {author} {\bibinfo {author} {\bibfnamefont {P.}~\bibnamefont
  {Oresta}}\ and\ \bibinfo {author} {\bibfnamefont {A.}~\bibnamefont
  {Prosperetti}},\ }\bibfield  {title} {\bibinfo {title} {Effects of particle
  settling on {R}ayleigh-{B}\'enard convection},\ }\href
  {https://doi.org/10.1103/PhysRevE.87.063014} {\bibfield  {journal} {\bibinfo
  {journal} {Phys. Rev. E}\ }\textbf {\bibinfo {volume} {87}},\ \bibinfo
  {pages} {063014} (\bibinfo {year} {2013})}\BibitemShut {NoStop}%
\bibitem [{\citenamefont {Oresta}\ \emph {et~al.}(2014)\citenamefont {Oresta},
  \citenamefont {Fornarelli},\ and\ \citenamefont
  {Prosperetti}}]{oresta2014multiphase}%
  \BibitemOpen
  \bibfield  {author} {\bibinfo {author} {\bibfnamefont {P.}~\bibnamefont
  {Oresta}}, \bibinfo {author} {\bibfnamefont {F.}~\bibnamefont {Fornarelli}},\
  and\ \bibinfo {author} {\bibfnamefont {A.}~\bibnamefont {Prosperetti}},\
  }\bibfield  {title} {\bibinfo {title} {Multiphase {R}ayleigh-{B}\'enard
  convection},\ }\href {https://doi.org/https://doi.org/10.1299/mer.2014fe0003}
  {\bibfield  {journal} {\bibinfo  {journal} {Mech. Eng. Rev.}\ }\textbf
  {\bibinfo {volume} {1}},\ \bibinfo {pages} {FE0003} (\bibinfo {year}
  {2014})}\BibitemShut {NoStop}%
\bibitem [{\citenamefont {Gereltbyamba}\ and\ \citenamefont
  {Lee}(2019)}]{gereltbyamba2019flow}%
  \BibitemOpen
  \bibfield  {author} {\bibinfo {author} {\bibfnamefont {B.}~\bibnamefont
  {Gereltbyamba}}\ and\ \bibinfo {author} {\bibfnamefont {C.}~\bibnamefont
  {Lee}},\ }\bibfield  {title} {\bibinfo {title} {Flow modification by inertial
  particles in a differentially heated cubic cavity},\ }\href
  {https://doi.org/https://doi.org/10.1016/j.ijheatfluidflow.2019.108445}
  {\bibfield  {journal} {\bibinfo  {journal} {Int. J. Heat Fluid Flow}\
  }\textbf {\bibinfo {volume} {79}},\ \bibinfo {pages} {108445} (\bibinfo
  {year} {2019})}\BibitemShut {NoStop}%
\bibitem [{\citenamefont {Demou}\ \emph {et~al.}(2022)\citenamefont {Demou},
  \citenamefont {Ardekani}, \citenamefont {Mirbod},\ and\ \citenamefont
  {Brandt}}]{demou2022turbulent}%
  \BibitemOpen
  \bibfield  {author} {\bibinfo {author} {\bibfnamefont {A.~D.}\ \bibnamefont
  {Demou}}, \bibinfo {author} {\bibfnamefont {M.~N.}\ \bibnamefont {Ardekani}},
  \bibinfo {author} {\bibfnamefont {P.}~\bibnamefont {Mirbod}},\ and\ \bibinfo
  {author} {\bibfnamefont {L.}~\bibnamefont {Brandt}},\ }\bibfield  {title}
  {\bibinfo {title} {Turbulent {R}ayleigh-{B}\'enard convection in
  non-colloidal suspensions},\ }\href {https://doi.org/10.1017/jfm.2022.534}
  {\bibfield  {journal} {\bibinfo  {journal} {J. Fluid Mech.}\ }\textbf
  {\bibinfo {volume} {945}},\ \bibinfo {pages} {A6} (\bibinfo {year}
  {2022})}\BibitemShut {NoStop}%
\bibitem [{\citenamefont {Du}\ and\ \citenamefont
  {Yang}(2023)}]{Yuhang&YangPF2023}%
  \BibitemOpen
  \bibfield  {author} {\bibinfo {author} {\bibfnamefont {Y.}~\bibnamefont
  {Du}}\ and\ \bibinfo {author} {\bibfnamefont {Y.}~\bibnamefont {Yang}},\
  }\bibfield  {title} {\bibinfo {title} {Effects of the gravitational force on
  the convection turbulence driven by heat-releasing point particles},\ }\href
  {https://doi.org/10.1063/5.0158055} {\bibfield  {journal} {\bibinfo
  {journal} {Phys. Fluids}\ }\textbf {\bibinfo {volume} {35}},\ \bibinfo
  {pages} {075142} (\bibinfo {year} {2023})}\BibitemShut {NoStop}%
\bibitem [{\citenamefont {Sun}\ \emph {et~al.}(2024)\citenamefont {Sun},
  \citenamefont {Wan},\ and\ \citenamefont {Sun}}]{SunWanSunPF2024}%
  \BibitemOpen
  \bibfield  {author} {\bibinfo {author} {\bibfnamefont {D.-F.}\ \bibnamefont
  {Sun}}, \bibinfo {author} {\bibfnamefont {Z.-H.}\ \bibnamefont {Wan}},\ and\
  \bibinfo {author} {\bibfnamefont {D.-J.}\ \bibnamefont {Sun}},\ }\bibfield
  {title} {\bibinfo {title} {Modulation of {R}ayleigh–{B}\'enard convection
  with a large temperature difference by inertial nonisothermal particles},\
  }\href {https://doi.org/10.1063/5.0185314} {\bibfield  {journal} {\bibinfo
  {journal} {Phys. Fluids}\ }\textbf {\bibinfo {volume} {36}},\ \bibinfo
  {pages} {017135} (\bibinfo {year} {2024})}\BibitemShut {NoStop}%
\bibitem [{\citenamefont {Zamansky}\ \emph {et~al.}(2016)\citenamefont
  {Zamansky}, \citenamefont {Coletti}, \citenamefont {Massot},\ and\
  \citenamefont {Mani}}]{zamansky2016turbulent}%
  \BibitemOpen
  \bibfield  {author} {\bibinfo {author} {\bibfnamefont {R.}~\bibnamefont
  {Zamansky}}, \bibinfo {author} {\bibfnamefont {F.}~\bibnamefont {Coletti}},
  \bibinfo {author} {\bibfnamefont {M.}~\bibnamefont {Massot}},\ and\ \bibinfo
  {author} {\bibfnamefont {A.}~\bibnamefont {Mani}},\ }\bibfield  {title}
  {\bibinfo {title} {Turbulent thermal convection driven by heated inertial
  particles},\ }\href {https://doi.org/10.1017/jfm.2016.630} {\bibfield
  {journal} {\bibinfo  {journal} {J. Fluid Mech.}\ }\textbf {\bibinfo {volume}
  {809}},\ \bibinfo {pages} {390} (\bibinfo {year} {2016})}\BibitemShut
  {NoStop}%
\bibitem [{\citenamefont {Gu}\ \emph {et~al.}(2018)\citenamefont {Gu},
  \citenamefont {Takeuchi},\ and\ \citenamefont {Kajishima}}]{gu2018influence}%
  \BibitemOpen
  \bibfield  {author} {\bibinfo {author} {\bibfnamefont {J.}~\bibnamefont
  {Gu}}, \bibinfo {author} {\bibfnamefont {S.}~\bibnamefont {Takeuchi}},\ and\
  \bibinfo {author} {\bibfnamefont {T.}~\bibnamefont {Kajishima}},\ }\bibfield
  {title} {\bibinfo {title} {Influence of {R}ayleigh number and solid volume
  fraction in particle-dispersed natural convection},\ }\href
  {https://doi.org/10.1016/j.ijheatmasstransfer.2017.12.020} {\bibfield
  {journal} {\bibinfo  {journal} {Int. J. Heat Mass Transf.}\ }\textbf
  {\bibinfo {volume} {120}},\ \bibinfo {pages} {250} (\bibinfo {year}
  {2018})}\BibitemShut {NoStop}%
\bibitem [{\citenamefont {Gereluchi}\ \emph {et~al.}(2019)\citenamefont
  {Gereluchi}, \citenamefont {Miyamori}, \citenamefont {Gu},\ and\
  \citenamefont {Kajishima}}]{takeuchi2019flow}%
  \BibitemOpen
  \bibfield  {author} {\bibinfo {author} {\bibfnamefont {S.}~\bibnamefont
  {Gereluchi}}, \bibinfo {author} {\bibfnamefont {Y.}~\bibnamefont {Miyamori}},
  \bibinfo {author} {\bibfnamefont {J.}~\bibnamefont {Gu}},\ and\ \bibinfo
  {author} {\bibfnamefont {T.}~\bibnamefont {Kajishima}},\ }\bibfield  {title}
  {\bibinfo {title} {Flow reversals in particle-dispersed natural convection in
  a two-dimensional enclosed square domain},\ }\href
  {https://doi.org/10.1103/PhysRevFluids.4.084304} {\bibfield  {journal}
  {\bibinfo  {journal} {Phys. Rev. Fluids}\ }\textbf {\bibinfo {volume} {4}},\
  \bibinfo {pages} {084304} (\bibinfo {year} {2019})}\BibitemShut {NoStop}%
\bibitem [{\citenamefont {Chen}\ and\ \citenamefont
  {Prosperetti}(2024)}]{ChenPropseretti2024}%
  \BibitemOpen
  \bibfield  {author} {\bibinfo {author} {\bibfnamefont {X.}~\bibnamefont
  {Chen}}\ and\ \bibinfo {author} {\bibfnamefont {A.}~\bibnamefont
  {Prosperetti}},\ }\bibfield  {title} {\bibinfo {title} {Particle-resolved
  multiphase {R}ayleigh-{B}\'enard convection},\ }\href
  {https://doi.org/https://doi.org/10.1103/PhysRevFluids.9.054301} {\bibfield
  {journal} {\bibinfo  {journal} {Phys. Rev. Fluids}\ }\textbf {\bibinfo
  {volume} {9}},\ \bibinfo {pages} {054301} (\bibinfo {year}
  {2024})}\BibitemShut {NoStop}%
\bibitem [{\citenamefont {Prakhar}\ and\ \citenamefont
  {Prosperetti}(2021)}]{prakhar2021linear}%
  \BibitemOpen
  \bibfield  {author} {\bibinfo {author} {\bibfnamefont {S.}~\bibnamefont
  {Prakhar}}\ and\ \bibinfo {author} {\bibfnamefont {A.}~\bibnamefont
  {Prosperetti}},\ }\bibfield  {title} {\bibinfo {title} {Linear theory of
  particulate {R}ayleigh-{B}\'enard instability},\ }\href
  {https://doi.org/10.1103/PhysRevFluids.6.083901} {\bibfield  {journal}
  {\bibinfo  {journal} {Phys. Rev. Fluids}\ }\textbf {\bibinfo {volume} {6}},\
  \bibinfo {pages} {083901} (\bibinfo {year} {2021})}\BibitemShut {NoStop}%
\bibitem [{\citenamefont {Climent}\ and\ \citenamefont
  {Magnaudet}(1999)}]{ClimentPRL1999}%
  \BibitemOpen
  \bibfield  {author} {\bibinfo {author} {\bibfnamefont {E.}~\bibnamefont
  {Climent}}\ and\ \bibinfo {author} {\bibfnamefont {J.}~\bibnamefont
  {Magnaudet}},\ }\bibfield  {title} {\bibinfo {title} {Large-scale simulations
  of bubble-induced convection in a liquid layer},\ }\href
  {https://doi.org/10.1103/PhysRevLett.82.4827} {\bibfield  {journal} {\bibinfo
   {journal} {Phys. Rev. Lett.}\ }\textbf {\bibinfo {volume} {82}},\ \bibinfo
  {pages} {4827} (\bibinfo {year} {1999})}\BibitemShut {NoStop}%
\bibitem [{\citenamefont {Mathai}\ \emph
  {et~al.}(2020{\natexlab{b}})\citenamefont {Mathai}, \citenamefont {Lohse},\
  and\ \citenamefont {Sun}}]{MathaiARFM2020}%
  \BibitemOpen
  \bibfield  {author} {\bibinfo {author} {\bibfnamefont {V.}~\bibnamefont
  {Mathai}}, \bibinfo {author} {\bibfnamefont {D.}~\bibnamefont {Lohse}},\ and\
  \bibinfo {author} {\bibfnamefont {C.}~\bibnamefont {Sun}},\ }\bibfield
  {title} {\bibinfo {title} {Bubbly and buoyant particle–laden turbulent
  flows},\ }\href
  {https://doi.org/https://doi.org/10.1146/annurev-conmatphys-031119-050637}
  {\bibfield  {journal} {\bibinfo  {journal} {Annual Review of Condensed Matter
  Physics}\ }\textbf {\bibinfo {volume} {11}},\ \bibinfo {pages} {529}
  (\bibinfo {year} {2020}{\natexlab{b}})}\BibitemShut {NoStop}%
\bibitem [{\citenamefont {Barletta}(2023)}]{Barletta-anomalous}%
  \BibitemOpen
  \bibfield  {author} {\bibinfo {author} {\bibfnamefont {A.}~\bibnamefont
  {Barletta}},\ }\bibfield  {title} {\bibinfo {title} {{R}ayleigh–{B}\'enard
  instability in a horizontal porous layer with anomalous diffusion},\ }\href
  {https://doi.org/10.1063/5.0174432} {\bibfield  {journal} {\bibinfo
  {journal} {Phys. Fluids}\ }\textbf {\bibinfo {volume} {35}},\ \bibinfo
  {pages} {104114} (\bibinfo {year} {2023})}\BibitemShut {NoStop}%
\bibitem [{\citenamefont {Maxey}\ and\ \citenamefont
  {Riley}(1983)}]{maxey-riley-1983}%
  \BibitemOpen
  \bibfield  {author} {\bibinfo {author} {\bibfnamefont {M.~R.}\ \bibnamefont
  {Maxey}}\ and\ \bibinfo {author} {\bibfnamefont {J.~J.}\ \bibnamefont
  {Riley}},\ }\bibfield  {title} {\bibinfo {title} {Equation of motion for a
  small rigid sphere in a nonuniform flow},\ }\href
  {https://doi.org/https://doi.org/10.1063/1.864230} {\bibfield  {journal}
  {\bibinfo  {journal} {Phys. Fluids}\ }\textbf {\bibinfo {volume} {26}},\
  \bibinfo {pages} {883} (\bibinfo {year} {1983})}\BibitemShut {NoStop}%
\bibitem [{\citenamefont {Gatignol}(1983)}]{gatignol-1983}%
  \BibitemOpen
  \bibfield  {author} {\bibinfo {author} {\bibfnamefont {R.}~\bibnamefont
  {Gatignol}},\ }\bibfield  {title} {\bibinfo {title} {The {F}ax\'en formulae
  for a rigid particle in an unsteady non-uniform stokes flow},\ }\href@noop {}
  {\bibfield  {journal} {\bibinfo  {journal} {J. Mécanique th\'eorique et
  appliqu\'ee}\ }\textbf {\bibinfo {volume} {1}},\ \bibinfo {pages} {143}
  (\bibinfo {year} {1983})}\BibitemShut {NoStop}%
\bibitem [{\citenamefont {Nakamura}\ \emph {et~al.}(2020)\citenamefont
  {Nakamura}, \citenamefont {Yoshikawa}, \citenamefont {Tasaka},\ and\
  \citenamefont {Murai}}]{PhysRevE.102.053102}%
  \BibitemOpen
  \bibfield  {author} {\bibinfo {author} {\bibfnamefont {K.}~\bibnamefont
  {Nakamura}}, \bibinfo {author} {\bibfnamefont {H.~N.}\ \bibnamefont
  {Yoshikawa}}, \bibinfo {author} {\bibfnamefont {Y.}~\bibnamefont {Tasaka}},\
  and\ \bibinfo {author} {\bibfnamefont {Y.}~\bibnamefont {Murai}},\ }\bibfield
   {title} {\bibinfo {title} {Linear stability analysis of bubble-induced
  convection in a horizontal liquid layer},\ }\href
  {https://doi.org/10.1103/PhysRevE.102.053102} {\bibfield  {journal} {\bibinfo
   {journal} {Phys. Rev. E}\ }\textbf {\bibinfo {volume} {102}},\ \bibinfo
  {pages} {053102} (\bibinfo {year} {2020})}\BibitemShut {NoStop}%
\bibitem [{\citenamefont {Ali~Amar}\ \emph {et~al.}(2022)\citenamefont
  {Ali~Amar}, \citenamefont {Hirata},\ and\ \citenamefont
  {Ouarzazi}}]{ali2022soret}%
  \BibitemOpen
  \bibfield  {author} {\bibinfo {author} {\bibfnamefont {K.}~\bibnamefont
  {Ali~Amar}}, \bibinfo {author} {\bibfnamefont {S.~C.}\ \bibnamefont
  {Hirata}},\ and\ \bibinfo {author} {\bibfnamefont {M.~N.}\ \bibnamefont
  {Ouarzazi}},\ }\bibfield  {title} {\bibinfo {title} {{S}oret effect on the
  onset of viscous dissipation thermal instability for {P}oiseuille flows in
  binary mixtures},\ }\href {https://doi.org/https://doi.org/10.1063/5.0115663}
  {\bibfield  {journal} {\bibinfo  {journal} {Phys. Fluids}\ }\textbf {\bibinfo
  {volume} {34}},\ \bibinfo {pages} {114101} (\bibinfo {year}
  {2022})}\BibitemShut {NoStop}%
\bibitem [{\citenamefont {Hirata}\ \emph {et~al.}(2015)\citenamefont {Hirata},
  \citenamefont {Alves}, \citenamefont {Delenda},\ and\ \citenamefont
  {Ouarzazi}}]{hirata2015convective}%
  \BibitemOpen
  \bibfield  {author} {\bibinfo {author} {\bibfnamefont {S.~C.}\ \bibnamefont
  {Hirata}}, \bibinfo {author} {\bibfnamefont {L.~S. d.~B.}\ \bibnamefont
  {Alves}}, \bibinfo {author} {\bibfnamefont {N.}~\bibnamefont {Delenda}},\
  and\ \bibinfo {author} {\bibfnamefont {M.~N.}\ \bibnamefont {Ouarzazi}},\
  }\bibfield  {title} {\bibinfo {title} {Convective and absolute instabilities
  in {R}ayleigh–{B}\'enard–{P}oiseuille mixed convection for viscoelastic
  fluids},\ }\href {https://doi.org/10.1017/jfm.2014.721} {\bibfield  {journal}
  {\bibinfo  {journal} {J. Fluid Mech.}\ }\textbf {\bibinfo {volume} {765}},\
  \bibinfo {pages} {167} (\bibinfo {year} {2015})}\BibitemShut {NoStop}%
\bibitem [{\citenamefont {Alves}\ \emph {et~al.}(2019)\citenamefont {Alves},
  \citenamefont {Hirata}, \citenamefont {Schuabb},\ and\ \citenamefont
  {Barletta}}]{Alvesetal}%
  \BibitemOpen
  \bibfield  {author} {\bibinfo {author} {\bibfnamefont {L.~S.~B.}\
  \bibnamefont {Alves}}, \bibinfo {author} {\bibfnamefont {S.~C.}\ \bibnamefont
  {Hirata}}, \bibinfo {author} {\bibfnamefont {M.}~\bibnamefont {Schuabb}},\
  and\ \bibinfo {author} {\bibfnamefont {A.}~\bibnamefont {Barletta}},\
  }\bibfield  {title} {\bibinfo {title} {Identifying linear absolute
  instabilities from differential eigenvalue problems using sensitivity
  analysis},\ }\href {https://doi.org/10.1017/jfm.2019.275} {\bibfield
  {journal} {\bibinfo  {journal} {J. Fluid Mech.}\ ,\ \bibinfo {pages} {941}}
  (\bibinfo {year} {2019})}\BibitemShut {NoStop}%
\bibitem [{\citenamefont {Mazzitelli}\ and\ \citenamefont
  {Lohse}(2009)}]{MazzitelliPRE2009}%
  \BibitemOpen
  \bibfield  {author} {\bibinfo {author} {\bibfnamefont {I.~M.}\ \bibnamefont
  {Mazzitelli}}\ and\ \bibinfo {author} {\bibfnamefont {D.}~\bibnamefont
  {Lohse}},\ }\bibfield  {title} {\bibinfo {title} {Evolution of energy in flow
  driven by rising bubbles},\ }\href
  {https://doi.org/10.1103/PhysRevE.79.066317} {\bibfield  {journal} {\bibinfo
  {journal} {Phys. Rev. E}\ }\textbf {\bibinfo {volume} {79}},\ \bibinfo
  {pages} {066317} (\bibinfo {year} {2009})}\BibitemShut {NoStop}%
\bibitem [{\citenamefont {Nakamura}\ \emph {et~al.}(2021)\citenamefont
  {Nakamura}, \citenamefont {Yoshikawa}, \citenamefont {Tasaka},\ and\
  \citenamefont {Murai}}]{Nakamura_Yoshikawa_Tasaka_Murai_2021}%
  \BibitemOpen
  \bibfield  {author} {\bibinfo {author} {\bibfnamefont {K.}~\bibnamefont
  {Nakamura}}, \bibinfo {author} {\bibfnamefont {H.~N.}\ \bibnamefont
  {Yoshikawa}}, \bibinfo {author} {\bibfnamefont {Y.}~\bibnamefont {Tasaka}},\
  and\ \bibinfo {author} {\bibfnamefont {Y.}~\bibnamefont {Murai}},\ }\bibfield
   {title} {\bibinfo {title} {Bifurcation analysis of bubble-induced convection
  in a horizontal liquid layer: role of forces on bubbles},\ }\href
  {https://doi.org/10.1017/jfm.2021.601} {\bibfield  {journal} {\bibinfo
  {journal} {J. Fluid Mech.}\ }\textbf {\bibinfo {volume} {923}},\ \bibinfo
  {pages} {R4} (\bibinfo {year} {2021})}\BibitemShut {NoStop}%
\end{thebibliography}
%

\end{document}